\pdfoutput=1 
\documentclass[a4paper, 11pt]{article}
\PassOptionsToPackage{table}{xcolor}

\usepackage{jheppub} 

\usepackage[T1]{fontenc} 
\usepackage{tangocolors}
\usepackage{tikz}
\usetikzlibrary{arrows, decorations,backgrounds, patterns}
\usetikzlibrary{decorations.pathreplacing ,decorations.markings}
\usepackage{amssymb}
\usetikzlibrary{decorations.pathmorphing,backgrounds,shapes,arrows,shadows}
\usetikzlibrary{patterns.meta}
\usetikzlibrary{patterns,decorations.pathmorphing}
\tikzset{
    snake it/.style={decorate, decoration=snake}
}
\usepackage{pgfplots}
\pgfplotsset{compat=1.11}
\usepgfplotslibrary{fillbetween}
\usetikzlibrary{intersections}
\pgfdeclarelayer{bg}
\pgfsetlayers{bg,main}
\tikzset{zigzag/.style={decorate,decoration=zigzag}}
\tikzset{snake it/.style={decorate, decoration=snake}}
\makeatletter
\def\@hex@@Hex#1%
 {\if a#1A\else \if b#1B\else \if c#1C\else \if d#1D\else
  \if e#1E\else \if f#1F\else #1\fi\fi\fi\fi\fi\fi \@hex@Hex}
\makeatother

\usepackage[all]{xy}
\usepackage[percent]{overpic}
\usepackage{slashed}
\usepackage{wrapfig}
\usepackage{tabu}
\usepackage{diagbox}
\usepackage{mathrsfs,amsmath,amssymb,amsthm,amsfonts,tikz,graphicx,accents,hyperref, color}
\usepackage{dsfont,epiolmec, latexsym, stmaryrd, comment}
\usepackage{slashed,ccaption}
\usepackage{mathrsfs, calligra}
\usepackage{leftidx}
\usepackage{import}
\usepackage{multirow}
\usepackage{amsfonts}
\usepackage{pifont}
\usepackage{tabularx}
\usepackage{cancel}
\usepackage[utf8]{inputenc}
\usetikzlibrary{intersections,calc}
\usepackage{ifthen}
\usepackage{amsmath}
\usepackage{cancel}
\usepackage{caption} 
\usepackage{subcaption}

\usetikzlibrary{patterns.meta}
\usepackage{array}
\hypersetup{ linktoc=all,
    colorlinks, linkcolor={palatinateblue},
    citecolor={brightpink}, urlcolor={amaranth}
}

\graphicspath{{Images/}}

\renewcommand{\d}[1]{\ensuremath{\operatorname{d}\!{#1}}}

\def\sideremark#1{\ifvmode\leavevmode\fi\vadjust{\vbox to0pt{\vss
 \hbox to 0pt{\hskip\hsize\hskip1em
 \vbox{\hsize2cm\tiny\raggedright\pretolerance10000
 \noindent #1\hfill}\hss}\vbox to8pt{\vfil}\vss}}}%
\DeclareSymbolFont{extraup}{U}{zavm}{m}{n}
\DeclareMathSymbol{\varheart}{\mathalpha}{extraup}{86}
\DeclareMathSymbol{\vardiamond}{\mathalpha}{extraup}{87}
\makeatletter
\renewcommand*{\@fnsymbol}[1]{\ensuremath{\ifcase#1\or \clubsuit \or \vardiamond \or \varheart\or
    \spadesuit\or \mathparagraph\or \|\or **\or \dagger\dagger
    \or \ddagger\ddagger \else\@ctrerr\fi}}
\makeatother

\definecolor{rosy}{RGB}{230,235,252}
\definecolor{myframetitle}{RGB}{90,89,170}
\definecolor{myblocktitle}{RGB}{140,185,249}
\definecolor{mytitle}{RGB}{10,80,26}

\definecolor{darkgreen}{RGB}{27,130,45}
\definecolor{darkblue}{rgb}{0,0,0.3}
\definecolor{darkred}{rgb}{0.7,0,0}

\definecolor{light gray}{RGB}{220,220,220}
\definecolor{dark purple}{RGB}{108,0,217}
\definecolor{pink}{RGB}{190,20,100}
\definecolor{orang}{RGB}{193,63,0}
\definecolor{green}{RGB}{11,98,17}
\definecolor{darkpink}{RGB}{153,0,76}
\definecolor{bluegreen}{RGB}{0,102,102}
\definecolor{greenlagan}{RGB}{0,102,0}
\definecolor{redgreen}{RGB}{102,102,0}
\definecolor{Redgreen}{RGB}{153,76,0}
\definecolor{vividviolet}{rgb}{0.62, 0.0, 1.0}
\definecolor{amaranth}{rgb}{0.9, 0.17, 0.31}
\definecolor{palatinateblue}{rgb}{0.15, 0.23, 0.89}
\definecolor{brightpink}{rgb}{1.0, 0.0, 0.5}
\definecolor{cornflowerblue}{rgb}{0.39, 0.58, 0.93}
\definecolor{deepcarminepink}{rgb}{0.94, 0.19, 0.22}
\definecolor{radicalred}{rgb}{1.0, 0.21, 0.37}

\newcommand{\bTh}{\boldsymbol{\Theta }}

\newcommand{\bO}{\boldsymbol{\Omega}}

\newcommand{\cc}{{\mathcal C }}

\DeclareFontFamily{OT1}{rsfs}{}

\DeclareFontShape{OT1}{rsfs}{m}{n}{ <-7> rsfs5 <7-10> rsfs7 <10->rsfs10}{} 

\DeclareMathAlphabet{\mycal}{OT1}{rsfs}{m}{n}

\newcommand{\be}{\begin{equation}}
\newcommand{\ee}{\end{equation}}
\newcommand{\bea}{\begin{eqnarray}}
\newcommand{\eea}{\end{eqnarray}}
\makeatletter \@addtoreset{equation}{section}

\begin{document}

\newcommand{\mytitle}{\begin{center}{\LARGE{\textbf{Hydro \& Thermo Dynamics at Causal Boundaries, }}} \\ 
{\LARGE{\textbf{Examples in  $3d$ Gravity}}}
\end{center}}

\title{{\mytitle}}

\author[a,b]{H.~Adami}
\author[c]{, A.~Parvizi}
\author[c]{, M.M.~Sheikh-Jabbari}
\author[c,d]{, V.~Taghiloo}
\author[b]{, H.~Yavartanoo}
\affiliation{$^a$ Yau Mathematical Sciences Center, Tsinghua University, Beijing 100084, China}
\affiliation{$^b$ Beijing Institute of Mathematical Sciences and Applications (BIMSA), \\ Huairou District, Beijing 101408, P. R. China}
\affiliation{$^c$ School of Physics, Institute for Research in Fundamental
Sciences (IPM),\\ P.O.Box 19395-5531, Tehran, Iran}
\affiliation{$^d$ Department of Physics, Institute for Advanced Studies in Basic Sciences (IASBS),\\ 
P.O. Box 45137-66731, Zanjan, Iran}
\emailAdd{hamed.adami@bimsa.cn, a.parvizi@ipm.ir,
jabbari@theory.ipm.ac.ir, v.taghiloo@iasbs.ac.ir,  yavar@bimsa.cn 
}

\abstract{We study 3{-}dimensional gravity on a spacetime bounded by a generic 2{-}dimensional causal surface. We review the  solution phase space  specified by 4 generic functions over the causal boundary,  construct the symplectic form over the solution space and the 4 boundary charges and their algebra. The boundary charges label boundary degrees of freedom. Three of these charges extend and generalize the  Brown-York charges to the generic causal boundary,  are  canonical conjugates of  boundary metric components and naturally give rise to a fluid description at the causal boundary. Moreover, we show that the  boundary charges besides the causal boundary hydrodynamic description, also admit a thermodynamic description with a natural (geometric) causal boundary temperature and angular velocity. When the causal boundary is the asymptotic boundary of the $3d$  AdS or flat space, the hydrodynamic description respectively recovers an extension of the known conformal or conformal-Carrollian asymptotic hydrodynamics. When the causal boundary is a generic null surface, we recover the null surface thermodynamics of \cite{Adami:2021kvx} which is an extension of the usual black hole thermodynamics description. 
 
}
\maketitle
\section{Introduction}\label{sec:Intro}

It is now a textbook knowledge that typical physical systems in  continuum coarse-grained limit exhibit a fluid description governed by hydrodynamic equations. Landau's seminal work \cite{Landau:1953gs} extended this to particle physics, proposing that the collective behavior of colliding high{-}energy and high{-}density particles can be described by hydrodynamic equations.
In more recent reincarnations, Landau's picture has been reformulated within the AdS/CFT setup \cite{Maldacena:1997re, Aharony:1999ti}, in fluid/gravity correspondence \cite{Bhattacharyya:2007vjd, Hubeny:2011hd, Bhattacharyya:2008mz}. In the fluid/gravity correspondence, the hydrodynamic limit of a (strongly interacting) conformal field theory which is residing at the boundary of dual asymptotically AdS geometry is related to a gravity dual picture. 

The fluid/gravity correspondence for $1+1$ dimensional fluids both within AdS3/CFT2 \cite{Banerjee:2013qha, Bhattacharyya:2008mz, Haack:2008cp, Campoleoni:2018ltl, Ciambelli:2020ftk, Ciambelli:2020eba} and in $3d$ flat space holography contexts \cite{Penna:2017vms, Penna:2017bdn, Ciambelli:2018xat, Ciambelli:2018wre, Ruzziconi:2019pzd, Freidel:2022vjq} is a well{-}developed topic. The $3d$ gravity setting is a particularly interesting playground as there are no bulk graviton modes and we have a much better computational handle on both sides of the duality see e.g. \cite{Campoleoni:2018ltl, Ciambelli:2020eba}. For the AdS$_3$ case, the field theory side is a $1+1$ dimensional conformal field theory (CFT)  at the Brown-Henneaux central charge \cite{Brown:1986nw} which admits a relativistic conformal fluid description. For the asymptotic flat case, the dual fluid system is a conformal Carrollian fluid residing at the asymptotic null infinity ${\cal I}$. See the above references and references therein for more discussions.

In this work, we reconsider the $1+1$ dimensional fluids with a $3d$ dual gravity description and extend the fluid/gravity description in some different ways. We establish that the fluid/gravity description for both AdS$_3$ or flat space cases can be extended to the bulk, where the boundary of spacetime is an arbitrary $2d$ causal (timelike or null) surface. To this end, one needs to formulate $3d$ gravity in spacetime with an arbitrary causal boundary. This problem has been explored and well formulated in \cite{Adami:2020ugu, Adami:2022ktn, Geiller:2021vpg}. The system is in general described by 4 boundary fields (boundary degrees of freedom) for a time-like boundary and by 3 boundary degrees of freedom for the null case.  The description developed in \cite{Adami:2020ugu, Adami:2022ktn} compared to the other works in the literature cited above has the advantage that it is not relying on asymptotic expansion or falloff behavior of metric components in a particular coordinate system. 

Building on the results of \cite{Adami:2020ugu, Adami:2022ktn}, we establish that boundary degrees of freedom for a generic timelike boundary on flat or AdS gravity cases admit a natural  relativistic, but non-conformal hydrodynamic description. This hydrodynamic description in general involves all 4 boundary degrees of freedom.  For the cases where the boundary is the asymptotic boundary of the spacetime, however, we show that the fluid description (a relativistic conformal fluid for the AdS$_3$ case and a conformal Carrollian fluid for the $3d$ flat space) naturally arises as the description of a part of the boundary degrees of freedom. Explicitly, we show that 2 of the 4 (or 3) boundary degrees of freedom appear in the fluid description for the timelike (or null) boundary cases. 
We show that the asymptotic hydrodynamic description can be obtained as a limit of our results for general boundaries, when the boundary is taken to infinity and that in this limit one retains the conformal (for AdS) and conformal Carrollian (for flat) fluid description which had been extensively studied in the literature. Finally, for a generic null boundary case, for both flat or AdS cases, we show that there is a hydrodynamic description, as well as an equivalent thermodynamic description, worked out in \cite{Adami:2021kvx, Sheikh-Jabbari:2022mqi}. 

Besides the hydrodynamic descriptions, we also revisit possible thermodynamic description at a generic causal boundary. In \cite{Adami:2021kvx} and as an extension of the usual black hole thermodynamic description, see Chapter 5 of \cite{Grumiller:2022qhx} for a review, it has been argued that the boundary degrees of freedom at a generic null boundary (not necessarily a black hole horizon) are governed by a generalization of the first law of thermodynamics (which is a local equation at the null boundary). In this work, we show that the null surface thermodynamics can be extended to a generic causal boundary.  

\paragraph{Outline of the paper.}  In section \ref{sec2:soln-space-review} we review the construction of the solution phase space presented in \cite{Adami:2022ktn} and introduce and fix our conventions and notations. We compute conserved charges and their corresponding algebra and set the stage for our hydrodynamic and thermodynamic analyses of the following sections. In section \ref{sec:HCB-AdS} we give a hydrodynamics description of AdS$_3$ or $3d$ flat space bounded by a generic timelike boundary. 
In section \ref{sec:Three-Special-cases} we explore the hydrodynamic description of three cases. In section \ref{sec:NBH-finite-r} we work out boundary hydrodynamic description for AdS$_3$ or $3d$ flat space bounded by a generic null boundary. In section \ref{Hydrodynamics at infinity:AdS-case} we study the hydrodynamic description of the effective conformal fluid at the causal boundary of AdS$_3$ where recover an extension of the results of \cite{Campoleoni:2018ltl, Ciambelli:2020ftk, Ciambelli:2020eba} in that only half  of the boundary degrees of freedom in our solution space appear in the asymptotic hydrodynamic description. Moreover, we show the hydrodynamics  at the causal boundary of AdS$_3$ may be recovered from the limit of general results in section \ref{sec:HCS--finite-r-AdS} by taking the timelike boundary to infinity. The details of the limit are  presented in  appendix \ref{Appendix}.  In section \ref{Hydrodynamics at infinity:flat-case} we explore how  a conformal Carrollian hydrodynamic description appears at the asymptotic null boundary of $3d$ flat space. In section \ref{sec:thermodynamic-causal-boundary} we construct ``thermodynamics at the causal boundary'', i.e. we show boundary degrees of freedom at a generic causal boundary satisfy a local version of the first law of thermodynamics. We close by discussions and outlook in section \ref{sec:discussion}.

\section{Solution space, a review}\label{sec2:soln-space-review}

In this section, we present some basic notations and conventions and review the construction of the most general solution space in a Gaussian-null-type coordinate system in the presence of a given generic causal boundary. More details may be found in \cite{Adami:2022ktn}.

\subsection{Spacetime 2+1 decomposition}\label{sec:2+1-decom}

Given a causal boundary, it is appropriate to consider a $2+1$ decomposition and adopt the radial coordinate $r$ such that the boundary resides at constant $r$ and choose the range $r\geq 0$. That is, the coordinate $r$ emanates from the causal boundary $\mathcal{C}_r$ and foliates spacetime into causal hypersurfaces which are parameterized by the advanced time $v$ and the periodic coordinate $\phi$, $\phi \sim \phi + 2 \pi$. See Figs.~\ref{fig:ADS3-timelike}, \ref{fig:flat-timelike} for timelike boundaries in AdS and flat space and and  Figs.~\ref{fig:ADS3-null}, \ref{fig:flat-null} for null boundaries in AdS and flat space. The most general form of the metric in these coordinates takes the form, 
\begin{equation}\label{metric}
    \d s^2= g_{\mu\nu}\d x^\mu \d x^\nu= -V \d v^2 + 2 \eta \d v \d r + {\cal R}^2 \left( \d \phi + U \d v \right)^2\, , \qquad x^\mu=\{v, r, \phi \}\, .
\end{equation}
where $V, U$ and $\cal R$ are functions of $v,r , \phi$ while $\eta$ is a function of $v, \phi$.  The induced metric on ${\cal C}_r$ is then,  
\begin{equation}\label{Boundary-metric}
\d \sigma^2 :=\gamma_{ab} \d x^a \d x^b = - V \d v^2 +\mathcal{R}^2 \left( \d \phi + U  \d v\right)^2 \, ,\qquad x^a=\{v, \phi \}.
\end{equation}
We restrict ourselves to a part of spacetime for which $V \geq 0$. $V>0$ ensures that the vector field normal to the boundary is outward pointing and is consistent with the choice of $v$ to be the advanced time coordinate on ${\cal C}_r$ and for this choice, $\eta>0$.
We raise and lower lowercase Latin indices by $\gamma^{ab}$ and $\gamma_{ab}$ respectively, where $\gamma^{ac}\gamma_{cb}=\delta^a_b$. We also note that $\sqrt{- g}={\eta}{\cal R}$ and $\gamma=\det(\gamma_{ab})=-{\cal R}^2{V}$.

Let  $s$ denote the vector field perpendicular to ${\cal C}_r$,
\begin{equation}
    s_{\mu} \d x^\mu:=\frac{\eta}{\sqrt{V}}\d r \, , \hspace{1 cm}  s^{\mu}\partial_{\mu}=\frac{1}{\sqrt{V}}\left(\partial_{v}+\frac{V}{\eta}\partial_{r}-U \partial_{\phi}\right)\, ,
\end{equation}
which is normalized as $s\cdot s = 1$. Then the induced metric on $\mathcal{C}_r$ is given by  
\be
\gamma_{\mu\nu}= g_{\mu \nu}-  s_\mu s_\nu\, .
\ee
The  induced metric  $\gamma_{\mu\nu}$ can be written in terms of unit timelike vector field $t^{\mu}$ and  spacelike vector field $k^{\mu}$,
\be  
\gamma_{\mu\nu} = -t_\mu t_\nu + k_\mu k_\nu\, ,
\ee 
where
\begin{equation}\label{t-k-zweibein}
    k_{\mu} \d x^{\mu}:=\mathcal{R}\left( \d \phi +U\d v\right)\, , \hspace{1 cm} k^{\mu}\partial_{\mu}=\frac{1}{\mathcal{R}} \, \partial_{\phi}\, ,
\end{equation}
\begin{equation}
    t_{\mu} \d x^\mu:=-\sqrt{V}\left(\d v-\frac{\eta}{V}\d r\right)\, , \hspace{1 cm} t^{\mu}\partial_{\mu}=\frac{1}{\sqrt{V}}(\partial_{v}-U \partial_{\phi})\, .
\end{equation}
The timelike vector field $t^\mu$ is normalized as $t^2=-1$ and the  spacelike vector field $k^\mu$ as $k^2=1$. We also need to define a metric on the transverse surface 
\begin{equation}
     q_{\mu\nu}= g_{\mu \nu}-  s_\mu s_\nu + t_\mu t_\nu = k_\mu k_\nu\, .
\end{equation}
In the $3d$ case, the non-zero eigenvalue of $q_{\mu\nu}$ is essentially the $\phi\phi$ component of the metric.

The two spacelike and timelike vector fields $s, t$ may be written in terms of  linear combinations of two normalized null vector fields $l,n$, 
\begin{subequations}\label{BT}
\begin{align}
    t=  \frac{1}{\sqrt{V}} \, l + \frac{1}{2}\, \sqrt{V} \, n \, &, \qquad s= \frac{1}{\sqrt{V}} \, l - \frac{1}{2}\, \sqrt{V} \, n,\\
l=\frac{\sqrt{V}}{2}(t+s) \, &,\qquad n=\frac{1}{\sqrt{V}} (t-s),  \label{null-vectors}
\end{align}
\end{subequations}
with $l^2=n^2=0,$ and $ l\cdot n=-1$, explicitly,
\begin{equation}\label{null-basis'}
\begin{split}
    l_{\mu} \d x^{\mu} &= -\frac{1}{2} V \d v  + \eta \d r \, , \qquad\qquad
    n_{\mu} \d x^{\mu} = -\d v \, , \\
    l^{\mu}\partial_{\mu} &=\partial_{v}+\frac{V}{2\eta}\partial_{r}-U\partial_{\phi}\, , \hspace{1 cm} n^{\mu}\partial_{\mu}=-\frac{1}{\eta}\partial_{r}\, .    
\end{split}
\end{equation}
Equation \eqref{BT} also makes it clear that $\ln(\sqrt{V})$ may be viewed as a boost speed which acts on $l,n$ like scaling by $\sqrt{V}, 1/\sqrt{V}$, respectively.

We will also be dealing with the ``asymptotic boundary'' ${\cal C}_{\infty}={\cal C}_{r\to\infty}$, which is a null (timelike) cylinder for asymptotic flat (AdS$_3$) cases, see Fig.~\ref{fig:asympt-ads-flat}. By $\text{D}_v$ and ${\cal D}_v$ we  denote the derivatives along the $v$ respectively on $\cc_r$ and $\cc_\infty$,
\be\label{Dv-s}
\text{D}_v:=\partial_v-{\cal L}_U,\qquad {\cal D}_v:=\text{D}_v|_{_{r\to\infty}}=\partial_v-{\cal L}_{\cal U}, \qquad {\cal U}:=U(r\to\infty)\, , \ee
where ${\cal L}_X$ denotes a differential operators which is defined as {
\begin{equation}
    \mathcal{L}_{X} O_w:= X \partial_\phi O_w + w O_w \partial_\phi X \, ,
\end{equation}
where $O_w$ is an arbitrary function in spacetime with weight $w$. Weights of  different quantities can be found in Table \ref{Table-1}.
\begin{table}[t]
\centering
\begin{tabular}{ |l|l| }
  \hline
  $w= -1$  & $U$, $\mathcal{U}$ , $Y$ , $\hat{Y}$ , $\hat{T}$\\
  $w= 0$  & $V$, $\eta$ , $\theta_{s}$ , $\kappa_t$ , $\theta_l$ , $\theta_n$ , $\kappa$ , $\omega_l$ , $\Pi$ , $T$ , $W$ , $Z$ , $\hat{W}$ , $\partial_{v}$\\
  $w= 1$  & $\mathcal{R}$ , $\Omega$  , $\lambda$ , $\mathcal{P}$ , $\omega_s$ , $\hat{Z}$ , $\partial_{\phi}$ \\
  $w= 2$  & $\hat{\mathcal{M}}$ , $\hat{\Upsilon}$ \\
  \hline
\end{tabular}
\caption{Weight $w$ for various quantities defined and used in this section and the following sections.}\label{Table-1}
\label{table:weight}
\end{table}}

Independent components of the covariant derivative of the boundary generating spacelike  vector field $s$ along the boundary are given through
\begin{equation} \label{rel:scalarsS}
    \begin{split}
    &\theta_s:=q^{\alpha\beta}\nabla_{\alpha}s_{\beta}= \frac{1}{{\sqrt{V}}{\cal R}}\left(\text{D}_v{\cal R}
    +\frac{{V}}{\eta}\partial_{r}\mathcal{R}\right)\, , \\
    &\omega_s:=-k^{\alpha}t^{\beta}\nabla_{\alpha}s_{\beta}=-\frac{1}{2\mathcal{R}}\left(\frac{\mathcal{R}^2}{\eta}\partial_{r}U+\frac{\partial_{\phi}V}{V}-\frac{\partial_{\phi}\eta}{\eta}\right)\, ,\\
    &\kappa_{t}:= t^{\beta}t^{\alpha} \nabla_{\alpha}s_{\beta}=\frac{1}{2\sqrt{V}}\left(\frac{\text{D}_v V}{V}
    -\frac{\partial_r {V}}{\eta}-2\frac{\text{D}_v\eta}{\eta}\right)\, .
    \end{split}
\end{equation}
Similarly, for the two null vectors $l,n$ the expansions $\theta_l,\theta_n$, the angular velocity, $\omega_l$, and non-affinity parameter, $\kappa$, are given by
\begin{equation}\label{kappa-thetal-thetan}
\begin{split}
     \kappa &:=-l^{\alpha} n^{\beta}\nabla_{\alpha}l_{\beta} = \frac{\text{D}_v\eta}{\eta}+\frac{\partial_r V}{2\eta}, \\ 
     \omega_l &:={-} k^{\mu}n^{\nu}\nabla_{\mu}l_{\nu}={-}\frac{1}{2{\cal R}}\left(-\frac{\partial_\phi\eta}{\eta}+\frac{{\cal R}^2}{\eta} \partial_r U\right), \\ 
     \theta_l&:=q_{\alpha\beta}\nabla^{\alpha} l^\beta=\frac{\text{D}_v {\cal R}}{{\cal R}}
    +\frac{V}{2\eta}\frac{\partial_{r}\mathcal{R}}{\mathcal{R}}\, ,\\
  \theta_n &:=q_{\alpha\beta}\nabla^{\alpha} n^\beta=-\frac{1}{\eta}\frac{\partial_{r}\mathcal{R}}{\mathcal{R}}\, .
    \end{split}
\end{equation}

\subsection{Equations of motion}\label{sec:EoM}
Field equations for Einstein-$\Lambda$ theory are
\begin{equation}\label{EOM}
    \mathcal{E}_{\mu \nu} = G_{\mu\nu} + \Lambda g_{\mu \nu}=0\, .
\end{equation}
Straightforward computations show that one can solve for the $r$-dependence of the 3 functions in the metric \eqref{metric} (recall that $\eta$ is $r$-independent) obtained to be \cite{Adami:2022ktn}: 
\begin{subequations}\label{metric-components}
    \begin{align}
    & U={\cal U} +  \frac{1}{\lambda \, {\cal R}}\, \frac{\partial_\phi \eta}{ \eta }  + \frac{\Upsilon}{2 \lambda {\cal R}^2} \, , \qquad\qquad \mathcal{R}= \Omega + r \, \eta \, \lambda \, ,\\
         & V= \frac{1}{\lambda^2}\left( - \Lambda \mathcal{R}^2 -\mathcal{M}  + \frac{\Upsilon^2}{4  {\cal R}^2} - \frac{2\mathcal{R}}{\eta }   \mathcal{D}_v ( \eta \lambda ) +\frac{\Upsilon}{\mathcal{R}}\, \frac{\partial_\phi \eta}{ \eta } \right) \, ,
    \end{align}
\end{subequations}
where $\Omega, \lambda, \eta, \Upsilon, \mathcal{U}, \mathcal{M}$ are six functions of $v, \phi$ and ${\cal D}_v$ is defined in \eqref{Dv-s}. Einstein equations yield 2 more constraints/relations among the 6 codimension one functions of $v,\phi$. To analyze these equations we divide them into three parts, 
\begin{equation}
\mathcal{E}_{ss}:=s^\mu s^\nu \mathcal{E}_{\mu \nu} , \quad \mathcal{E}_{sa}:=s^\mu \gamma^\nu_a\mathcal{E}_{\mu \nu},\quad \mathcal{E}_{ab}:= \gamma^\mu_a \gamma^\nu_b \mathcal{E}_{\mu \nu}. 
\end{equation}
Among them, $\mathcal{E}_{sa}=0$  yield 
\begin{subequations}\label{eq:constraints-r}
    \begin{align}
        &{\text{D}}_{v}({\mathcal{R}^2\omega_s})+{\mathcal{R}}\partial_{\phi}(\sqrt{V} \kappa_t) +{\mathcal{R}}\theta_s \partial_{\phi}{\sqrt{V}}=0\, ,\\
        &{\text{D}}_{v}({\mathcal{R}\theta_s})+\kappa_t{{\text{D}}_{v}\mathcal{R}}+\frac{1}{\sqrt{V}}\partial_{\phi}(V \omega_s)=0\, .
    \end{align}
\end{subequations}
The other equations $\mathcal{E}_{ss}=0, \mathcal{E}_{ab}=0$ are readily satisfied once \eqref{metric-components} and \eqref{eq:constraints-r} hold. It should be noted that the equations \eqref{eq:constraints-r} consist of two first-order time ($v$) derivative equations, which are linear in the variables $\theta_s$, $\omega_s$, and $\kappa_t$. These equations are completely defined at the boundary ${\cal C}_r$. Upon these equations, the solution space is completely specified by 4 functions over ${\cal C}_r$. 

\subsection{Asymptotic metric, asymptotic boundary}\label{sec:asymptotic-metric}
The asymptotic behavior in the absence or presence of cosmological constant $\Lambda$  is different. So we analyze the two asymptotic flat, $\Lambda=0$, and asymptotic AdS$_3$, $\Lambda=-1/\ell^2$, cases separately.

\paragraph{Asymptotic AdS$_3$ case.} 
The asymptotic value of metric functions (at large $r$ limit) is given by 
\begin{equation}
    U=\mathcal{U}+\mathcal{O}(\mathcal{R}^{-1})\, , \hspace{1 cm}  V=\frac{\mathcal{R}^2}{\ell^2\lambda^2}+\mathcal{O}(\mathcal{R})\, .
\end{equation}
The asymptotic form of the line element is 
\begin{equation}
    \d s^2= -\frac{\mathcal{R}^2}{\lambda^2\ell^2} \d v^2 + 2 \eta \d v \d r + {\cal R}^2 \left( \d \phi + \mathcal{U} \d v \right)^2+\mathcal{O}(\mathcal{R})\, .
\end{equation}
From this, one can read the asymptotic boundary metric as
\begin{equation}\label{Asymptotic-AdS3-boundary-metric}
    \d \sigma^2|_{r\to\infty}= r^2{\cal P}^2\left[-\frac{1}{\ell^2\lambda^2} \d v^2 +  \left( \d \phi + \mathcal{U} \d v \right)^2\right]+\mathcal{O}({r})\,, \qquad {\cal P}(v,\phi) := \eta\lambda, 
\end{equation}
The above is the metric of $1+1$ dimensional cylinder of unit radius  up to the conformal factor $\mathcal{P}^2 \, r^2$.  The asymptotic (conformal/causal) boundary $\mathcal{C}_{\infty}$ is specified by 3 functions of $v,\phi$, {namely} $\lambda, \eta, {\cal U}$.  ${\lambda (v,\phi)^{-1}}$ is the local scale of asymptotic time and ${\cal U}(v,\phi)$ specifies the local angular velocity of the frame at the boundary. Moreover,  note that ${\cal P}$ sets a local scale of $r$. 

The asymptotic expression of geometric quantities are 
\begin{equation}\label{thetas-omegas-kappat}
    \begin{split}
        \theta_{s}=& \frac{1}{\ell}-\frac{\ell}{2r^2\mathcal{P}^2}\left[\hat{\mathcal{M}}-\left(\frac{\ell\mathcal{D}_{v}\mathcal{P}}{\eta}\right)^2+2\mathcal{S}[\int \mathcal{P},\phi]\right] +\mathcal{O}(r^{-3})\, , \\
        \omega_{s}=& \frac{1}{2r^2 \mathcal{P}^2}\left[\hat{\Upsilon}+2\ell^2\mathcal{P}\partial_{\phi}\left(\frac{\mathcal{D}_{v}\mathcal{P}}{\eta\mathcal{P}}\right)\right]+\mathcal{O}(r^{-3})\, , \\
        {\kappa_{t}-\theta_{s}=}&{-\frac{2}{\ell}+\frac{\ell\, \lambda }{r^2 \, \mathcal{P}^2}\left[-\ell^2 \mathcal{D}_{v}\left(\frac{\mathcal{D}_{v}\mathcal{P}}{\eta}\right)+\partial_{\phi}\left(\frac{\partial_{\phi}\eta}{\mathcal{P}}\right)\right]+\mathcal{O}(r^{-3})}\, .
    \end{split}
\end{equation}
where
\begin{equation}\label{hatM-hatUsp''}
    \begin{split}
       &  \hat{\Upsilon}={\Upsilon+\Omega\partial_{\phi}\Pi}\, , \qquad {\Pi :=\ln \left( \frac{\eta \lambda}{\Omega} \right)^2}=2\ln\frac{{\cal P}}{{\Omega}}\, ,\\
       & \hat{\mathcal{M}}= \mathcal{M} + {\lambda\Omega\mathcal{D}_{v}\Pi}+ \left(\frac{\partial_{\phi}\eta}{\eta}\right)^2-2\mathcal{S}[\int \lambda,\phi]\, ,
    \end{split}
\end{equation}
and $ \mathcal{S}[X;\phi]$ is the Schwarzian derivative of the quantity $X$,
\begin{equation}
    \mathcal{S}[X;\phi]:=\frac{\partial_{\phi}^3 X}{\partial_{\phi}X}-\frac{3}{2}\left(\frac{\partial_{\phi}^2X}{\partial_{\phi}X}\right)^2\, .
\end{equation}
We note that Einstein equations \eqref{eq:constraints-r} lead to \cite{Adami:2022ktn}
\begin{subequations}\label{M-Upsilon-EoM''}
\begin{align}
    &\mathcal{D}_{v}\hat{\mathcal{M}}+\Lambda\lambda\partial_{\phi}\left(\frac{\hat{\Upsilon}}{\lambda^2}\right)+2\partial_{\phi}^3\mathcal{U}=0\, , \label{EOM-M}\\
    &\mathcal{D}_{v}\hat{\Upsilon}-\lambda\partial_{\phi}\left(\frac{\hat{\mathcal{M}}}{\lambda^2}\right)+2\partial_{\phi}^3(\lambda^{-1})=0 \, .\label{EOM-Upsilon}
\end{align}
\end{subequations}
Third derivatives terms in equations \eqref{M-Upsilon-EoM''} involve $\lambda^{-1}, {\cal U}$ which are components of metric \eqref{Asymptotic-AdS3-boundary-metric}. Finally, we also note the asymptotic expression of null geometric quantities
\begin{equation}\label{kappa-thetal-thetan-large-r}
    \begin{split}
        \kappa=& \frac{r\eta}{\ell^2 }+\frac{\Omega-\ell^2\mathcal{D}_{v}\lambda}{\ell^2\lambda}-\frac{1}{r\eta}\left(\frac{\partial_{\phi}\eta}{\eta\lambda}\right)^2+\mathcal{O}(r^{-2})\, , \\
        \theta_{l}=&\frac{r\eta}{2\ell^2}+\frac{\Omega}{2\ell^2\lambda}-\frac{\eta}{2r\mathcal{P}^2}\left[\hat{\mathcal{M}}+2\mathcal{S}[\int \mathcal{P};\phi]\right]+\mathcal{O}(r^{-2})\, \\
    {\omega_l=}&{\frac{\partial_{\phi}\eta}{\eta}\frac{1}{\mathcal{P} r} +\frac{1}{2\mathcal{P}^2 r^2}\left(\Upsilon-2\Omega\frac{\partial_{\phi}\eta}{\eta}\right)+\mathcal{O}(r^{-3})}\\         
    \end{split}
\end{equation}

\paragraph{Asymptotic flat $\Lambda=0$ case.}\footnote{We should note that in $3d$ gravity asymptotically flat or AdS means (locally) flat or AdS, as Riemann curvature is specified through Ricci tensor.}
For the flat case, we have a Carrollian  geometry at the asymptotic null boundary. 
(See \cite{Duval:2014uoa, Duval:2014uva, Duval:2014lpa, Henneaux:1979vn, Ciambelli:2019lap} for more discussions on Carrollian geometries.) It is described by a degenerate metric, 
\begin{equation}\label{metric-flat-asymptotic}
    \d \sigma^2|_{r\to\infty}= \mathcal{P}^2r^2\left( \d \phi + \mathcal{U} \d v \right)^2+\mathcal{O}(r):=r^2\hat{q}_{\mu\nu} \d{}x^\mu\d x^\nu+\mathcal{O}(r)\, ,
\end{equation}
with kernel 
\begin{equation}\label{l-flat-asymptotic}
    \hat{l}^{\mu}\partial_{\mu}=\partial_{v}-\mathcal{U}\partial_{\phi}\, , \qquad  \hat{q}_{\mu\nu} \hat{l}^{\mu}=0\, .
\end{equation}
Recalling \eqref{metric-components}, we see that  $V= -\frac{2{\cal D}_v{\cal P}}{\lambda}r+ {\cal O}(r^0)$. Therefore, in order $V>0$ we should have ${\cal D}_v{\cal P}<0$ (Note that we have already assumed $\eta,\lambda>0$ and hence ${\cal P}>0$.) Moreover, to ensure $V>0$ in the region of spacetime of interest, the function $\mathcal{P}$ should be monotonically decreasing with respect to advanced time. We assume that this condition is satisfied.

The asymptotic expression of geometric quantities are 
\begin{equation}
    \begin{split}
        &\theta_s={\cal Q}\ r^{-\frac{1}{2}}+\mathcal{O}(r^{-\frac{3}{2}})\, ,\qquad {\cal Q}^2:={\frac{-\mathcal{D}_{v}\mathcal{P}}{2\eta\mathcal{P}}}\, ,\\
        &\omega_s=-\frac1{{\cal P}}\partial_{\phi}\ln{\cal Q}\  r^{-1}+\mathcal{O}(r^{-2})\, ,\\
        &\kappa_t=-\frac{\mathcal{D}_{v}({\cal P} {\cal Q})}{\mathcal{D}_{v}\mathcal{P}}\ r^{-\frac12}+\mathcal{O}(r^{-\frac{3}{2}})\, .
    \end{split}
\end{equation}
and
\begin{equation}\label{kappa-theta--omegal-large-r}
    \begin{split}
        \kappa=&-\frac{\mathcal{D}_{v}\lambda}{\lambda}-\frac{1}{r\eta}\left(\frac{\partial_{\phi}\eta}{\eta\lambda}\right)^2+\mathcal{O}(r^{-2})\, , \\
        \theta_{l}=&-\frac{\eta}{2r\mathcal{P}^2}\left[\hat{\mathcal{M}}+2\mathcal{S}[\int \mathcal{P};\phi]\right]+\mathcal{O}(r^{-2})\, ,\\
        {\omega_l=}&{\frac{\partial_{\phi}\eta}{\eta}\frac{1}{\mathcal{P} r} +\frac{1}{2\mathcal{P}^2 r^2}\left(\Upsilon-2\Omega\frac{\partial_{\phi}\eta}{\eta}\right)+\mathcal{O}(r^{-3})}\\    
    \end{split}
\end{equation}
and $\hat{\mathcal{M}}$ still satisfies \eqref{M-Upsilon-EoM''} with $\Lambda=0$.

\subsection{Surface charges, general analysis}\label{sec:sym-charge-review}
This section summarizes charge analysis associated with a generic causal surface. More detailed analysis and discussion may be found in \cite{Adami:2022ktn}. 

\paragraph{Causal boundary symmetry.}
The metric of the form \eqref{metric}, with \eqref{metric-components}, is preserved by the infinitesimal diffeomorphism generated by the vector field \cite{Adami:2022ktn}
\begin{equation}\label{null-bondary-sym-gen}
\begin{split}
    \xi &=T\partial_{v}+\left[Z  - \frac{r }{2}\, W -\frac{\Upsilon}{2\eta\lambda^2\mathcal{R}} \, \partial_{\phi}T - \frac{1}{\eta^2\lambda }\partial_{\phi}\left(\frac{\eta\partial_{\phi}T}{\lambda}\right)\right]\partial_{r}+\left(Y+\frac{\partial_{\phi}T}{\lambda\mathcal{R}}\right)\partial_{\phi}\, ,
\end{split}
\end{equation}
where $T$, $Z$, $W$, and $Y$ are generic functions at codimension one constant $r$ surfaces, generic functions of $v, \phi$. $T$ generates supertranslation in $v$ direction, $Z$ supertranslation in $r$ direction, $W$ superscaling, and $Y$ superrotations.

Using the adjusted Lie bracket we have
\begin{equation}\label{3d-NBS-KV-algebra''}
    [\xi(  T_1, Z_1, W_1, Y_1), \xi( T_2, Z_2,  W_2, Y_2)]_{_{\text{adj. bracket}}}=\xi(  T_{12}, Z_{12}, W_{12}, Y_{12})\, ,
\end{equation}
where 
\begin{subequations}\label{W12-T12-Z12-Y12''}
\begin{align}
    &T_{12}=(T_{1}\partial_{v}+Y_{1}\partial_{\phi})T_{2}-(1\leftrightarrow 2)\, ,\\
    &Z_{12}=(T_{1}\partial_{v}+Y_{1}\partial_{\phi})Z_{2}+\frac{1}{2}W_{1}Z_{2}-(1\leftrightarrow 2)\, ,\\
    &W_{12}=(T_{1}\partial_{v}+Y_{1}\partial_{\phi})W_{2}-(1\leftrightarrow 2)\, ,\\
    &Y_{12}=(T_{1}\partial_{v}+Y_{1}\partial_{\phi})Y_{2}-(1\leftrightarrow 2)\, .
\end{align}
\end{subequations}
The above is a semi-direct sum of Diff(${\cal C}_r$) and supertranslations along $r$ (generated by $Z$) and superscalings along $r$ (generated by $W$). While the Diff(${\cal C}_r$) has a clear geometric meaning at the boundary ${\cal C}_r$, the $Z, W$ parts do not. In particular, we note that Weyl scaling on ${\cal C}_r$, transformations $\gamma_{ab} \to {\cal W} \gamma_{ab}$ with $\gamma_{ab}$ being the boundary metric \eqref{Boundary-metric}, are not generated by our symmetry generators.\footnote{That the $2d$ Weyl$\oplus$Diff (at the boundary ${\cal C}_r$) cannot be embedded into our boundary symmetry algebra, among other things, implies that the boundary theory which should have the causal boundary algebra as its gauge (local) symmetry, cannot be a $2d$ string theory. Of course, the situation is different for null boundaries, where the boundary symmetry algebra can be realized as the local symmetry of null strings, see \cite{Bagchi:2022iqb} and references therein for more discussions. \label{footnote-null-string}} Note, however, that as we discuss in section \ref{Hydrodynamics at infinity:AdS-case} Weyl scaling becomes a part of our symmetry algebra in $r\to\infty$ limit. 
Some of the relevant transformation laws are \cite{Adami:2022ktn}
\begin{subequations}
    \begin{align}
    \delta_{\xi}\Omega &=T {\cal D}_v\Omega + \partial_\phi\left( (Y+\mathcal{U}T)\Omega\right) + \eta\lambda Z, \\ 
\delta_{\xi}\Pi & =-W + T\partial_v \Pi -2e^{\Pi/2}Z + Y\partial_\phi \Pi\, ,\\     
\delta_{\xi} V 
        &=\xi^{v}\partial_{v}V+\xi^{\phi}\partial_{\phi}V+\xi^{r}\partial_{r}V+2V\text{D}_{v}\xi^{v}-2\eta\text{D}_{v}\xi^{r}\, , \label{delta-V}\\
        \delta_\xi \eta &= \mathcal{D}_v (T \eta )  + (Y+\mathcal{U}T) \partial_\phi \eta -\frac{1}{2}\eta W \, .
    \end{align}
\end{subequations}

\paragraph{Surface charge.}
We start from the three-dimensional gravity action in the presence of the cosmological constant 
\begin{equation}\label{3D-action}
    S= \frac{1}{16\pi G} \int \d{}^3 x\ \sqrt{-g}  \left( R -2\Lambda \right)+\int \d{}^3 x\ \partial_{\mu} L_{\text{\tiny b}}^{\mu}\, ,
\end{equation}
where $L_{\text{\tiny b}}^{\mu}$ is a boundary Lagrangian. The pre-symplectic potential for this Lagrangian can be read as,
\begin{equation}\label{Theta-mu-generic}
  \Theta^\mu[ g ; \delta g]:=\Theta_{\text{\tiny LW }}^\mu [ g ; \delta g] + \nabla_{\nu} Y^{\mu \nu}[ g ; \delta g]+\delta  L_{\text{\tiny b}}^{\mu}[ g ] \, ,
\end{equation}
where 
\begin{equation}\label{Theta-LW}
    \Theta^{\mu}_{_{\text{\tiny{LW}}}} [g; \delta g]:=\frac{\sqrt{-g}}{8 \pi G} \nabla^{[\alpha} \left( g^{\mu ] \beta} \delta g_{\alpha \beta} \right)\,,
\end{equation}
is Lee-Wald's pre-symplectic potential for the Einstein-Hilbert term and $Y^{\mu \nu}[ g; \delta g]$ is the $Y$-freedom, a covariant skew-symmetric tensor constructed out of metric and its variation.

 The Lee-Wald surface charge variation reads
\begin{equation}\label{surface-charge-LW}
\begin{split}
       \hspace{-0.3 cm}     \slashed{\delta} Q_{\text{\tiny{LW}}}(\xi)
&= {\frac{1}{16\pi G}}\oint_{\cc_{r,v}} \d \phi \ {\cal R}\epsilon_{\mu\nu}\left(h^{\lambda [ \mu} \nabla _{\lambda} \xi^{\nu]} - \xi^{\lambda} \nabla^{[\mu} h^{\nu]}_{\lambda} - \frac{1}{2} h \nabla ^{[\mu} \xi^{\nu]} + \xi^{[\mu} \nabla _{\lambda} h^{\nu] \lambda} - \xi^{[\mu} \nabla^{\nu]}h  \right)\\
 &={\frac{1}{16\pi G}}\oint_{\cc_{r,v}} \d \phi \left( 
            -\delta_{\xi}\Pi \delta\Omega +\delta_{\xi}\Omega  \delta \Pi + (Y+\mathcal{U}T)  \delta\hat{\Upsilon}  + \frac{T}{\lambda}  \delta\hat{\mathcal{M}} +\delta_{\xi}n^r\delta\sqrt{-g}-\delta n^r\delta_{\xi}\sqrt{-g}\right),
\end{split}
\end{equation}
where $\cc_{r,v}$ denotes a constant $v$ slice at the boundary ${\cal C}_r$. In the first line $\epsilon_{\mu\nu}$ is binormal to $\cc_{r,v}$, and $ h_{\mu\nu}=\delta g_{\mu\nu}$, its indices are raised and lowered with the background metric $g_{\mu\nu}$ and $h$ is its trace. The second line in \eqref{surface-charge-LW} shows that 
the $r$-dependence of the Lee-Wald surface charge arises from the last two terms in the above equation. This $r$-dependence can be removed by adding the following $Y$-term
\begin{equation}\label{Y-r-removing}
    Y^{\mu\nu}[\delta g; g]=\frac{\delta \sqrt{-g}}{8 \pi G}t^{[\mu}s^{\nu]}\, .
\end{equation}
This $Y$-term is covariant and is completely built out of geometric quantities defining the boundary ${\cal C}_r$. The charge expression then becomes 
\begin{subequations}\label{surface-charge-Y-infty'''}
\begin{align}
\slashed{\delta} Q_\xi  &={\frac{1}{16\pi G}}\oint_{\cc_{r,v}} \d \phi \left(  W \delta\Omega + 2Z \delta(\Omega \, e^{\Pi/2}) +Y\delta \Upsilon+ T \slashed\delta{\cal H} \right) \label{LW-charge-1}\\
            &=\frac{1}{16\pi G} \oint_{\cc_{r,v}} \d \phi \left(
            \hat{W}\delta\Omega +\hat{Z} \delta \Pi+\hat{Y} \delta\hat{\Upsilon}  + \hat{T}\delta\hat{\mathcal{M}} \right), \label{charge-integrable-slicing}
\end{align}
\end{subequations}
where 
\begin{equation}\label{delta-slash-H}
    \begin{split}
         \slashed\delta{\cal H}&= -\mathcal{D}_{v}\Pi \, \delta\Omega+{\mathcal{U}} \delta\Upsilon+\mathcal{D}_{v}\Omega\ \delta\Pi+\lambda^{-1}\delta{\hat{\mathcal{M}}}.
    \end{split}
\end{equation}
and 
\begin{equation}\label{hat-slicing''}
\begin{split}
    \hat{T}= \frac{T}{\lambda},  \qquad  \hat{Y}= Y+\mathcal{U} T \, , \qquad \hat{Z} =\delta_{\xi}\Omega,\qquad      \hat{W}=-\delta_{\xi}\Pi 
\end{split}
\end{equation}
As \eqref{LW-charge-1} shows, the charge is not integrable in the slicing that $T, Z, W, Y$ are field independent, i.e. when $\delta T=\delta Z=\delta W=\delta Y=0$. Explicitly,  while the charges of $W, Z, Y$ are integrable, the charge associated with $T$ is not. However, as is made manifest in \eqref{charge-integrable-slicing}, the charge expression \eqref{charge-integrable-slicing} becomes integrable if we assume the new hatted generators are field-independent. That is, change of slicing to the hatted generators makes the charges integrable.\footnote{See \cite{Adami:2020ugu, Adami:2021sko, Taghiloo:2022kmh, Grumiller:2022qhx} for more discussions on the change of slicing.} Recalling the asymptotic boundary metric \eqref{Asymptotic-AdS3-boundary-metric}, we observe that $\hat{Y}, \hat{T}$ are  generators of translations in normalized $\phi, v$ directions, respectively. We will explore this point more closely in the next sections when we discuss the hydrodynamic description.

The charge variation \eqref{charge-integrable-slicing} shows that our solution phase space is well parametrized with four codimension-one surface charges, $\{\hat{\mathcal{M}},\hat{\Upsilon},\Omega,\Pi\}$. We crucially note that  while these four charges are $r$-independent, the surface charge is computed by integrating over an arbitrary constant $r,v$ surfaces.  We will show in sections \ref{sec:HCB-AdS} and \ref{sec:HCB-flat} that the first two charges lead to a hydrodynamic description at infinity while  $\Omega, \Pi$ do not enter the boundary hydrodynamic description.

\paragraph{Surface charge algebra in the direct-sum slicing.} 
In general the algebra of charges is the same as the algebra of symmetry generators, up to possible central terms, i.e.
\begin{equation}\label{3d-BT-Bracket-02''}
    \left\{Q^{\text{I}}_{\xi_{1}},Q^{\text{I}}_{\xi_{2}}\right\}_{{{\text{\tiny{MB}}}}} =\, Q^{\text{I}}_{[\xi_{1},\xi_{2}]_{{\text{adj. bracket}}}}+K_{\xi_1,\xi_2}\, .
\end{equation}
In the direct-sum hatted slicing of previous page, 
\begin{equation}
    K_{\xi_1,\xi_2}=\frac{1}{8\pi G} \oint_{\cc_{r,v}} \d \phi \, (\hat{T}_{2}\partial_{\phi}^3 \hat{Y}_{1}-\hat{T}_{1}\partial_{\phi}^3 \hat{Y}_{2})+\frac{1}{16\pi G} \oint_{\cc_{r,v}} \d \phi \, (\hat{Z}_{2}\hat{W}_{1}-\hat{Z}_{1}\hat{W}_{2})\, .
\end{equation}
The explicit form of non-zero commutation relations of the algebra are given by
\begin{subequations}\label{Heisenberg-direct-sum-algebra''}
    \begin{align}
        &\{\Omega(v,\phi),\Pi(v,\phi')\}=16\pi G\ \delta(\phi-\phi')\, ,\\
        &\{\hat{\Upsilon}(v,\phi),\hat{\Upsilon}(v,\phi')\}=16\pi G\left(\hat{\Upsilon}(v,\phi')\partial_{\phi}-\hat{\Upsilon}(v,\phi)\partial_{\phi'}\right)\delta(\phi-\phi')\, ,\\
        &\{\hat{\mathcal{M}}(v,\phi),\hat{\mathcal{M}}(v,\phi')\}=-16\pi G\Lambda\left(\hat{\Upsilon}(v,\phi')\partial_{\phi}-\hat{\Upsilon}(v,\phi)\partial_{\phi'}\right)\delta(\phi-\phi')\, , \label{Heisenberg-direct-sum-MM}\\
        &\{\hat{\Upsilon}(v,\phi),\hat{\mathcal{M}}(v,\phi')\}=16\pi G\left(\hat{\mathcal{M}}(v,\phi')\partial_{\phi}-\hat{\mathcal{M}}(v,\phi)\partial_{\phi'}-2\partial_{\phi}^3\right)\delta(\phi-\phi')\, .
    \end{align}
\end{subequations}
That is, the algebra is a direct sum of a Heisenberg algebra and a BMS$_3$ ($\Lambda=0$) or Virasoro$\oplus$Virasoro ($\Lambda<0$) algebra at Brown-Henneaux central charge \cite{Brown:1986nw, Adami:2022ktn}. Note that while the charges are formally $v$-dependent, the algebra of charges is the same at any arbitrary $v$.

\paragraph{Symplectic form.}
As pointed out, the solution space is equipped with a symplectic 2-form $\Omega [\delta g ,\delta g; g]$. The explicit form of the symplectic form depends on the $Y$-freedom. With the $Y$-term which removes the $r$-dependence \eqref{Y-r-removing}, the symplectic form is
\begin{equation}\label{r-independent-symplectic-form}
    \Omega [\delta g ,\delta g; g] =  \frac{1}{16 \pi G} \int_{\mathcal{C}_r} \d v \d{}\phi\,\left[ \delta(\lambda^{-1})\wedge\delta \hat{\mathcal{M}}+\delta\mathcal{U}\wedge\delta\hat{\Upsilon}+{\cal D}_v ( \delta \Omega \wedge \delta \Pi ) \right] \, .
\end{equation}
This symplectic form consists of 2 parts,  ``codimension 1 modes'' given by a codimension 1 integral  spanned by $\hat{\mathcal{M}},\hat{\Upsilon}$ and their respective canonical conjugates $\lambda^{-1}, {\cal U}$ and  ``codimension 2 modes'' given by a codimension 2 integral spanned by $\Omega, \Pi$ which are canonical conjugates of each other.\footnote{In the literature codimension 2 modes have also been called corner or edge modes, see e.g. \cite{Ciambelli:2022vot} and references therein. We, however, prefer to use the terminology codimension 1 and codimension 2 modes because depending on the slicing, the choice of $Y$-terms and explicitly imposing on-shell conditions codimension 1 and 2 modes can be transformed into each other, e.g. see \eqref{SF-null}.} We note that for the generic (e.g. higher dimensional gravity) cases besides the codimension 1 and codimension 2 modes, we also have bulk modes (gravitons in higher dimensional gravity), see e.g. \cite{Sheikh-Jabbari:2022mqi}. Symplectic form for the bulk modes is also given by a codimension one integral, taken over a (partial) Cauchy surface. Such bulk modes are absent in the $3d$ example we study here. 

We should also point out that the solution space is specified by 4 functions of $v,\phi$, whereas in \eqref{r-independent-symplectic-form} we have 6 functions. This is of course the off-shell symplectic form and the codimension 1  modes and their canonical conjugates, as in any Hamiltonian system, are related to each through evolution equations, which in our case are \eqref{M-Upsilon-EoM''}.

\section{Hydrodynamics at a generic timelike boundary}\label{sec:HCB-AdS}

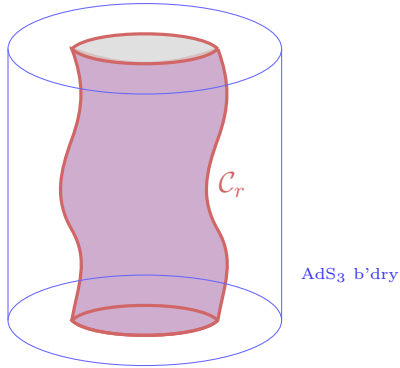
\begin{figure}[t]
\centering
\begin{tikzpicture}[scale=1.2]
    \node (v1) at (0.8,0) {};
    \node (v12) at (0.4,0.2) {};
    \node (v13) at (0,0.2) {};
    \node (v14) at (-0.4,0.2) {};
    \node (v16) at (-0.4,-0.2) {};
    \node (v17) at (0,-0.2) {};
    \node (v18) at (0.4,-0.2) {};
    \node (v2) at (0.8,-1) {};
    \node (v3) at (0.8,-2) {};
    \fill (v3)+(-0.1,0.5)  node [right,darkred!60] {$\cc_r$};
    \node (v4) at (0.8,-3) {};
    \fill (v4)+(0.8,0.5)  node [right,blue!80] {\tiny{AdS$_3$ \text{b'dry}}};
    \node (v5) at (-0.8,0) {};
    \node (v6) at (-0.8,-1) {};
    \node (v7) at (-0.8,-2) {};
    \node (v8) at (-0.8,-3) {};
\begin{scope}[fill opacity=0.8, darkred!60]
    \filldraw [fill=darkred!60!blue!40, very thick] ($(v1)+(0,0)$) 
        to[out=290,in=70] ($(v2) + (0,0)$) 
        to[out=250,in=120,thick] ($(v3) + (0,0)$)
        to[out=300,in=80,thick] ($(v4) + (0,0)$)
        to[out=225,in=315,looseness=0.5] ($(v8) + (0,0)$)
        to[out=80,in=300] ($(v7) + (0,0)$)
        to[out=120,in=250] ($(v6) + (0,0)$)
        to[out=70,in=290] ($(v5) + (0,0)$)
        to[out=340,in=200] ($(v1) + (0,0)$);
    \end{scope}
\begin{scope}[fill opacity=0.8, very thick, darkred!60] 
    \filldraw [fill=gray!30] ($(v1)+(0,0)$) 
        to[out=135,in=45, looseness=0.5] ($(v5) + (0,0)$)
        to[out=315,in=225,looseness=0.5] ($(v1) + (0,0)$);
    \end{scope}
\begin{scope}[fill opacity=0.8, very thick,darkred!60] 
    \draw ($(v4)+(0,0)$) 
        to[out=135,in=45, looseness=0.5] ($(v8) + (0,0)$)
        to[out=315,in=225,looseness=0.5] ($(v4) + (0,0)$);
    \end{scope}
\draw [blue!60](0,0) ellipse (1.5 and 0.5);
\draw [blue!60] (0,-3) ellipse (1.5 and 0.5);
\draw [blue!60] (-1.5,0) -- (-1.5,-3);
\draw [blue!60] (1.5,0) -- (1.5,-3);
\end{tikzpicture}
	\caption{\footnotesize{Portion of AdS$_3$ bounded by a generic timelike boundary $\cc_r$. We formulate physics in the shaded region. The wiggles on $\cc_r$ are to highlight the boundary degrees of freedom, where the Brown-York-type charges ${\cal T}^{ab}$ are canonical conjugates of boundary metric components $\gamma_{ab}$.}} \label{fig:ADS3-timelike}
\end{figure}

\begin{figure}[t]
\centering
\begin{tikzpicture}[scale=1.2]
    \node (v1) at (0,0) {};
    \fill (v1)+(0,0.2)  node [blue!70] {$i^{+}$};
    \node (v2) at (-2,-2) {};
    \node (v3) at (2,-2) {};
    \fill (v3)+(0.1,0)  node [right, blue!70] {$i^{0}$};
    \node (v4) at (0,-4) {};
    \fill (v4)+(0,-0.2)  node [blue!70] {$i^{-}$};
	\node (v5) at (0.5,-2) {};
    \fill (v5)+(0.3,0)  node [darkred!70] {$\mathcal{C}_r$};
	\node (v6) at (-0.5,-2) {};
 	\node (v7) at (0.3,-3) {};
	\node (v8) at (-0.3,-3) {};
 	\node (v9) at (0.3,-1) {};
	\node (v10) at (-0.3,-1) {};
 \node [right] at (1.2,-3) {$\textcolor{blue!70}{\mathcal{I}^{-}}$};
\node [right] at (1.2,-1) {$\textcolor{blue!70}{\mathcal{I}^{+}}$};
\begin{scope}[fill opacity=0.8, blue!70]
    \draw  ($(v1)+(0,0)$) 
        to[out=315,in=135] ($(v3) + (0,0)$) 
        to[out=225,in=45,thick] ($(v4) + (0,0)$)
        to[out=135,in=315,thick] ($(v2) + (0,0)$)
        to[out=45,in=225,thick] ($(v1) + (0,0)$);
        \end{scope}
\begin{scope}[fill opacity=0.8, very thick, black!60,dash pattern={on 2pt off 2pt}] 
    \draw ($(v2) + (0,0)$)
        to[out=25,in=155,looseness=1.3] ($(v3) + (0,0)$);
    \end{scope}
\begin{scope}[fill opacity=0.8, very thick, darkred!60] 
    \filldraw [fill=darkred!60!blue!40] ($(v1)+(0,0)$) 
        to[out=305,in=90, looseness=0.7] ($(v9) + (0,0)$)
        to[out=270,in=90,looseness=0.7] ($(v5) + (0,0)$)
        to[out=270,in=100,looseness=0.7] ($(v7) + (0,0)$)
        to[out=280,in=55,looseness=0.7] ($(v4) + (0,0)$)
        to[out=125,in=260,looseness=0.7] ($(v8) + (0,0)$)
        to[out=80,in=270,looseness=0.7] ($(v6) + (0,0)$)
        to[out=90,in=270,looseness=0.7] ($(v10) + (0,0)$)
        to[out=90,in=235,looseness=0.7] ($(v1) + (0,0)$);
    \end{scope}
\begin{scope}[fill opacity=0.8, very thick,black!60,dash pattern={on 2pt off 2pt}] 
    \filldraw [pattern=north west lines, pattern color=darkred!70, thick,dash pattern={on 7pt off 2pt}] ($(v5)+(0,0)$) 
        to[out=90,in=90, looseness=0.7] ($(v6) + (0,0)$);
    \end{scope}
\begin{scope}[fill opacity=0.8, very thick,black!60] 
    \filldraw [pattern=north west lines, pattern color=darkred!70, thick,dash pattern={on 7pt off 2pt}] ($(v6) + (0,0)$)
        to[out=270,in=270,looseness=0.7] ($(v5) + (0,0)$);
    \end{scope}
\begin{scope}[fill opacity=0.8, very thick, black!60] 
    \draw ($(v3)+(0,0)$) 
        to[out=205,in=335, looseness=1.3] ($(v2) + (0,0)$);
    \end{scope}
\end{tikzpicture}
	\caption{\footnotesize{A generic time-like boundary $\cc_r$ in flat space. The wiggles on $\cc_r$ depict the boundary degrees of freedom which are associated with the boundary metric and its geometry.}} \label{fig:flat-timelike}
\end{figure}
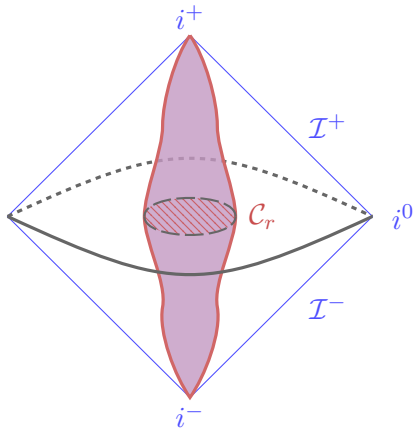

In this section, we show that the boundary degrees of freedom at a generic timelike boundary $\cc_r$ in AdS$_3$ (as depicted in Fig.~\ref{fig:ADS3-timelike}) or the flat spacetimes (as depicted in Fig.~\ref{fig:flat-timelike}) admit a $1+1$ dimensional relativistic (but non-conformal) hydrodynamic description. Since the analysis of flat space case can be obtained by simply setting $\Lambda=0$, we focus on the AdS case and in section \ref{sec:HCB-flat} we discuss the flat case. 

\paragraph{Causal boundary Brown-York stress tensor.} We start with extrinsic curvature of constant $r$ surfaces $K_{\mu \nu}$,
\begin{equation}\label{eq:extrinsicC}
        K_{\mu \nu}:= \frac{1}{2} \gamma^\alpha_{\mu} \gamma^\beta_{\nu} \mathcal{L}_s \gamma_{\alpha \beta} = \nabla_{(\mu} s_{\nu)}- s_{(\mu} s \cdot \nabla s_{\nu)}\, ,
\end{equation}
where  $\gamma^\alpha_{\mu}= g^{\alpha\nu} \gamma_{\mu\nu}$. 
The causal boundary Brown-York energy-momentum tensor \cite{Balasubramanian:1999re}
\begin{equation}\label{r-BY}
    \mathcal{T}^{\mu\nu}=\frac{1}{8\pi G} \left( K^{\mu\nu}-K \, \gamma^{\mu\nu}{+\frac{1}{\ell}\,  \gamma^{\mu\nu}} \right) \, ,\qquad \ell^2=-1/\Lambda,
\end{equation}
is by construction a symmetric tensor whose contraction with the normal vector to the causal boundary $s^\mu$ vanishes, $\mathcal{T}^{\mu\nu}s_\nu=0$. It can hence  be decomposed as
\begin{equation}\label{Brown-York-mu-nu}
    \begin{split}
        \mathcal{T}^{\mu\nu} =&  - \mathcal{E}\, (t^\mu t^\nu+k^{\mu}k^{\nu}) + 2 \mathcal{J}\ k^{(\mu} t^{\nu )}+\frac12\mathcal{T}\, (-t^\mu t^\nu+k^{\mu}k^{\nu})\, ,
    \end{split}
\end{equation}
where
\begin{equation}\label{Tmunu-BY-compt}
    {\mathcal{E}:=-\frac{1}{16 \pi G}\, \left( \theta_s+\kappa_t\right)}\, , \qquad \mathcal{T}:= \frac{1}{8 \pi G} \left(\kappa_t - \theta_s {+\frac{2}{\ell}} \right)\, ,  
    \qquad \mathcal{J}:= \frac{\omega_s}{8 \pi G}\, ,
\end{equation}
where $\mathcal{T}$ is the trace of the causal boundary Brown-York energy-momentum tensor and $\theta_s, \omega_s, \kappa_t$ are defined in \eqref{rel:scalarsS}. One can readily check that $\mathcal{T}^{\mu\nu}$ has only 3 non-zero components. These components are those along the constant $r$ surface and will be denoted by  $\mathcal{T}^{ab}=\gamma_\mu^a\gamma_\nu^b\mathcal{T}^{\mu\nu}$.

\paragraph{Symplectic potential.}
The symplectic potential on the causal boundary $\cc_r$, a constant $r$ hypersurface, is $\bTh_{_{\mathcal{C}}}:=\int_{_{\mathcal{C}_r}}  \Theta^\mu \, \d x_\mu $, with $\Theta^\mu$ defined in \eqref{Theta-mu-generic}. Its explicit form is
\begin{equation}\label{Theta-Spacelike}
    \begin{split}
      \bTh_{_{\mathcal{C}}}&=\int_{\mathcal{C}_r} \left\{- \frac{1}{2}\,  s^\mu\,\sqrt{-g}\mathcal{T}^{\alpha\beta}\delta \gamma_{\alpha \beta}+\delta \left[- \frac{\sqrt{-g}}{ 8\pi G} {\left( K -\frac{1}{\ell}\right)} s^\mu+{s} \cdot L_{\text{\tiny b}}\, s^\mu\right]\right\}\d{}^2 x_\mu\\
        &+\int_{\mathcal{C}_r}  \partial_\nu \left(\frac{\sqrt{-g}}{ 8\pi G}\, s^{[\mu}\delta s^{\nu]}+Y^{\mu\nu}[g; \delta g ]\right)\d{}^2 x_\mu\, .
    \end{split}
\end{equation}
One way to fix the two freedoms $ L_{\text{\tiny b}}^\mu, Y^{\mu\nu}$ is the well-known Gibbons-Hawking-York boundary term \cite{York:1972sj, Gibbons:1976ue, Wald:1984rg, Chandrasekaran:2021hxc}
\begin{equation}\label{chioce-Y-01}
\begin{split}
    L_{\text{\tiny b}}^\mu=L_{\text{\tiny GHY}}^\mu+L_{\circ}^\mu\, ,\qquad 
    Y^{\mu\nu}=Y_{\text{\tiny GHY}}^{\mu\nu}+Y^{\mu\nu}_{\circ}\, ,
\end{split} \end{equation}
where 
\begin{equation}\label{GHY-Lb-Y}
    L_{\text{\tiny GHY}}^\mu:=\frac{\sqrt{-g}}{ 8\pi G}{\left( K -\frac{1}{\ell}\right)} s^\mu \, , \hspace{1 cm} Y_{\text{\tiny GHY}}^{\mu\nu}:=-\frac{\sqrt{-g}}{ 8\pi G}\, s^{[\mu}\delta s^{\nu]} \, .
\end{equation}
With these choices the symplectic potential becomes
\begin{equation}\label{symplectic-potential-r}
      \bTh_{_{\mathcal{C}}}=\int_{\mathcal{C}_r}\,\d{}^2 x \,\left[- \frac{1}{2}\sqrt{-\gamma}\,\mathcal{T}^{ab}\delta \gamma_{ab}+\delta\left({s} \cdot L_{\circ}\, s^r\right)+\partial_a Y^{ra}_{\circ}[g; \delta g ]\right]\, .
\end{equation}
\paragraph{Symplectic form.} Using the above we learn that 
\begin{equation}\label{symplectic-form-T-gamma}
\begin{split}
     \bO_{_{\mathcal{C}}} &=\int_{\mathcal{C}_r}\,\d{}^2 x \, \left[- \frac{1}{2}\delta ( \sqrt{-\gamma}\,\mathcal{T}^{ab})\wedge \delta \gamma_{ab}+\partial_a \delta Y^{ra}_{\circ}[g; \delta g ]\right]\, \\
&= \int_{\mathcal{C}_r}\,\d{}v \d\phi \, \Biggl[\delta U \wedge \delta \left(\mathcal{R}^2\mathcal{J}\right)+\delta \left(\frac{\sqrt{V}}{{\cal R}}\right)\wedge \delta\left(\mathcal{R}^2\mathcal{E}\right)+ \, \frac12{\delta(\sqrt{V}}{{\cal R}})\wedge \delta \mathcal{T}+\partial_a \delta Y^{ra}_{\circ}[g; \delta g ]\Biggr]\, .
\end{split}
\end{equation}
As we can see, in the absence of $Y_{\circ}$, the \textit{off-shell} symplectic form consists of three causal boundary Brown-York charges $\mathcal{T}^{ab}$ which are canonically conjugate to the boundary metric $\gamma_{ab}$. This $3+3$ $(\gamma_{ab}, \mathcal{T}^{ab})$ decomposition of off-shell configuration space is different than $2+2$ ($\lambda^{-1}, \hat{\mathcal{M}}; {\cal U}, \hat\Upsilon$) plus $1+1$ ($\Omega, \Pi$) decomposition appearing in \eqref{r-independent-symplectic-form}. As we will see below this $3+3$ decomposition provides the setting for a hydrodynamic description at a generic timelike boundary $\cc_r$.

We stress again that $ \mathcal{T}^{ab}$ and their canonical conjugates $\gamma_{ab}$ are $r$-dependent. Moreover,  recalling \eqref{thetas-omegas-kappat}, observe that ${\cal E}, {\cal J}, {\cal T}$ at large $r$ behave as $\sim 1/r^2$ and therefore, combinations ${\cal R}^2{\cal E}, {\cal R}^2{\cal J}$ which appear in the symplectic form \eqref{symplectic-form-T-gamma} remain finite at large $r$ and noting that ${\cal R}\sqrt{V}\sim r^2$ at large $r$, we see that the symplectic form remains finite at large $r$, even without the inclusion of the $Y$-term. That the symplectic form without $Y^{ra}_{\circ}$ term is finite at large $r$, is a result of the addition of Gibbons-Hawking-York term \eqref{GHY-Lb-Y}. This point was also noted in \cite{Alessio:2020ioh, Geiller:2021vpg}. In the $r$-independent symplectic form \eqref{r-independent-symplectic-form} this contribution is removed by 
\begin{equation}\label{Y0-term}
    Y_{\circ}^{\mu\nu}[\delta g; g]=-Y_{\text{\tiny GHY}}^{\mu\nu}[\delta g; g]+\frac{\delta \sqrt{-g}}{8 \pi G}t^{[\mu}s^{\nu]}\, .
\end{equation}
This choice replaces Gibbons-Hawking-York $Y$-term with the one used in \cite{Adami:2022ktn}.

\subsection{Hydrodynamics description at finite distance: \texorpdfstring{AdS$_3$}{} case}\label{sec:HCS--finite-r-AdS}

One should note that the above symplectic form is off-shell and it is subject to  \eqref{eq:constraints-r}. These equations {take} the inspiring form, 
\begin{equation}\label{EMC-r}
\mathscr{D}_b \mathcal{T}^{ab}=0 \, ,
\end{equation}
where $\mathscr{D}_a$ is metric connection compatible with boundary metric $\gamma_{ab}$. Note that the $r$-dependence of $\mathcal{T}^{ab}$  is fixed within the solution space specified by \eqref{metric-components} and \eqref{EMC-r} holds at ${\cal C}_r$ for any $r$. 
These equations relate 2 out of 6 functions and hence, as expected and discussed, the solution phase space is governed by 4 functions over ${\cal C}_r$.

Recalling \eqref{symplectic-form-T-gamma} and \eqref{EMC-r}, the above description resembles a hydrodynamic description at the boundary $\cc_r$ with energy-momentum tensor ${\cal T}^{ab}$ which is canonically conjugate to boundary metric $\gamma_{ab}$ \eqref{Boundary-metric}. While ${\cal T}^{ab}$ is divergence-free, it is not traceless and the hydrodynamic system is not a conformal one.  To understand this, recall that \eqref{EMC-r} is a manifestation of the fact that our boundary symmetries \eqref{null-bondary-sym-gen} include general $2d$ diffeomorphisms at $\cc_r$. Nonetheless, as pointed out, Weyl scaling on $\cc_r$,  ${\gamma}_{ab} \to {\cal W}^{2} \, \gamma_{ab}$, is not a part of our boundary symmetries at generic $r$ and hence the effective relativistic hydrodynamic description at the boundary is not a conformal one. 
As we see all $3+3$ modes in the configuration space appear in our hydrodynamic description on generic $\cc_r$. As we will show in the next subsection in the $r\to\infty$ limit, where the boundary approaches the causal boundary of spacetime, we recover a conformal hydrodynamic description which only involves $2+2$ codimension 1 modes (cf. discussions below \eqref{r-independent-symplectic-form}. The $1+1$ codimension 2 modes associated with the $r$ supertranslations and superscalings, respectively generated by $Z, W$, are not relevant to our hydrodynamic description manifested in \eqref{EMC-r}.

We  note that the hydrodynamic at generic $r$ is not unique. We start with a general analysis of causal boundary Brown-York charges under a Weyl scaling. Consider two boundary metrics related by a Weyl scaling,
\begin{equation}\label{conformal-metric}
 \gamma_{ab} \to   \tilde{\gamma}_{ab}={\cal W}^{-2} \, \gamma_{ab} \, ,
\end{equation}
where $\mathcal{W}$ is a generic function on the spacetime and a scalar in the $\tilde \gamma$--frame. Then, one can readily verify that,
\begin{equation}\label{symplectic-form-Weyl-scaling}
\begin{split}
 \sqrt{-\gamma}\,\mathcal{T}^{ab}\ \delta \gamma_{ab}&= \sqrt{-\tilde\gamma}\,\tilde{\mathcal{T}}^{ab}\  \delta \tilde\gamma_{ab}+2\, \sqrt{-\tilde\gamma}\,\tilde{\mathcal{T}}\ \frac{\delta{\cal W}}{{\cal W}} \\
    \delta ( \sqrt{-\gamma}\,\mathcal{T}^{ab})\wedge \delta \gamma_{ab}
   &=  \delta ( \sqrt{-\tilde{\gamma}}\,\tilde{\mathcal{T}}^{ab})\wedge \delta \tilde{\gamma}_{ab}+2\, \delta (\sqrt{-\tilde{\gamma}} \ \tilde{\cal{T}})\wedge\frac{\delta {\cal W}}{{\cal W}}
\end{split}
\end{equation}
where 
\be
\tilde{{\cal T}}^{ab}={\cal W}^4 {\cal T}^{ab},\qquad \tilde{{\cal T}}:=\tilde{\gamma}_{ab}\tilde{{\cal T}}^{ab}={\cal W}^2 \gamma_{ab}{\cal T}^{ab}:={\cal W}^2 {\cal T}.
\ee
We raise and lower indices for tilde-quantities by $\tilde{\gamma}^{ab}$ and $\tilde{\gamma}_{ab}$ respectively, as such $\mathcal{T}_{ab}= \tilde{\mathcal{T}}_{ab}$. The divergence-free condition \eqref{EMC-r} can be written as,
\be\label{Divergence-tilde-T}
\tilde\nabla_b \tilde{{\cal T}}^{ab}=\frac12 \, {\cal T} \,  \tilde\nabla^a {\cal W}^2
\ee
where $\tilde\nabla_a$ is the covariant derivative w.r.t. $\tilde\gamma_{ab}$.  That is, in a generic Weyl-frame neither the divergence nor the trace of the energy-momentum tensor is zero. 

\paragraph{Divergence-free frames.} The above is true for an arbitrary  Weyl factor ${\cal W}$. One may  choose ${\cal W}=f({{\cal T}})$, where $f$ is an arbitrary function of ${\cal T}$. Then, one can construct a new divergence-free energy-momentum tensor ${\text{T}}^{ab}$, 
\be\label{Divergence-tilde-Ts}
\tilde\nabla_a {\text{T}}^{ab}= 0, \qquad {\text{T}}^{ab}:= \tilde{{\cal T}}^{ab}-\frac12\tilde{{\gamma}}^{ab} F({{\cal T}}) ,\qquad F'= 2{\cal T} f f',\qquad  \text{T}=\tilde{{\gamma}}_{ab}{\text{T}}^{ab}=\int^{{\cal T}} f^2 \d{\cal T},
\ee
where $prime$ denotes derivative w.r.t. the argument. The above also makes it clear that while ${\text{T}}^{ab}$ is divergence-free, it is not trace-less. One may also show,
\begin{subequations}\label{symplectic-form-Weyl-frame}
\begin{align}
\sqrt{-\gamma}\,\mathcal{T}^{ab}\ \delta \gamma_{ab}&= \sqrt{-\tilde\gamma}\,{\text{T}}^{ab}\  \delta \tilde\gamma_{ab}+\delta\left(\sqrt{-\tilde\gamma}\ F({\mathcal{T}}) \right) \, ,\label{presympl-forms-Weyl-frames}\\ 
  \delta ( \sqrt{-\gamma}\,\mathcal{T}^{ab})\wedge \delta \gamma_{ab}
  &= \delta ( \sqrt{-\tilde{\gamma}}\, {\text{T}}^{ab})\wedge \delta \tilde{\gamma}_{ab} \, . \label{sympl-forms-Weyl-frames}
\end{align}
\end{subequations}
To obtain the above we used, $\delta \sqrt{-\tilde{\gamma}}= \frac12 \sqrt{-\tilde{\gamma}}\  \tilde{\gamma}^{ab}\delta\tilde{\gamma}_{ab}$ and that $\delta(\sqrt{-\tilde{\gamma}} \tilde{\gamma}^{ab})\wedge \delta\tilde{\gamma}_{ab}=0$.

Eq.~\eqref{presympl-forms-Weyl-frames} makes it apparent that the Weyl scaling by ${\cal W}=f({\cal T})$, amounts to the addition of $\sqrt{-\tilde\gamma}\ F({\mathcal{T}})$ to the boundary Lagrangian and \eqref{sympl-forms-Weyl-frames} relates two divergence-free energy- momentum tensors which are canonical conjugates to two metrics which are related by the Weyl scaling. In other words and more explicitly, the subclass of Weyl scalings by ${\cal W}=f({\cal T})$ (together with \eqref{Divergence-tilde-Ts}) is a canonical transformation and hence is a (local) symmetry of  our solution space.\footnote{Recalling \eqref{Tmunu-BY-compt} and that ${\cal T}$ is constructed from metric and its derivatives, we note Weyl factor  ${\cal W}=f({\cal T})$ involves scalings of metric by generic functions of metric and its derivatives.} Moreover, dealing with divergence-free stress tensors, one may use either the original or tilde-frames and associated stress tensors, respectively ${\cal T}^{ab}, \text{T}^{ab}$, for the hydrodynamic description. That is, the hydrodynamic description is not unique and since $f({\cal T})$ is an arbitrary function, there are infinitely many such descriptions.\footnote{Notice that as \eqref{symplectic-form-Weyl-scaling} and \eqref{Divergence-tilde-T} show, a generic Weyl factor (not a function of ${\cal T}$) is neither a canonical transformation nor yields a hydrodynamical description.}

\subsection{Hydrodynamics at a generic timelike boundary: 3d flat case}\label{sec:HCB-flat}

While our general analysis of the previous subsection  generically works for $\Lambda<0$ and $\Lambda=0$, we focused more on  the AdS$_3$, corresponding to $\Lambda=-1/\ell^2$, case. In this section and for completeness we also present the 3d flat space case and consider a generic causal boundary at a constant finite $r$ surface in 3d flat spacetime. To this end, we need to take $\ell\to \infty$ limit of the results in the previous section. 

The only difference between flat and AdS cases at the level of the metric  is in the absence of ${\cal R}^2$ term in $V$ \eqref{metric-components}, 
\begin{equation}\label{V-lambda}
    V|_{\Lambda=0}= \frac{1}{\lambda^2}\left( -\mathcal{M}  + \frac{\Upsilon^2}{4  {\cal R}^2} - \frac{2\mathcal{R}}{\eta }   \mathcal{D}_v ( \eta \lambda ) +\frac{\Upsilon}{\mathcal{R}}\, \frac{\partial_\phi \eta}{ \eta } \right)
\end{equation}
and at the level of the boundary symmetry algebra, we have \eqref{Heisenberg-direct-sum-algebra''} with $\Lambda=0$ which is a direct sum of  BMS$_3$ and Heisenberg algebras. The other geometric quantities at a generic boundary $\cc_r$ discussed in section \ref{sec2:soln-space-review}, like the dribeins $t,s,k$ and $\theta_s, \omega_s, \kappa_t$ are still given by the same expressions with $V$ given in \eqref{V-lambda}.

The first step in our hydrodynamic description is working out the causal boundary Brown-York stress tensor ${\cal T}^{\mu\nu}$. For the flat spacetime, we need to make two different modifications, one in the equations \eqref{r-BY}-\eqref{GHY-Lb-Y}  where $\ell$ appears explicitly and we can simply take the flat  $\ell\to\infty$ limit  and the other one is in the implicit dependence on $\ell$ through $V.$ In the latter case, we only need to replace $V$ with \eqref{V-lambda}. Boundary Einstein equations are still given by \eqref{EMC-r}, with a hydrodynamic description as the AdS case. In this case, too, the trace of ${\cal T}^{ab}$ is nonzero and for generic $r$ we deal with a non-conformal fluid. One may readily check that analysis in section \ref{sec:HCS--finite-r-AdS} works verbatim for $\Lambda=0$. In particular, the hydrodynamic symplectic form still takes the same form as in \eqref{symplectic-form-T-gamma}. 

\section{Hydrodynamics at a generic null or asymptotic boundaries }\label{sec:Three-Special-cases}

\begin{figure}[t]
	\centering
\subcaptionbox{}{
\centering
\begin{tikzpicture}[scale=1.2]
    \node (v1) at (0,-2) {};
    \fill (v1)+(0,0.6)  node [darkred!90] {\tiny{$\cal N$}};
    \node (v2) at (1.5,{-2+1.5*tan(45)}) {};
    \node (v3) at (-1.5,{-2-1.5*tan(45)}) {};
    \node (v4) at (1.5,0) {};
    \node (v5) at (-1.5,0) {};
    \node (v6) at (1.5,-4) {};
    \fill (v6)+(0.3,0.2)  node [right,blue!80] {\tiny{AdS$_3$ \text{b'dry}}};
    \node (v7) at (-1.5,-4) {};
\filldraw [fill=darkred!70, rotate around={45:($(v1)$)}]($(v1)$) ellipse ({1.5/cos(45)} and 0.2);
\begin{scope}[fill opacity=0.8, very thick,black!80]  
    \filldraw [fill=darkred!40!blue!60, very thick]($(v2)+(0,0)$) 
        to[out=270,in=90] ($(v6) + (0,0)$)
        to[out=270,in=270, looseness=0.6] ($(v7) + (0,0)$)
        to[out=90,in=270] ($(v3) + (0,0)$)
        to[out=340,in=290,looseness=0.2] ($(v2) + (0,0)$);
    \end{scope}
\begin{scope}[fill opacity=0.8, very thick,black!60] 
    \filldraw [pattern=north west lines, pattern color=blue, thick,dash pattern={on 7pt off 2pt}] ($(v2) + (0,0)$)
        to[out=270,in=90] ($(v6) + (0,0)$)
        to[out=90,in=90, looseness=0.6] ($(v7) + (0,0)$)  
        to[out=90,in=270] ($(v3) + (0,0)$)
        to[out=110,in=160,looseness=0.2] ($(v2) + (0,0)$);
    \end{scope}
\draw [blue!60](0,0) ellipse (1.5 and 0.5);
\draw [blue!60] (0,-4) ellipse (1.5 and 0.5);
\draw [blue!60] (-1.5,0) -- (-1.5,-4);
\draw [blue!60] (1.5,0) -- (1.5,-4);
\end{tikzpicture}}
\subcaptionbox{}{
\centering $\hspace{1cm}$
\begin{tikzpicture}[scale=1.2]
    \node (v1) at (0,-2) {};
    \fill (v1)+(0,-0.6)  node [darkred!90] {\tiny{$\cal N$}};
    \node (v2) at (1.5,{-2-1.5*tan(45)}) {};
    \node (v3) at (-1.5,{-2+1.5*tan(45)}) {};
    \node (v4) at (1.5,0) {};
    \node (v5) at (-1.5,0) {};
    \node (v6) at (1.5,-4) {};
    \fill (v6)+(0.3,0.2)  node [right,blue!80] {\tiny{AdS$_3$ \text{b'dry}}};
    \node (v7) at (-1.5,-4) {};
\filldraw [fill=darkred!70, rotate around={135:($(v1)$)}]($(v1)$) ellipse ({1.5/cos(45)} and 0.2);
\begin{scope}[fill opacity=0.8, very thick,black!80]  
    \filldraw [fill=darkred!40!blue!60, very thick]($(v2)+(0,0)$) 
        to[out=90,in=270] ($(v4) + (0,0)$)
        to[out=266,in=274, looseness=0.6] ($(v5) + (0,0)$)
        to[out=270,in=90] ($(v3) + (0,0)$)
        to[out=20,in=0,looseness=0.2] ($(v2) + (0,0)$);
    \end{scope}
\begin{scope}[fill opacity=0.8, thick,black!80!blue!60] 
    \filldraw [pattern=north west lines, pattern color=blue, thick,dash pattern={on 7pt off 2pt}] ($(v2) + (0,0)$)
        to[out=90,in=270] ($(v4) + (0,0)$)
        to[out=90,in=90, looseness=0.56] ($(v5) + (0,0)$)  
        to[out=270,in=90] ($(v3) + (0,0)$)
        to[out=245,in=270,looseness=0.2] ($(v2) + (0,0)$);
    \end{scope}
\draw [blue!60](0,0) ellipse (1.5 and 0.5);
\draw [blue!60] (0,-4) ellipse (1.5 and 0.5);
\draw [blue!60] (-1.5,0) -- (-1.5,-4);
\draw [blue!60] (1.5,0) -- (1.5,-4);
\end{tikzpicture}}
\caption{\footnotesize{A generic null boundary in $AdS_3$. The null boundary ${\cal N}$ may be viewed as future (Left figure) or past (Right figure) Poincar\'e horizon on AdS. We formulate physics in the shaded region and focus on the boundary degrees of freedom residing at ${\cal N}$.}} \label{fig:ADS3-null}
\end{figure}
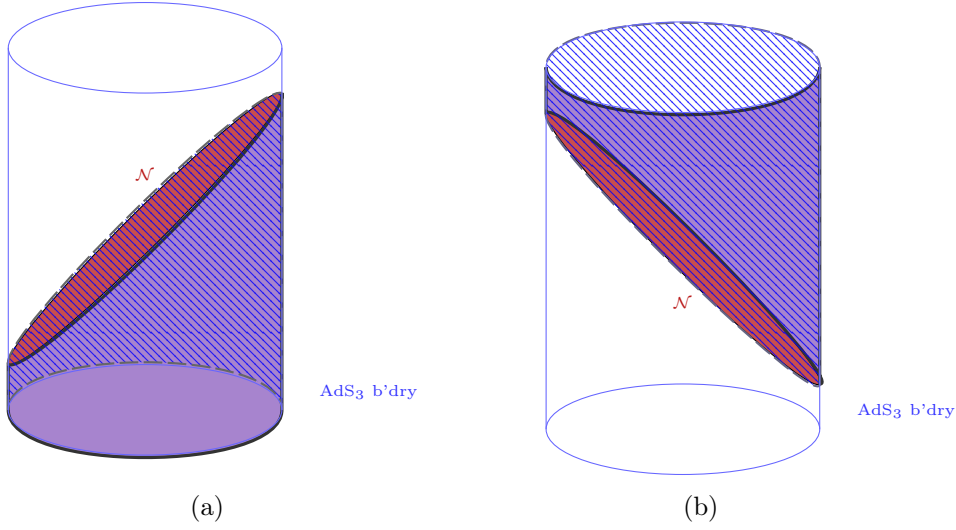

\begin{figure}[t]
	\centering
\subcaptionbox{}{
\centering
\begin{tikzpicture}[scale=1.2]
    \node (v1) at (0,0) {};
    \fill (v1)+(0,0.2)  node [blue!70] {$i^{+}$};
    \node (v2) at (-2,-2) {};
    \node (v3) at (2,-2) {};
    \fill (v3)+(0.1,0)  node [right, blue!70] {$i^{0}$};
    \node (v4) at (0,-4) {};
    \fill (v4)+(0,-0.2)  node [blue!70] {$i^{-}$};
	\node (v5) at (1,-2) {};
	\node (v51) at (0,-1) {};
    \fill (v51)+(-0.6,-0.3)  node [darkred!70] {${\cal N}$};
	\node (v6) at (-1,-2) {};
	\node (v52) at (0,-2.5) {};
 	\node (v7) at (1.5,-2.5) {};
	\node (v8) at (0.75,-3.25) {};
 	\node (v9) at (-0.75,-3.25) {};
	\node (v10) at (-1.5,-2.5) {};
    \node [right] at (1.2,-3) {$\textcolor{blue!70}{\mathcal{I}^{-}}$};
\node [right] at (1.2,-1) {$\textcolor{blue!70}{\mathcal{I}^{+}}$};
\begin{scope}[fill opacity=0.8, black!60,dash pattern={on 2pt off 2pt}] 
    \draw ($(v2) + (0,0)$)
        to[out=35,in=145,looseness=0.4] ($(v3) + (0,0)$)
        to[out=215,in=325,looseness=0.4] ($(v2) + (0,0)$);
    \end{scope}
\draw[fill opacity=0.8, black!60,dash pattern={on 15pt off 4pt}] ($(v1) + (0,0)$) to ($(v4) + (0,0)$);
\begin{scope}[fill opacity=0.8, ultra thick, darkred!60] 
    \filldraw [fill=darkred!60!blue!40] ($(v51) + (0,0)$)
        to[out=315,in=135] ($(v7) + (0,0)$)
        to[out=225,in=45] ($(v4) + (0,0)$)
        to[out=135,in=315] ($(v10) + (0,0)$)
        to[out=45,in=225] ($(v51) + (0,0)$);
    \end{scope}
\begin{scope}[fill opacity=0.8, very thick,black!60,dash pattern={on 2pt off 2pt}] 
    \filldraw [pattern=north west lines, pattern color=darkred!70, thick,dash pattern={on 7pt off 2pt}] ($(v5)+(0,0)$) 
        to[out=145,in=35, looseness=0.4] ($(v6) + (0,0)$)
        to[out=325,in=215,looseness=0.4] ($(v5) + (0,0)$);
    \end{scope}
\begin{scope}[fill opacity=0.8, very thick,black!60,dash pattern={on 2pt off 2pt}] 
    \filldraw [pattern=north west lines, pattern color=darkred!70, thick,dash pattern={on 7pt off 2pt}] ($(v8)+(0,0)$) 
        to[out=145,in=35, looseness=0.4] ($(v9) + (0,0)$)
        to[out=325,in=215,looseness=0.4] ($(v8) + (0,0)$);
    \end{scope}
\begin{scope}[fill opacity=0.8, thick,black!60,dash pattern={on 7pt off 2pt}] 
    \draw ($(v7)+(0,0)$) 
        to[out=145,in=35, looseness=0.4] ($(v10) + (0,0)$)
        to[out=325,in=215,looseness=0.4] ($(v7) + (0,0)$);
    \end{scope}
\begin{scope}[fill opacity=0.8, thick,blue!60,] 
    \draw ($(v1)+(0,0)$) 
        to[out=315,in=135] ($(v3) + (0,0)$) 
        to[out=225,in=45,thick] ($(v4) + (0,0)$)
        to[out=135,in=315,thick] ($(v2) + (0,0)$)
        to[out=45,in=225,thick] ($(v1) + (0,0)$);
        \end{scope}
\end{tikzpicture}}
\subcaptionbox{$\hspace{-1cm}$}{
\centering $\hspace{1cm}$
\begin{tikzpicture}[scale=1.2]
    \node (v1) at (0,0) {};
    \fill (v1)+(0,0.2)  node [blue!70] {$i^{+}$};
    \node (v2) at (-2,-2) {};
    \node (v3) at (2,-2) {};
    \fill (v3)+(0.1,0)  node [right, blue!70] {$i^{0}$};
    \node (v4) at (0,-4) {};
    \fill (v4)+(0,-0.2)  node [blue!70] {$i^{-}$};
	\node (v5) at (1,-2) {};
	\node (v51) at (0,-3) {};
    \fill (v51)+(-0.65,0.3)  node [darkred!70] {${\cal N}$};
	\node (v6) at (-1,-2) {};
	\node (v52) at (0,-2.5) {};
 	\node (v7) at (1.5,-1.5) {};
	\node (v8) at (0.75,-0.75) {};
 	\node (v9) at (-0.75,-0.75) {};
	\node (v10) at (-1.5,-1.5) {};
    \node [right] at (1.2,-3) {$\textcolor{blue!70}{\mathcal{I}^{-}}$};
\node [right] at (1.2,-1) {$\textcolor{blue!70}{\mathcal{I}^{+}}$};
\begin{scope}[fill opacity=0.8, black!60,dash pattern={on 2pt off 2pt}] 
    \draw ($(v2) + (0,0)$)
        to[out=35,in=145,looseness=0.4] ($(v3) + (0,0)$)
        to[out=215,in=325,looseness=0.4] ($(v2) + (0,0)$);
    \end{scope}
\draw[fill opacity=0.8, black!60,dash pattern={on 15pt off 4pt}] ($(v1) + (0,0)$) to ($(v4) + (0,0)$);
\begin{scope}[fill opacity=0.8, ultra thick, darkred!70] 
    \filldraw [fill=darkred!60!blue!40] ($(v1) + (0,0)$)
        to[out=315,in=135] ($(v7) + (0,0)$)
        to[out=225,in=45] ($(v51) + (0,0)$)
        to[out=135,in=315] ($(v10) + (0,0)$)
        to[out=45,in=225] ($(v1) + (0,0)$);
    \end{scope}
\begin{scope}[fill opacity=0.8, very thick,black!60,dash pattern={on 2pt off 2pt}] 
    \filldraw [pattern=north west lines, pattern color=darkred!70, thick,dash pattern={on 7pt off 2pt}] ($(v5)+(0,0)$) 
        to[out=145,in=35, looseness=0.4] ($(v6) + (0,0)$)
        to[out=325,in=215,looseness=0.4] ($(v5) + (0,0)$);
    \end{scope}
\begin{scope}[fill opacity=0.8, very thick,black!60,dash pattern={on 2pt off 2pt}] 
    \filldraw [pattern=north west lines, pattern color=darkred!70, thick,dash pattern={on 7pt off 2pt}] ($(v8)+(0,0)$) 
        to[out=145,in=35, looseness=0.4] ($(v9) + (0,0)$)
        to[out=325,in=215,looseness=0.4] ($(v8) + (0,0)$);
    \end{scope}
\begin{scope}[fill opacity=0.8, thick,black!60,dash pattern={on 7pt off 2pt}] 
    \draw ($(v7)+(0,0)$) 
        to[out=145,in=35, looseness=0.4] ($(v10) + (0,0)$)
        to[out=325,in=215,looseness=0.4] ($(v7) + (0,0)$);
    \end{scope}
\begin{scope}[fill opacity=0.8,thick,blue!60,] 
    \draw ($(v1)+(0,0)$) 
        to[out=315,in=135] ($(v3) + (0,0)$) 
        to[out=225,in=45,thick] ($(v4) + (0,0)$)
        to[out=135,in=315,thick] ($(v2) + (0,0)$)
        to[out=45,in=225,thick] ($(v1) + (0,0)$);
        \end{scope}
\end{tikzpicture}}
\caption{Null boundaries ${\cal N}$ in a flat $3d$ space. The left and right figures respectively show a future or past null boundary. We are interested in formulating physics in the shaded regions bounded by ${\cal N}$.}\label{fig:flat-null}
\end{figure}
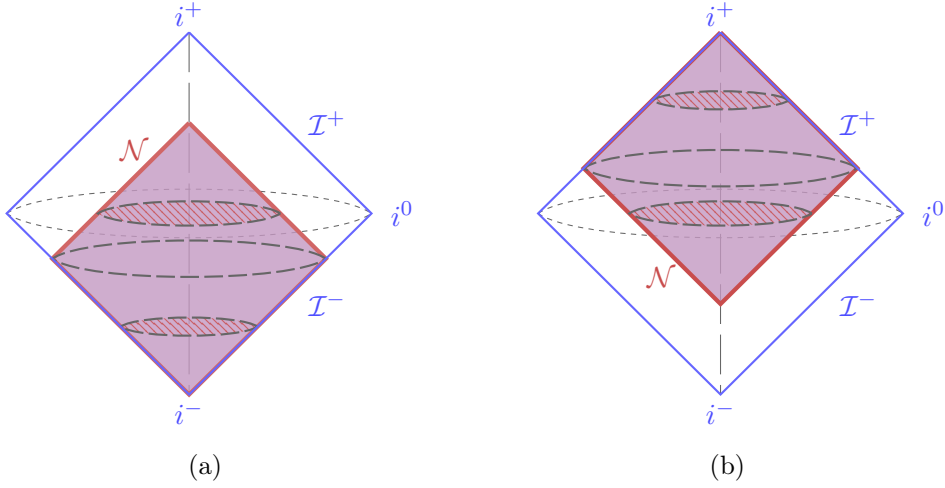

The causal boundary Brown-York stress tensor ${\cal T}^{ab}$ defined at  $\cc_r$ for finite $r$ laid the basis for our hydrodynamic description of the previous section. This stress tensor is constructed out of geometric quantities like $\theta_s, \kappa$ and $\omega_s$ as well as the boundary metric $\gamma_{ab}$. These geometric quantities need to be reconsidered and redefined in two special cases, when we consider asymptotic $r\to \infty$ and when the boundary is a null surface. The latter can be at finite $r$ in AdS or flat cases as depicted in Fig.~\ref{fig:ADS3-null} and Fig.~\ref{fig:flat-null}, like the cases discussed in \cite{Adami:2020ugu}, or can be the asymptotic boundary of $3d$ flat space ${\cal I}$, as depicted in the Right figure in Fig.~\ref{fig:asympt-ads-flat}. In this section, we explore hydrodynamic description of the boundary theory in these three different cases.

\subsection{Hydrodynamics on a null surface at finite distance}\label{sec:NBH-finite-r}

In this section, we consider a generic null boundary in a spacetime with $\Lambda\leq 0$. Moreover, we restrict the spacetime to $V \geq 0$, which we associate with $r\geq 0$ in our adopted coordinate system. Requiring that $\cc_r$ at finite $r$ is a null surface {that} amounts to having $V=0$ at the position of the boundary. Requiring the null boundary ${\cal N}$ to be located at  $r=0$  yields
\begin{equation}\label{V=0-null}
    V(r=0)=0 \hspace{.5 cm} \Rightarrow \hspace{.5 cm} \mathcal{M} =  -\Lambda\Omega^2 + \frac{\Upsilon^2}{4  \Omega^2} - \frac{2\,\Omega}{\eta }   \mathcal{D}_v ( \eta \lambda ) +\frac{\Upsilon}{\Omega}\, \frac{\partial_\phi \eta}{ \eta } \, .
\end{equation}
$V=0$ may be viewed as a second-class constraint on a generic causal boundary solution phase space. This yields the null surface solution space analyzed in detail in  \cite{Adami:2020ugu} which is described by three codimension 1 functions, compared to 4 functions for  the cases with a timelike boundary \cite{Adami:2022ktn}. That is, when we take the null limit of the timelike boundary $\cc_r$,  we lose one of the charges and its canonical conjugate.

The 2-dimensional null boundary ${\cal N}$ is a Carrollian geometry which is described by an enhanced Carrollian structure \cite{Henneaux:2021yzg,Bagchi:2022eav,Campoleoni:2022ebj}, a 1-dimensional metric, a kernel, and relevant connections and covariant derivatives. In  two subsections  \ref{sec:NBH-finite-r} and \ref{Hydrodynamics at infinity:flat-case}, we introduce the Carrollian structure that emerges at the null limit for a boundary at finite $r$ and asymptotic region of $3d$ flat spacetime.

\paragraph{Basics of Carrollian geometry.} The two-dimensional metric \eqref{Boundary-metric} becomes degenerate in the null limit where $V$ vanishes. Constraint \eqref{V=0-null} induces the following degenerate metric on the boundary
\begin{equation}\label{gammabar}
\d{} \bar{\gamma}^2 :=\bar{\gamma}_{ab} \d x^a \d x^b =  \Omega^2 \left( \d \phi + \bar{U}  \d v\right)^2 \, .
\end{equation}
where $\bar{U}=U(r=0)$. From now on, we use the barred notion for quantity $X$ to indicate it computed at $r=0$, namely, $\bar{X}=X(r=0)$. The null boundary is  a Carrollian geometry which besides the degenerate metric 
\begin{equation}
    \bar{\gamma}_{ab}:=
    \bar{k}_{a}\bar{k}_{b}\, , \hspace{1 cm}   \bar{k}_{a}\d{}x^a=\Omega \left( \d \phi + \bar{U}  \d v\right)\, , \qquad \bar{k}^a\partial_{a}=\Omega^{-1}\partial_{\phi} \, ,
\end{equation}
is equipped with the kernel
\begin{equation}
    \bar{l}^{a}\partial_{a}:=
    \partial_{v}-\bar{U}\partial_{\phi}\, , \qquad \bar{\gamma}_{ab}\bar{l}^{a}=0\, ,
\end{equation}
$\bar{l}^a$ is a null vector that generates the null boundary from the $3d$ embedding space viewpoint, cf. discussions in section \ref{sec:2+1-decom}. Since the metric is degenerate, we need to track the index placement of the geometric quantities. In this regard, we define the one-form dual of $\bar{l}^{a}$ as follows
\begin{equation}
    \bar{n}_{a}\d x^a:=- \d v \, , \qquad  \bar{l}^{a}\bar{n}_{a}=-1\, .
\end{equation}
This one-form field allows us to define the Ehreshmann connection \cite{Bekaert:2015xua, Ciambelli:2019lap}. The one-forms $\{ \bar{n}_a, \bar{k}_a \}$ are linearly independent and they form a basis of the cotangent space. Accordingly, the dual basis of the tangent space is $\{\bar{l}^a \, ,\bar{k}^a \} $ and they fulfill the relation
\begin{equation}\label{Identity}
   - \bar{l}^a\bar{n}_b +\bar{k}^a \bar{k}_b = \delta^a_b \, .
\end{equation}
The vector field $\bar{l}^a$ and the one-form $\bar{n}_a$ are  respectively pullbacks of the vector field $l^\mu$ and the one-form $n_\mu$ defined in \eqref{null-vectors} to the null surface $r=0$, up to proportionality factors.
The identity \eqref{Identity} allows us to define a projector onto horizontal forms
\begin{equation}
    \bar{\gamma}^a{}_b =  \delta_b^a + \bar{l}^a \,\bar{n}_b \,  ,
\end{equation}
and $\bar{\gamma}^{ab}= \bar{k}^a \bar{k}^b$ defines a partial inverse of the degenerate metric. Taking the null vector $n^\mu$ defined in section \ref{sec:2+1-decom} to be a null rigging vector, we can define a projector for spacetime vectors onto the null
surface
\begin{equation}
    \text{P}^\mu{}_\nu =  \delta^\mu_\nu +l^\mu \, n_\nu \, .
\end{equation}
{Using this projection operator, we define a rigged connection on the null surface \cite{Mars:1993mj}}
\begin{equation}\label{rigged-connection}
    \mathbb{D}_a X^b := \text{P}^\mu{}_{a} \text{P}^b{}_{\nu} \nabla_\mu X^\nu \, , \qquad \mathbb{D}_a X_b := \text{P}^\mu{}_{a} \text{P}^\nu{}_{b} \nabla_\mu ( \text{P}^\lambda{}_{\nu} X_\lambda)\, ,
\end{equation}
where $X^a=  \text{P}^a{}_{\mu} X^\mu$ and $X_a= \text{P}^\mu{}_{a} X_\mu$ for any vector $X^\mu$. 

\paragraph{Field equations and solution space.} One may rewrite \eqref{eq:constraints-r} in terms of variables more appropriate to the null boundary ${\cal N}$. To this end, we note that from \eqref{rel:scalarsS} and \eqref{kappa-thetal-thetan} we learn, 
\be\label{time-like--vs-null}
\sqrt{V}\kappa_t=-\kappa +\frac{\text{D}_v V}{2V}\, ,  \qquad \sqrt{V}\theta_s=\theta_l +\frac{\lambda V}{2{\cal R}}\, ,\qquad  \omega_s=\omega_l -\frac{\partial_\phi V}{2{\cal R}V}\, .
\ee
Replacing the above into \eqref{eq:constraints-r} and using constraint $V=0$ on the null boundary ${\cal N}$, we arrive at following equations which yields the desired null field equations
\begin{subequations}\label{eom-null}
    \begin{align}
       &{\bar{D}_{v}(\Omega^2\bar{\omega}_l)-{\Omega}\partial_{\phi}\bar{\kappa}=0\, ,} \label{Raych-eq}\\
        & \bar{D}_{v} \bar{\theta}_l+(\bar{\theta}_l-\bar{\kappa})\bar{\theta}_{l}=0\, ,\label{Damour-eq}
    \end{align}
\end{subequations}
where $\bar\kappa, \bar{\omega}_l, \bar\theta_l$ are obtained from $\kappa,{\omega}_l, \theta_l$ in \eqref{kappa-thetal-thetan} computed at $r=0$. \footnote{{Here we extend the results presented in \cite{Donnay:2019jiz}, which demonstrated that the geometry of a black hole horizon can be described by a Carrollian geometry emerging from an ultra-relativistic limit, in two ways: First, this applies to a generic null surface in the bulk. Second, the ultra-relativistic limit can be obtained by sending the boost parameter $\ln{\sqrt{V}}$ to infinity. This result is consistent with the recent observation \cite{Bagchi:2023ysc} that any null fluid exhibits Carrollian structure.}} {In the following, we show that the above equations can be derived from the conservation of an energy-momentum tensor.}  
We also note that \eqref{Damour-eq} and \eqref{Raych-eq} yield some useful relations that will be used in the future,
\begin{equation}\label{Null-EoM-treated}
    \bar\kappa=\bar{D}_{v} \ln(\Omega\bar\theta_l)\, , \qquad \bar{D}_{v}\bigg(\Omega^2\bar{\omega}_l- \Omega \partial_\phi \ln\bar\theta_l\bigg) =0\, .
\end{equation}

\paragraph{Null boundary Brown-York stress tensor.} To construct the null version of the Brown-York stress tensor we start from the definition of the shape operator or Weingarten map,
\begin{equation}
    \mathbb{W}^a{}_{b} := \bar{\theta}_l \, \bar{k}^a \bar{k}_{b} +\bar{l}^a \left(\bar{\omega}_{l}\, \bar{k}_b -\bar{\kappa} \, \bar{n}_b\right)\, , \hspace{.5 cm}  \mathbb{W}:= \mathbb{W}^a{}_{a}=\bar{\theta}_l+\bar{\kappa}\, .
\end{equation}
{Following the construction presented in \cite{Chandrasekaran:2021hxc}, we define the null boundary energy-momentum tensor as
\begin{equation}
    \mathbb{T}^a{}_b := -\frac{1}{8\pi G} \left( \mathbb{W}^a{}_{b} - \mathbb{W}\, \delta^a_{b}\right)\, .
\end{equation}
{This is analogous to the Brown-York energy-momentum tensor \eqref{r-BY} for a null boundary.} The explicit form of this null energy-momentum tensor is given by
\begin{equation}\label{Tab-null}
     \mathbb{T}^a{}_b 
     =\frac{1}{8\pi G}\left[\bar{\kappa}\, \bar{k}^{a}\bar{k}_{b} - \bar{\omega}_{l}\,\bar{l}^a \bar{k}_{b}-\bar{\theta}_{l}\,\bar{l}^a\bar{n}_{b}\right]\, , \hspace{.5 cm}  \mathbb{T}:= \mathbb{T}^a{}_{a}=\frac{1}{8\pi G}(\bar{\theta}_l+\bar{\kappa})\, .
\end{equation}

{To study its conservation through the rigged connection \eqref{rigged-connection}, we need the spacetime uplift of this tensor. In this regard, we define the spacetime version of the null energy-momentum tensor \eqref{Tab-null} as follows}
\begin{equation}
     \mathbb{T}^\mu{}_{\nu}:=\frac{1}{8\pi G}\left[\kappa \, \delta^{\mu}_{\nu} - \omega_{l}\, l^{\mu}\, k_{\nu}+(\kappa-\theta_{l})\, l^{\mu}\, n_{\nu}\right]\, .
\end{equation}
It is easy to check that $ \mathbb{T}^a{}_b=\text{P}^{a}{}_{\mu}\text{P}^{\nu}{}_{b}\mathbb{T}^\mu{}_{\nu}$ and $ {8\pi G \,\mathbb{T}^\mu{}_{\nu}l^\nu=\theta_l l^\mu}$. We now have all materials to consider the conservation equation associated with this energy-momentum tensor, namely,
\begin{equation}
    \mathbb{D}_{a} \mathbb{T}^a{}_b=\text{P}^{\nu}{}_{b}\text{P}^{\alpha}{}_{\mu}\nabla_{\alpha}\mathbb{T}^\mu{}_\nu\, .
\end{equation}
The projection of this equation along $\bar{l}^a$ and $\bar{k}^b$ gives
\begin{subequations}
    \begin{align}
        &\bar{l}^b\mathbb{D}_{a} \mathbb{T}^a{}_b {|_{_{{\cal N}}}}=\frac{1}{8\pi G}\left[\bar{l}^{\mu}\partial_{\mu}\bar{\theta}_{l}+(\bar{\theta}_l-\bar{\kappa})\bar{\theta}_l\right]=0 \, ,\label{omegal-EoM}\\
        & \bar{k}^b\mathbb{D}_{a} \mathbb{T}^a{}_b {|_{_{{\cal N}}}}=-\frac{1}{8\pi G}\left[\bar{l}^{\mu}\partial_{\mu}\bar{\omega}_{l} + 2\bar{\omega}_l\, \bar{\theta}_l - \bar{k}^{\mu}\partial_{\mu}\bar{\kappa}\right]=0 \, .\label{thetal-EoM}
    \end{align}
\end{subequations}
The above yield the Raychaudhuri and Damour equations \eqref{eom-null}, once we recall that $\bar{l}^{\mu}\partial_{\mu}= \bar{D}_v$ and $\Omega\bar\theta_l=\bar{D}_v\Omega$. 

To summarize this part, $ \mathbb{D}_{b} \mathbb{T}^b{}_a=0$ and therefore $\mathbb{T}^b{}_a$ may be viewed as energy-momentum tensor associated with the null surface hydrodynamics. Note also that $\mathbb{T}^b{}_a \bar{l}^b=\frac{\bar{\theta}_{l}}{8\pi G} \bar{l}^a$, i.e. the expansion of null boundary $\bar{\theta}_{l}$ is the eigenvalue of the energy-momentum tensor along the null vector $\bar{l}^a$. This eigenvalue equation and $\mathbb{D}_{b} \mathbb{T}^b{}_a=0$ may be viewed as properties defining the energy-momentum tensor $\mathbb{T}^b{}_a$.

\paragraph{Symmetry generators at null boundary.} To have a consistent solution space in which the null boundary is fixed at $r=0$, we need to impose further $\delta V(r=0)=0=\delta_{\xi}V(r=0)$. Then from \eqref{delta-V} we get 
\begin{equation}\label{deltaV0}
    \delta_{\xi} V|_{r=0}=\xi_{0}^{r}V_1-2\eta\bar{D}_{v}\xi^{r}_{0}=0\, .
\end{equation}
where $\bar{D}_{v}:=\partial_{v}-\mathcal{L}_{\bar{U}}$. A simple and natural solution to \eqref{deltaV0} is $\xi_{0}^{r}=0$. This condition in terms of the causal boundary symmetries \eqref{null-bondary-sym-gen} yields
\begin{equation}
    Z  =\frac{\Upsilon}{2\eta\lambda^2\Omega} \, \partial_{\phi}T + \frac{1}{\eta^2\lambda }\partial_{\phi}\left(\frac{\eta\partial_{\phi}T}{\lambda}\right)\, ,
\end{equation}
which fixes $Z$ in terms of $T$ and its derivatives. We note that this is a field-dependent condition for the $Z$ generator. So, at the null boundary ${\cal N}$ we remain with symmetry generators $T, W, Y$ and the associated transformations and charges. It is clear that $T,Y$ do not change ${\cal N}$ at $r=0$ and that $W$, being proportional to $r\partial_r$ does not change $r=0$ surface either. See \cite{Adami:2020ugu, Adami:2021nnf} for more details. 

\paragraph{Hydrodynamic symplectic potential and symplectic forms.}
It is easy to see that the Gibbons-Hawking-York freedoms \eqref{GHY-Lb-Y} are not appropriate for the null case. 
One appropriate choice, as discussed in \cite{Adami:2020ugu, Adami:2021kvx, Sheikh-Jabbari:2022xix} is to consider the Lee-Wald symplectic potential for the null case, explicitly, 
\begin{equation}\label{Y0-null}
  L_{\circ}^\mu=-L_{\text{\tiny GHY}}^\mu\, , \hspace{1 cm} Y_{\circ}^{\mu\nu}[\delta g; g]=-Y_{\text{\tiny GHY}}^{\mu\nu}[\delta g; g]\, .
\end{equation}
{With this choice,} the Lee-Wald symplectic potential is given by
\begin{equation}\label{SP-null}
\begin{split}
      \bar{\bTh}_{_{\mathcal{N}}} 
           &= {\frac{1}{8\pi G}\int_{\mathcal{N}}\,\d{}^2 x \,\left[\bar{\kappa}\delta\Omega - \Omega^2\bar{\omega}_l \delta \bar{U}+ \frac12\bar{D}_v \left(\Omega\frac{\delta\eta}{\eta}\right)-\delta\left(\Omega(\bar{\theta}_l+\bar\kappa)\right)\right]}\\
      & = \frac{1}{8\pi G}\int_{\mathcal{N}}\,\d{}^2 x \, \bar{D}_v\left( - \mathbb{J} \delta \bar{u}+ \frac12 \Omega{\delta\mathbb{P}}-\delta\Omega\right) \\ &=\frac{1}{8\pi G} \oint \d{}\phi \left[ \bar{u}\delta\mathbb{J}+ \frac12 \Omega{\delta\mathbb{P}}-\delta(\Omega +\bar{u}\mathbb{J})\right]_{v_0}^v\, ,
          \end{split}
\end{equation}
where $v_0$ is an arbitrary initial value, while the following definitions
\be
\begin{split}
\mathbb{P}:=\ln(\frac{\eta}{\bar\theta_l^2})\ , \qquad \mathbb{J}:=\Omega^2\omega_l - \Omega\partial_\phi \ln\bar\theta_l\ , \\ \delta \bar{U}:= \bar{D}_v \delta \bar{u} \quad \Longrightarrow \quad \bar{u}=\int_{[\gamma]}^v \d{}\ell\  \bar{U}\, ,
\end{split}
\ee
and the on-shell relations \eqref{Null-EoM-treated} have been used in the third line.  In the above $[\gamma]$ denotes a particular field-dependent path in the $v,\phi$ plane with $\delta[\gamma]\neq 0$ and $\d{}\ell$ is the length element along the path which is field-dependent. One may argue that there exists a field dependent path $[\gamma]$  such that the above equality is satisfied. A similar symplectic potential has also been discussed in \cite{Freidel:2022vjq}.

The Lee-Wald symplectic form is 
\begin{equation}\label{SF-null}
\begin{split}
      \bar{\bO}_{_{\mathcal{N}}} & = {\frac{1}{8\pi G}\int_{\mathcal{N}}\,\d{}^2 x \,\left[\delta \bar{\kappa}\wedge\delta\Omega + \delta \bar{U}\wedge \delta(\Omega^2\bar{\omega}_l)+ \frac12\bar{D}_v \left(\delta\Omega\wedge \delta\ln{\eta}\right)\right]} \\
 & = \frac{1}{8\pi G}\int_{\mathcal{N}}\,\d{}^2 x \,\left[ \bar{D}_v \left(\delta \bar{u}\wedge \delta\mathbb{J}\right)+ \frac12\bar{D}_v \left(\delta\Omega\wedge \delta\mathbb{P}\right)\right] \\ &= \frac{1}{8\pi G} \oint \d{}\phi \left[\delta \bar{u}\wedge \delta\mathbb{J}+\frac12 \delta\Omega\wedge \delta\mathbb{P} \right]_{v_0}^v .
\end{split}
\end{equation}
In view of symplectic potential and symplectic form expressions above, some comments are in order:
 \begin{enumerate}
     \item The last term in the first line of \eqref{SP-null} is a total variation and hence do not contribute to the symplectic form, as is explicitly seen from the first line in \eqref{SF-null}. This term in $\bar{\bTh}_{_{\mathcal{N}}}$  is proportional to the trace of stress tensor $\mathbb{T}$ and may be absorbed into a $W$-freedom, a boundary Lagrangian.
     \item The first two terms in the first line of \eqref{SF-null} are proportional to variations of the trace-free part of stress tensor $\mathbb{T}^a{}_b$, namely $\bar\kappa, \bar\omega_l$. These two terms may be compared with the first term in \eqref{symplectic-potential-r}. 
     \item The third term in the first line of \eqref{SP-null} which is a total $v$ derivative, may be absorbed into a $Y$-freedom. This term yields the third term in the first line of \eqref{SF-null}. 
     \item The first two terms in the first line of \eqref{SF-null} are the ``thermodynamical symplectic form'' discussed in \cite{Adami:2021kvx, Sheikh-Jabbari:2022xix}, once we view $\bar\kappa$ as the temperature, $\Omega$ as the entropy, $\Omega^2\bar\omega_l$ as angular momentum and $\bar{U}$ as angular velocity. 
     \item The first line of \eqref{SF-null} and recalling the above comments, reinforces the fact that null surface thermodynamics \cite{Adami:2021kvx} and the Carrollian limit of causal surface hydrodynamics are closely related. More precisely, they provide a complementary information about the null surface.   
     \item Interestingly and importantly, the second line in \eqref{SF-null} shows that the ``on-shell'' symplectic form, i.e. when  field equations \eqref{Null-EoM-treated} are imposed on the first line of \eqref{SF-null}, becomes a total time derivative. This renders the on-shell symplectic form as a surface integral (an integral over a codmiension 2 surface, the integral over $\phi$ in our case). This is a $3d$ realization of the general analysis in  \cite{Sheikh-Jabbari:2022mqi}.
     \item The last equality in \eqref{SF-null} makes it clear that $\Omega, \mathbb{P}$ and $\bar{u}, \mathbb{J}$ are canonical conjugate of each other, the last equality in \eqref{SF-null} is in Darboux basis and yields the following Poisson brackets on the solution phase space:
     \begin{equation}\label{Null-Poisson-bracket}
         \{ \Omega(v, \phi), \mathbb{P}(v,\phi')\}= 16\pi G\ \delta(\phi-\phi')\, ,   \qquad  \{ \bar{u}(v, \phi), \mathbb{J}(v,\phi')\}= 8\pi G\ \delta(\phi-\phi')\, .
     \end{equation}
\end{enumerate}

\subsection{Hydrodynamics at asymptotic causal boundary of \texorpdfstring{AdS$_3$}{} space}\label{Hydrodynamics at infinity:AdS-case}

The above hydrodynamical description on $\cc_r$ at a generic $r$, becomes more interesting when we take $r\to \infty$ and take $\cc_\infty$ to be the usual AdS$_3$ causal (asymptotic) boundary. This is schematically depicted in Fig.~\ref{fig:asympt-ads-flat}, Left figure. While Weyl scaling at $\cc_r$ is not among our symmetry generators \eqref{null-bondary-sym-gen}, as we will show below, it becomes one of the boundary symmetries at infinity, i.e. the classical symmetry algebra at $\cc_\infty$ includes Weyl$\oplus$Diff.\footnote{{It is demonstrated in \cite{Fiorucci:2020xto, Ruzziconi:2020wrb} that the symmetry of a timelike boundary in three dimensions, located at infinity in both Fefferman-Graham and Bondi gauges, can be solely represented by Weyl$\oplus$Diff symmetries. Here we have an enhanced asymptotic symmetry algebra that includes Weyl$\oplus$Diff as a subalgebra. A similar enhancement, in the so-called Bondi-Weyl gauge, but in a different slicing has been reported in \cite{Geiller:2021vpg}.}} Among other things, this leads to a conformally invariant hydrodynamical description. As is well known, however, due to anomaly in either of Diff or Weyl parts of the symmetry algebra, the boundary stress tensor can be made either divergence-free or traceless, not both simultaneously. 

We will establish below that asymptotically there are a continuum of the effective hydrodynamical descriptions of the dual $2d$ CFT  residing at the asymptotic causal boundary. These continuum of descriptions are specified by choices of boundary Lagrangians ($W$-freedom). In particular, we discuss two trace-free and divergence-free hydrodynamic slicings of the solution phase space and  explicitly show  how the anomalies associated with Weyl and diffeomorphisms at the asymptotic $2d$ cylinder  can be transformed between these two slicings through choice of boundary Lagrangians and associated change of slicing. In appendix \ref{Appendix} we derive the asymptotic hydrodynamic description directly from taking the large $r$ limit of what we have in  subsection \ref{sec:HCS--finite-r-AdS}.

\begin{figure}[t]
	\centering
\subcaptionbox{}{
\centering
\begin{tikzpicture}[scale=1.2]
    \node (v1) at (1.45,0) {};
    \node (v2) at (1.45,-1) {};
    \node (v3) at (1.45,-2) {};
    \fill (v3)+(0,0.5)  node [right,darkred!60] {$\cc_r$};
    \node (v4) at (1.45,-3) {};
    \fill (v4)+(0.1,0.3)  node [right,blue!80] {\tiny{AdS$_3$ \text{b'dry}}};
    \node (v5) at (-1.45,0) {};
    \node (v6) at (-1.45,-1) {};
    \node (v7) at (-1.45,-2) {};
    \node (v8) at (-1.45,-3) {};
    \node (v11) at (1.5,0) {};
    \node (v12) at (1.5,-1) {};
    \node (v13) at (1.5,-2) {};
    \node (v14) at (1.5,-3) {};
    \node (v15) at (-1.5,0) {};
    \node (v16) at (-1.5,-1) {};
    \node (v17) at (-1.5,-2) {};
    \node (v18) at (-1.5,-3) {};
\begin{scope}[fill opacity=0.8, darkred!60]
    \filldraw [fill=darkred!60!blue!40, very thick] ($(v1)+(0,0)$) 
        to[out=290,in=70] ($(v2) + (0,0)$) 
        to[out=250,in=120,thick] ($(v3) + (0,0)$)
        to[out=300,in=80,thick] ($(v4) + (0,0)$)
        to[out=270,in=270,looseness=0.5] ($(v8) + (0,0)$)
        to[out=80,in=300] ($(v7) + (0,0)$)
        to[out=120,in=250] ($(v6) + (0,0)$)
        to[out=70,in=290] ($(v5) + (0,0)$)
        to[out=270,in=270,looseness=0.5] ($(v1) + (0,0)$);
    \end{scope}
\begin{scope}[fill opacity=0.8, thick, darkred!60] 
    \filldraw [fill=gray!30] ($(v1)+(0,0)$) 
        to[out=90,in=90, looseness=0.5] ($(v5) + (0,0)$)
        to[out=270,in=270,looseness=0.5] ($(v1) + (0,0)$);
    \end{scope}
\begin{scope}[fill opacity=0.8, thick, black!60,dash pattern={on 2pt off 2pt}] 
    \draw ($(v4)+(0,0)$) 
        to[out=90,in=90, looseness=0.5] ($(v8) + (0,0)$);
    \end{scope}
\begin{scope}[fill opacity=0.8, thick, blue!60] 
    \draw ($(v11)+(0,0)$) 
        to[out=90,in=90, looseness=0.57] ($(v15) + (0,0)$);
    \end{scope}
\begin{scope}[fill opacity=0.8, thick, black!60,dash pattern={on 2pt off 2pt}] 
    \draw ($(v14)+(0,0)$) 
        to[out=90,in=90, looseness=0.57] ($(v18) + (0,0)$);
    \end{scope}
\begin{scope}[fill opacity=0.8,thick, blue!60]
    \draw  ($(v11)+(0,0)$) 
        to[out=290,in=70] ($(v12) + (0,0)$) 
        to[out=250,in=120,thick] ($(v13) + (0,0)$)
        to[out=300,in=80,thick] ($(v14) + (0,0)$)
        to[out=270,in=270,looseness=0.57] ($(v18) + (0,0)$)
        to[out=80,in=300] ($(v17) + (0,0)$)
        to[out=120,in=250] ($(v16) + (0,0)$)
        to[out=70,in=290] ($(v15) + (0,0)$)
        to[out=270,in=270,looseness=0.57] ($(v11) + (0,0)$);
    \end{scope}
\end{tikzpicture}}
\subcaptionbox{}{
\centering
\begin{tikzpicture}[scale=1.2]
    \node (v1) at (0,0) {};
    \fill (v1)+(0,0.2)  node [blue!70] {$i^{+}$};
    \node (v2) at (-2,-2) {};
    \fill (v2)+(0.2,0.5)  node [darkred!70] {$\mathcal{C}_r$};
    \node (v3) at (2,-2) {};
    \fill (v3)+(0.1,0)  node [right, blue!70] {$i^{0}$};
    \node (v4) at (0,-4) {};
    \fill (v4)+(0,-0.2)  node [blue!70] {$i^{-}$};
	\node (v5) at (1.8,-2) {};
	\node (v6) at (-1.8,-2) {};
 	\node (v7) at (0.7,-3.25) {};
 	\node (v71) at (1.4,-2.5) {};
	\node (v8) at (-0.7,-3.3) {};
	\node (v81) at (-1.4,-2.6) {};
 	\node (v9) at (0.7,-0.75) {};
 	\node (v91) at (1.4,-1.5) {};
	\node (v10) at (-0.7,-0.7) {};
	\node (v11) at (-1.4,-1.4) {};
    \node [right] at (1.2,-3) {$\textcolor{blue!70}{\mathcal{I}^{-}}$};
\node [right] at (1.2,-1) {$\textcolor{blue!70}{\mathcal{I}^{+}}$};
\begin{scope}[fill opacity=0.8, black!60,dash pattern={on 2pt off 2pt}] 
    \draw ($(v2) + (0,0)$)
        to[out=40,in=140,looseness=0.5] ($(v3) + (0,0)$);
    \end{scope}
\draw[fill opacity=0.8, black!60,dash pattern={on 15pt off 4pt}] ($(v1) + (0,0)$) to ($(v4) + (0,0)$);
\begin{scope}[fill opacity=0.8, ultra thick, darkred!70] 
    \filldraw [fill=darkred!60!blue!40] ($(v1)+(0,0)$) 
        to[out=305,in=90,looseness=0.2] ($(v5) + (0,0)$)
        to[out=270,in=55,looseness=0.2] ($(v4) + (0,0)$)
        to[out=125,in=270,looseness=0.2] ($(v6) + (0,0)$)
        to[out=90,in=235,looseness=0.2] ($(v1) + (0,0)$);
    \end{scope}
\begin{scope}[fill opacity=0.8, very thick,black!60,dash pattern={on 2pt off 2pt}] 
    \filldraw [pattern=north west lines, pattern color=darkred!70, thick,dash pattern={on 7pt off 2pt}] ($(v5)+(0,0)$) 
        to[out=140,in=40, looseness=0.4] ($(v6) + (0,0)$);
    \end{scope}
\begin{scope}[fill opacity=0.8, very thick,black!60] 
    \filldraw [pattern=north west lines, pattern color=darkred!70, thick,dash pattern={on 7pt off 2pt}] ($(v6) + (0,0)$)
        to[out=320,in=220,looseness=0.4] ($(v5) + (0,0)$);
    \end{scope}
\begin{scope}[fill opacity=0.8, black!60,dash pattern={on 2pt off 2pt}] 
    \draw ($(v3)+(0,0)$) 
        to[out=220,in=320, looseness=0.5] ($(v2) + (0,0)$);
    \end{scope}
\begin{scope}[fill opacity=0.8,thick, blue!70]
    \draw  ($(v1)+(0,0)$) 
        to[out=315,in=135] ($(v3) + (0,0)$) 
        to[out=225,in=45,thick] ($(v4) + (0,0)$)
        to[out=135,in=315,thick] ($(v2) + (0,0)$)
        to[out=45,in=225,thick] ($(v1) + (0,0)$);
        \end{scope}
\end{tikzpicture}}
\caption{Asymptotic boundary of AdS$_3$ (Left figure) and $3d$ flat space (Right figure). These asymptotic boundaries may be viewed as asymptotic limits of respective timelike boundaries. }\label{fig:asympt-ads-flat}
\end{figure}
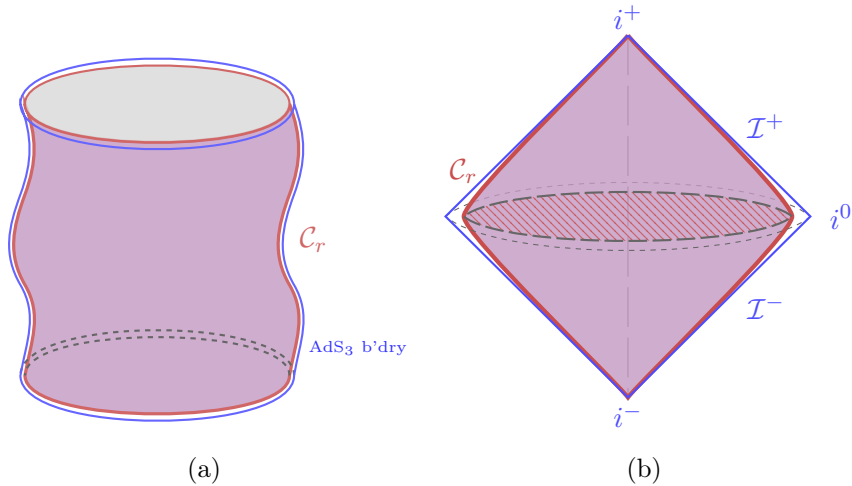

\paragraph{Recovering Weyl symmetry at infinity.} Consider the asymptotic boundary metric \eqref{Asymptotic-AdS3-boundary-metric}. It is clear that a scaling in $r$ yields a Weyl transformation on this metric. Therefore, the $W$ symmetry generator (cf.\eqref{null-bondary-sym-gen}) is generating Weyl transformations and $T,Y,W$ generate Weyl$\oplus$Diff. The $Z$ transformation ($r$ supertranslations) do not change this boundary metric at the leading order. Explicitly, using variations in metric coefficients given  in \cite{Adami:2022ktn}, one may show that
\begin{equation}\label{bdry-metric-Weyl-recovery-infinity}
    \delta_{W}\sigma_{ab}=-W\sigma_{ab}\, , \hspace{1 cm}  \delta_{Z}\sigma_{ab}=0\, ,
\end{equation}
where $\sigma_{ab}$ is the boundary metric \eqref{Asymptotic-AdS3-boundary-metric}. Therefore, unlike the boundary at generic boundary $\cc_r$ which is only Diff invariant,  the asymptotic boundary metric is  manifestly  Weyl$\oplus$Diff.  Some further comments are in order: 
\begin{enumerate}
     \item Weyl$\oplus$Diff is a subset of our algebra generated by $W,T,Y$ at $\cc_\infty$. 
    \item The $Z$ transformation, while not acting on the leading order boundary metric, is not a trivial (pure gauge) transformation, and is still a symmetry generator once one considers the subleading terms in the boundary metric. The charges associated to the $Z, W$ transformations (denoted by $\Omega, \Pi$ respectively, cf. \eqref{surface-charge-Y-infty'''}) are Heisenberg conjugate of each other, either commutator is $(8\pi G)^{-1}$.
    \item  Recalling the comment in footnote \ref{footnote-null-string}, besides the asymptotic boundary, when the boundary is a null surface is another interesting special case where the boundary symmetries coincides with that of a null string.  
   \item The ``enhanced AdS$_3$ asymptotic symmetries''  considered in \cite{Troessaert:2013fma} is a subset of our symmetry algebra generated by $v$-independent $W, T,Y$. Explicitly, our symmetry algebra \eqref{Heisenberg-direct-sum-algebra''} extends the algebra in \cite{Troessaert:2013fma} in two different ways:  Our symmetry generators are generic functions of $v,\phi$ (and not only $\phi$ considered in \cite{Troessaert:2013fma}). Moreover, we also have the $Z$ generator which act at subleading order. The charge associated to $Z$ (denoted by $\Pi$) is Heisenberg conjugate to the charge associated with the Weyl scaling (denoted by $\Omega$) generated by $W$. In this sense, our work at asymptotic infinity recovers \cite{Geiller:2021vpg}. 
   \end{enumerate}

 \paragraph{Hydrodynamic descriptions, preliminaries.} To uncover the hydrodynamical descriptions, we rewrite the asymptotic geometric quantities of the previous section in a particular conformal frame with metric,
\begin{equation}\label{metric:infy}
\hat{\gamma}_{a b}=\lim_{r \to \infty}\left(\frac{\gamma_{ab}}{\mathcal{P}^2 \, r^2}\right)\, ,
\end{equation}
which its explicit form is
\begin{equation}\label{rel:metric-asy}
    \d {\hat{s}}^{2}=\hat{\gamma}_{ab}\d x^{a}\d x^{b}={-\frac{1}{\ell^2\, \lambda^2}  \d v^2}+(\d\phi+\mathcal{U}\d v)^2\, .
\end{equation}
We note that $\sqrt{-\hat{\gamma}}=(\ell\, \lambda)^{-1}$.
Metric at infinity can be expressed in terms of zwibein,
\begin{subequations}
\begin{align}
\hat{\gamma}_{ab}  = -  \hat{t}_{a} \hat{t}_{b}+\hat{k}_{a} \hat{k}_{b},\qquad &\hat{t}^2= -1,\ \hat{k}^2=1,\ \hat{t}\cdot \hat{k}=0. \\ 
\hat{t}^{a}\partial_{a}
       =\ell \, \lambda \, (\partial_{v}-\mathcal{U}\partial_{\phi})\, , \qquad  & \qquad \hat{t}_{a}\d x^a =- \frac{1}{\ell \lambda} \d v \, , \\
   \hat{k}^{a}\partial_{a}
   =\partial_{\phi}\, , \qquad & \qquad \hat{k}_{a}\d x^a =   \d \phi+\mathcal{U} \d v\, , 
\end{align}
\end{subequations}
These two vector fields can also be obtained by scaling two vector fields $t^a$ and $k^a$ \eqref{t-k-zweibein} by $\mathcal{R}$ and inducing them to the boundary at infinity $\cc_\infty$. The deviation tensors associated with these vectors are
\begin{subequations}\label{hat-theta}
    \begin{align}
        & \hat{\nabla}_a \hat{t}_{b}=-\hat{\theta}_{k}\, \hat{t}_{a}\hat{k}_{b}+\hat{\theta}_{t}\, \hat{k}_{a}\hat{k}_{b}\, , \qquad \hat{\nabla}_a \hat{k}_{b}=- \hat{\theta}_{k}\, \hat{t}_{a}\hat{t}_{b}+ \hat{\theta}_{t}\, \hat{k}_{a}\hat{t}_{b} \, ,\\
    &\hat{\theta}_{k}:=\hat{\nabla}_{a}\hat{k}^{a}=-\lambda^{-1}\partial_{\phi}\lambda\, , \hspace{1 cm} \hat{\theta}_{t}:=\hat{\nabla}_{a}\hat{t}^{a}=-\ell\, \lambda\, \partial_{\phi}\mathcal{U}\, ,
    \end{align}
\end{subequations}
where $\hat{\nabla}_{a}$ is the covariant derivative with respect to boundary metric $\hat{\gamma}_{ab}$.

\paragraph{Trace-free hydrodynamic slicing.} 
Motivated by the surface charge expression \eqref{charge-integrable-slicing} and the symplectic form \eqref{r-independent-symplectic-form}, we suggestively define a symmetric, trace-less energy-momentum tensor suitable for the asymptotically AdS spacetimes with the boundary metric given by \eqref{rel:metric-asy} as 
\begin{equation}\label{EM-asymp}
    \hat{\mathcal{T}}^{ab}:=- \hat{\mathcal{E}}\, \hat{t}^a \hat{t}^b + 2\hat{\mathcal{J}}\hat{k}^{(a} \hat{t}^{b )}-\hat{\mathcal{E}}\, \hat{k}^{a} \hat{k}^{b} \, , \qquad \hat{\gamma}_{ab}\hat{\mathcal{T}}^{ab}=0,
\end{equation}
where
\begin{equation}\label{M-hat-J-hat}
    \hat{\mathcal{E}}:=\frac{\ell\hat{\mathcal{M}}}{16\pi G}\, ,  
    \qquad  \hat{\mathcal{J}}:=\frac{\hat{\Upsilon}}{16\pi G}\, .
\end{equation}
While the proposed energy-momentum tensor is trace-free, it is not divergence-free. The main obstacle in making it divergence-free is the anomalous third derivative terms in the equations of motion \eqref{M-Upsilon-EoM''}. To take into account these terms, we gather them in a symmetric ``anomalous energy-momentum tensor'' $\mathcal{A}^{ab}$. Then equations of motion \eqref{M-Upsilon-EoM''} can be recast as follows
\begin{equation}\label{eom-cons-T}
\hat{\nabla}_{a}\hat{\mathcal{T}}^{ab}=\hat{\nabla}_{a}\mathcal{A}^{ab}\, ,
\end{equation}
where the anomalous energy-momentum tensor is
\begin{equation}\label{anomalous-T}
   \mathcal{A}_{ab}:=-\frac{c}{24 \pi }\left[\hat{\theta}_{t}^2 \, ( \hat{t}_{a}\hat{t}_{b}+\hat{k}_{a}\hat{k}_{b})+4\,  (\hat{k}\cdot\hat{\nabla}\hat{\theta}_{t})\, \hat{t}_{(a}\hat{k}_{b)}-\hat{R} \, \hat{k}_{a}\hat{k}_{b}\right]\,,\qquad {\cal A}:=\hat\gamma^{ab}\mathcal{A}_{ab}=\frac{c }{24 \pi }\hat{R}\, .
\end{equation}
Here $c=3\ell/(2G)$ is the Brown-Henneaux central charge \cite{Brown:1986nw} and 
\begin{equation}\label{R-hat}
    \hat{R}=2 \left( \hat{t} \cdot \hat{\nabla} \hat{\theta}_{t} + \hat{\theta}_{t}^2 -\hat{k}\cdot \hat{\nabla} \hat{\theta}_{k} -\hat{\theta}_{k}^2 \right) \, ,
\end{equation}
is the Ricci scalar of the boundary metric $\hat{\gamma}_{ab}$. Observe that of the anomalous energy-momentum tensor \eqref{anomalous-T} is constructed from the second derivatives of the conformal boundary metric \eqref{rel:metric-asy} and hence its divergence in \eqref{eom-cons-T} yields the third derivative terms in \eqref{M-Upsilon-EoM''}.

\paragraph{Divergence-free hydrodynamic slicing.} 
Alternatively, \eqref{eom-cons-T} suggests that one can define a symmetric, divergence-free energy-momentum tensor,
\begin{equation}\label{T-to-T'} 
\hat{\text{T}}_{ab}:=\hat{\mathcal{T}}_{ab}-\mathcal{A}_{ab}, \qquad \hat{\text{T}}=\hat{\gamma}^{ab}\hat{\text{T}}_{ab}=-\frac{c}{24\pi}\, \hat{R}\, ,\qquad \hat\nabla^a\hat{\text{T}}_{ab}=0\, .
\end{equation}
Decomposing this energy-momentum tensor in terms of the basis on $\cc_\infty$, we find that
\begin{equation}\label{EM-tensor-hattext}
   \hat{\text{T}}^{ab}=-\hat{\text{E}}\, (\hat{t}^a \hat{t}^b+\hat{k}^a \hat{k}^b) + 2\, \hat{\text{J}}\, \hat{k}^{(a} \hat{t}^{b )}+\frac{\hat{\text{T}}}{2}(-\hat{t}^a \hat{t}^b+ \hat{k}^{a} \hat{k}^{b}) \, ,
\end{equation}
where
\begin{equation}\label{D-F-E-J}
    \hat{\text{E}}:=\hat{\mathcal{E}}-\frac{c}{24 \pi} \, \hat{\theta}_{t}^2+\frac{c}{48\pi}\, \hat{R}
    \, , \qquad  \hat{\text{J}}:=\hat{\mathcal{J}}+\frac{c}{12\pi} \,\hat{k} \cdot \hat{\nabla} \hat{\theta}_{t}
    \, .
\end{equation}
Note that $\hat\nabla^a\hat{\text{T}}_{ab}=0$ yields equations of motion \eqref{M-Upsilon-EoM''}. See appendix \ref{Appendix} for derivation of the above from the asymptotic large $r$ limit of hydrodynamic description at a generic $r$ on $\cc_r$, discussed in section \ref{sec:HCS--finite-r-AdS}.

\paragraph{Symplectic potential and symplectic form.} 
Given the above  divergence and trace free energy-momentum tensors,  it is instructive to rewrite the symplectic potential in terms of these tensors:
\begin{equation}
    \begin{split}
        \bTh_{_{\mathcal{C}}}&=\int_{\mathcal{C}_{r}}\d{}^2 x\left[-\frac{1}{2}\sqrt{-\hat{\gamma}}\hat{\mathcal{T}}^{ab}\delta \hat{\gamma}_{a b}+\frac{\mathcal{D}_{v}(\Omega\delta\Pi)}{16\pi G}\right]
        \\
        &=\int_{\mathcal{C}_{r}}\d{}^2 x\left[-\frac{1}{2}\sqrt{-\hat{\gamma}}\hat{\text{T}}^{ab}\delta \hat{\gamma}_{a b}+\frac{\mathcal{D}_v(\Omega\delta\Pi)}{16\pi G}\right]-\int_{\mathcal{C}_{r}}\delta(
        \sqrt{|\hat{\gamma}|}\hat{t}^{a}\hat{t}^{b}\mathcal{A}_{ab})\, .
    \end{split}
\end{equation}
{where we chose $L_\circ^r$ 
\begin{equation}
     L_{\circ}^r= -L_{\text{\tiny GHY}}^r-\frac{1}{16\pi G}\left(\frac{2\Lambda}{\lambda}\mathcal{R}^2+3\mathcal{D}_v\mathcal{R}+\lambda^{-1}\hat{\mathcal{M}}\right)\, .
\end{equation}
}
{Interestingly, the change of frame shows itself as a $W$-freedom. In other words, by changing the location of the anomaly (trace to divergence and vice versa), an anomalous term induces on the boundary Lagrangian.

Since the symplectic potential changes by a $W$-term, the symplectic form \eqref{r-independent-symplectic-form} written in terms of divergence and trace free energy-momentum tensors remains intact:
\begin{equation}\label{symplectic-form-hydrodynamic}
    \begin{split}
        \bO_{_{\mathcal{C}}}&=\int_{\mathcal{C}_{r}}\d{}^2 x\left[-\frac{1}{2}\delta(\sqrt{-\hat{\gamma}}\hat{\mathcal{T}}^{ab})\wedge\delta\hat{\gamma}_{a b}+\frac{1}{16\pi G}{{\cal D}_{v}(\delta\Omega\wedge\delta\Pi)}\right]\\
        &=\int_{\mathcal{C}_{r}}\d{}^2 x\left[-\frac{1}{2}\delta(\sqrt{-\hat{\gamma}}\hat{\text{T}}^{ab})\wedge\delta\hat{\gamma}_{a b}+\frac{1}{16\pi G}{{\cal D}_{v}}{(\delta\Omega\wedge\delta\Pi)}\right]\, .
    \end{split}
\end{equation}
While the above is written over a generic boundary $\cc_r$, is also true for the asymptotic boundary $\cc_\infty$. 

\paragraph{Change of slicing between Trace-free and Divergence-free frames.} The above discussion, and in particular \eqref{D-F-E-J}, suggests that the trace-free and divergence-free frames are related by a change of slicing. To this end, we note that in the trace-free hydrodynamic slicing the symmetry generators of the $r$-independent charge (in the direct sum integrable slicing) \eqref{hat-slicing''}, $\hat{T}, \hat{Y}$, can be written as 
\begin{equation}
    \hat{T}=- \ell \lim_{r \to \infty}(\hat{t}\cdot {\xi})\, , \qquad \hat{Y}=\lim_{r \to \infty}(\hat{k}\cdot {\xi})\, ,
\end{equation}
with $\xi$ given in \eqref{null-bondary-sym-gen}. 
The charge variation in the two slicings can be written as 
\begin{equation}
\begin{split}
\slashed{\delta} Q_\xi  &= \frac{1}{16\pi G}\oint_{\cc_{r,v}} \d \phi \Biggl[ \, \hat{W}\delta\Omega +\hat{Z} \delta \Pi+16\pi G\, \hat{\hat{\text{Y}}} \delta \hat{\text{J}} +16\pi G\, \frac{\hat{\hat{\text{T}}}}{\ell}\,  \delta \hat{\text{E}}
\Biggr] \\ 
 &= \oint_{\cc_{r,v}} \d \phi \Biggl[  \, \hat{W}\delta(\frac{\Omega}{16\pi G}) +\hat{Z} \delta(\frac{\Pi}{16\pi G})+ \hat{Y} \delta \hat{\mathcal{J}} +\frac{\hat{T}}{\ell}\,  \delta \hat{\mathcal{E}}  \Biggr]\, .
\end{split}
\end{equation}
Recalling \eqref{D-F-E-J}, the above makes it clear that upon change of slicing, 
\be
\hat{Y} =\frac{\delta \hat{\text{J}} }{\delta \hat{\mathcal{J}} } \hat{\hat{\text{Y}}}=\left(1-\frac{c\ell}{12\pi}\frac{\delta }{\delta \hat{\mathcal{J}} }(\lambda\partial_\phi{\cal U})\right)\hat{\hat{\text{Y}}}\, ,\qquad \hat{T}=\frac{\delta \hat{\text{E}}}{\delta \hat{\mathcal{E}}} \hat{\hat{\text{T}}} =\left(1- \frac{c\ell^2}{24\pi}\frac{\delta }{\delta \hat{\mathcal{E}}}(\lambda\partial_\phi{\cal U})^2\right) \hat{\hat{\text{T}}}\, ,
\ee
the divergence and trace-free slicings are mapped onto each other. Note that to compute the above one should view $\lambda, {\cal U}$ as functions of $\hat{\mathcal{J}}, \hat{\mathcal{E}}$, as specified through \eqref{M-Upsilon-EoM''}. 
The above  analysis and in particular \eqref{symplectic-form-hydrodynamic} shows that  hydrodynamic description at infinity only involves 2 of the 4 charges and more interestingly, the 2 charges appearing in the hydrodynamical analysis are the ``codimension 1'' modes and the ``codimension 2'' modes $\Omega, \Pi$, which are generated by $\hat{Z}, \hat{W}$,  do not enter the hydrodynamical description. These modes do not have a direct appearance in the unit circle cylinder at the boundary \eqref{rel:metric-asy}. 

We close this subsection by the comment that  similar energy-momentum tensors have been proposed in \cite{Campoleoni:2018ltl, Ciambelli:2020ftk, Campoleoni:2022wmf}, following the fluid/gravity correspondence through a derivative expansion. However, their construction is more restrictive than ours, as it is limited to  specific classes of spacetimes. However, here we bypassed this restriction by basing our construction on the surface charge and the symplectic form analyses of the theory.

\subsection{Hydrodynamics at asymptotic null boundary of 3d flat space}\label{Hydrodynamics at infinity:flat-case}

The asymptotic boundary of $3d$ flat space $\mathcal{I}$ is a null surface and analysis of  section \ref{sec:NBH-finite-r} is expected to extend over to this case. However, as in the AdS$_3$ case discussed in section \ref{Hydrodynamics at infinity:AdS-case}, one may get extra enhancements in the asymptotic region. So, we study this case separately. Let us start with the geometry at $\mathcal{I}$. It is a Carrollian geometry described by \eqref{metric-flat-asymptotic} and \eqref{l-flat-asymptotic}. One can readily observe that both $W, Z$ symmetry generators keep the form of asymptotic metric and the kernel vector and one-form to the leading order in $r$. We note in particular that for asymptotic flat case, the $V=0$ condition \eqref{V=0-null} is not required. Therefore, unlike the generic null boundary which discussed in the previous section and like the asymptotic AdS$_3$ case, the symmetry group contains 4 generators, explicitly,
\begin{equation}\label{asympt-null-bondary-sym-gen}
\begin{split}
    \xi &=T\partial_{v}+\left[Z - \frac{1}{\eta^2\lambda }\partial_{\phi}\left(\frac{\eta\partial_{\phi}T}{\lambda}\right)  - \frac{r }{2}\, W  \right]\partial_{r}+Y\partial_{\phi}+\cdots 
\end{split}
\end{equation}
where $\cdots$ denotes the subleading terms. As pointed out around eq.~\eqref{metric-flat-asymptotic}, in order to asymptote to $\mathcal{I}$, ${\cal D}_v {\cal P}<0$, recalling that  $\delta_{Z}\mathcal{D}_{v}\Omega=\mathcal{D}_{v}\mathcal{P}$, this means  $\delta_{Z}\mathcal{D}_{v}\Omega<0$. Note that $\Omega$ does not explicitly appear in the geometric information at $\mathcal{I}$.

There are associated 4 charges which in the slicing discussed in section \ref{sec:sym-charge-review} may be taken to be $\hat{\mathcal{M}}, \hat\Upsilon, \Omega, \Pi$ which satisfy BMS$_3\oplus$Heisenberg algebra, \eqref{Heisenberg-direct-sum-algebra''} with $\Lambda=0$ \cite{Adami:2022ktn}.\footnote{See \cite{Geiller:2021vpg} for a similar algebra in a different slicing.} One may use the same symplectic form \eqref{r-independent-symplectic-form}, however, this slicing and the associated $Y$-term used, obscure the hydrodynamic picture which we would like to uncover below. 
\paragraph{Carroll structure at null infinity.}
The conformal boundary structure is given by
\begin{equation}
    \hat{\gamma}_{ab}:=
    \hat{k}_{a}\hat{k}_{b}\, , \hspace{1 cm}  \hat{k}_{a}\d{} x^a:=\d \phi + \mathcal{U} \d v \, , \hspace{1 cm} \hat{k}^{a}\partial_{a}=\partial_{\phi}\, ,
\end{equation}
where the null surface is described by $\hat{k}_{a}$ and kernel $\hat{l}^{a}$ as follows
\begin{equation}\label{kernel-inf}
    \hat{l}^{a}\partial_{a}:=
    \lambda(\partial_{v}-\mathcal{U}\partial_{\phi})\, , \hspace{1 cm} \hat{\gamma}_{ab}\hat{l}^{a}=0\, .
\end{equation}
This structure has the following Ehresmann connection which is dual to the Carrollian vector $\hat{l}^{a}$
\begin{equation}\label{connection-inf}
    \hat{n}:=\hat{n}_{a} \d{}x^a=-\frac{1}{\lambda}\d{}v\, , \qquad  \hat{l}^{a}\hat{n}_{a}=-1\, .
\end{equation}
These vectors are related to the normal vector of AdS boundary $\hat{t}$, through the following scaling
\begin{equation}
    \hat{n}_{a}=\ell \hat{t}_{a}\, , \hspace{1 cm} \hat{l}^a=\ell^{-1}\hat{t}^a\, .
\end{equation}
\paragraph{Field equations at null infinity.} Our solution space has well-defined null dynamics, which can be derived simply by setting $\Lambda$ to zero. These are equations of motion \eqref{M-Upsilon-EoM''} for $\Lambda=0$, explicitly, 
\begin{subequations}\label{eom-null-infty}
\begin{align}
&{\cal D}_v \hat{\epsilon} + \frac{1}{8\pi G} \partial^3_\phi {\cal U}=0\, ,\qquad \hat{\epsilon}:=\frac{\hat{{\mathcal{M}}}}{16\pi G}\, ,\label{eom-null-infty-epsilon}\\
&{\cal D}_v \hat{\mathcal{J}}-\lambda \partial_\phi \left(\frac{\hat{\epsilon}}{\lambda^2}\right)+ \frac{1}{8\pi G} \partial^3_\phi \lambda^{-1}=0\, .\label{eom-null-infty-J}
\end{align}
\end{subequations} 
As we see the equation for $\hat{\epsilon}$  \eqref{eom-null-infty-epsilon} does not explicitly involve the other charge $\hat{\mathcal{J}}$. This is reminiscent of the fact that two $\hat{\epsilon}$ charges commute (cf. \eqref{Heisenberg-direct-sum-MM}), i.e. they are supertranslation charges. 

\paragraph{Hydrodynamics at null infinity.} As in section \ref{Hydrodynamics at infinity:AdS-case}  we construct two energy-momentum tensors, a trace-free and a divergent-free one. We may start with the results in section \ref{Hydrodynamics at infinity:AdS-case}  and take the appropriate large $\ell$ limit. {This parallels the studies in \cite{Barnich:2012aw, Barnich:2013yka, Duval:2014uva, Duval:2014lpa, Barnich:2006av, Bagchi:2012cy} (see also \cite{Geiller:2021vpg} for further discussions)} where BMS$_3$ algebra was obtained as a contraction of Vir$\oplus$Vir algebra, corresponding to taking a flat space $\ell\to\infty$ limit of AdS$_3$ geometry. 
We can define a trace-free and a divergence-free energy-momentum tensor at null infinity, respectively $\hat{\mathcal{T}}_{_{{\cal I}}}{}^{a}{}_b$ and $\hat{\text{T}}_{_{{\cal I}}}{}^{a}{}_b$ starting from \eqref{EM-asymp}, \eqref{M-hat-J-hat} and \eqref{EM-tensor-hattext}, \eqref{D-F-E-J},
\begin{equation}\label{T-scrI-trace-free}
   \hat{\mathcal{T}}_{_{{\cal I}}}{}^{a}{}_b:=\ell^{-1}\hat{\mathcal{T}}^{a}{}_b=-\hat{\mathcal{E}}_{_{{\cal I}}}\, (\hat{l}^a \hat{n}_b+\hat{k}^a \hat{k}_b) + \hat{\mathcal{J}}\, (\ell^{-2}\hat{k}^{a} \hat{n}_{b }+\hat{l}^{a} \hat{k}_{b })\, ,\qquad \hat{\mathcal{E}}_{_{{\cal I}}}:=\ell^{-1}\hat{\mathcal{E}}=\frac{\hat{\mathcal{M}}}{16\pi G}\, ,
   \end{equation}
with $\hat{\mathcal{T}}_{_{{\cal I}}}{}^{a}{}_a=0$ and 
\begin{equation}\label{T-scrI-div-free}
   \hat{\text{T}}_{_{{\cal I}}}{}^{a}{}_b:=\ell^{-1}\hat{\text{T}}^{a}{}_b=-\hat{\text{E}}_{_{{\cal I}}}\, (\hat{l}^a \hat{n}_b+\hat{k}^a \hat{k}_b) + \hat{\text{J}}_{_{{\cal I}}}\, (\ell^{-2}\hat{k}^{a} \hat{n}_{b }+\hat{l}^{a} \hat{k}_{b })+\frac{\hat{\text{T}}_{_{{\cal I}}}}{2}(-\hat{l}^a \hat{n}_b+ \hat{k}^{a} \hat{k}_{b}) \, ,
\end{equation}
 where
\begin{equation}\label{EM-div-free-null-inf}
    \begin{split}
        &\hat{\text{E}}_{_{{\cal I}}}:=\ell^{-1}\hat{\text{E}}=\frac{1}{16\pi G}\left[\hat{\mathcal{M}}-(\hat{k}\cdot \partial \hat{\theta}_k+\hat{\theta}_{k}^2)+\ell^{2} \hat{l}\cdot\partial \hat{\theta}_l\right]\, , \\
        &\hat{\text{J}}_{_{{\cal I}}}:=\hat{\text{J}}=\frac{1}{16\pi G}\left[\hat{\Upsilon}+2\ell^2 \hat{k}\cdot \partial \hat{\theta}_l\right]\, ,\\
        &\hat{\text{T}}_{_{{\cal I}}}:=\ell^{-1}\hat{\text{T}}=\frac{1}{8\pi G}\left[\hat{\theta}_{k}^2+\hat{k}\cdot\partial \hat{\theta}_k-\ell^2(\hat{\theta}_{l}^2+\hat{l}\cdot\partial \hat{\theta}_l)\right]\, \, \\
&\hat{\theta}_{l}:=\lim_{\ell\to\infty}(\ell^{-1}\hat{\theta}_{t})=-\, \lambda\, \partial_{\phi}\mathcal{U},
    \end{split}
\end{equation}
more definitions can be found in Eqs.~\eqref{hat-theta}. It is then straightforward to verify that the divergence-free condition $\hat\nabla_a \hat{\text{T}}_{_{{\cal I}}}{}^{a}{}_b=0$, yields the on-shell equations \eqref{eom-null-infty}.

Given the expressions above some comments are in order:
\begin{enumerate}
    \item The expressions for $\hat{\text{E}}_{_{{\cal I}}}, \hat{\text{J}}_{_{{\cal I}}}$ have $\ell^2$ and $\ell^0$ contributions. The $\ell^2$ terms, which are divergent in the flat $\ell\to\infty$ limit, are of the form $\hat{l}\cdot\partial \hat{\theta}_l, \hat{k}\cdot \partial \hat{\theta}_l$, i.e. they are components of gradient of the expansion at ${\cal I}$.  These $\ell^2$ terms may be understood noting that unlike a generic null boundary discussed in previous subsection \ref{sec:NBH-finite-r}, we allow for our null boundary at infinity $\mathcal{I}$ to also fluctuate and that $\hat{\theta}_{l}$ is the expansion of the null boundary $\mathcal{I}$.
    \item  The $\ell^2$ terms in the trace of the divergence-free energy-momentum tensor $\hat{\text{T}}_{_{{\cal I}}}$ are nothing but the extrinsic curvature of $\mathcal{I}$ and represent the anomaly term, which appears as the central extension term in the BMS$_3$ algebra. 
    \item One may directly verify that the $\ell^2$ terms do not contribute to $\hat\nabla_a \hat{\text{T}}_{_{{\cal I}}}{}^{a}{}_b=0$ and hence to on-shell equations.
  \item The angular momentum $\hat{\text{J}}_{_{{\cal I}}}$ do not have an $\ell^0$ contribution, reflecting the fact that angular momentum part of the algebra (the Virasoro subalgebra of BMS$_3$) has no anomaly. 
\item The energy $\hat{\text{E}}_{_{{\cal I}}}$ has an $\ell^0$ contribution, a reflection of the fact that $\hat{\text{E}}_{_{{\cal I}}}$ is not a scalar under diffeomorphisms at the null boundary. 
\end{enumerate}
We close this part noting that one may carry out the analysis of the ``hydrodynamic symplectic form'' as was done in section \ref{sec:HCB-AdS} and obtain an analogue of \eqref{symplectic-form-hydrodynamic}. In a similar way one may show that the divergence-free and trace-free cases are related by a change of slicing. Since the analysis is very similar to what was done in section \ref{sec:HCB-AdS} we do not repeat it again.

\section{Thermodynamic description}\label{sec:thermodynamic-causal-boundary}

In the previous sections, we developed hydrodynamical descriptions at a generic causal boundary $\cc_r$, which essentially stems from the diffeomorphism invariance of the effective field theory residing at the boundary. In this section, we show that at each boundary we also have a thermodynamical description which also involves the ``codimension 2'' modes. To work out the thermodynamical description we follow the ideas for ``null surface thermodynamics'' discussed in \cite{Adami:2021kvx, Sheikh-Jabbari:2022xix, Taghiloo:2022hxc} modulo two important differences: (1) Here we seek a thermodynamical description for a generic causal surface and (2) we address the $Y$-freedom dependence of the description in \cite{Adami:2021kvx}.

As in \cite{Adami:2021kvx}, our starting point is the expression for charge variation \eqref{LW-charge-1}, which we rewrite for having it at hand,
\begin{equation}\label{surface-charge-Thermo}
\slashed{\delta} Q_\xi = {\frac{1}{16\pi G}}\oint_{\cc_{r,v}} \d \phi \left\{
            W \delta\Omega + 2Z \delta {\cal P} +Y\delta \Upsilon+ T \slashed\delta{\cal H} \right\} \, ,
\end{equation}
where
\begin{equation}\label{delta-slash-H-thermo}
    \begin{split}
         \slashed\delta{\cal H}&= -\mathcal{D}_{v}\Pi \, \delta\Omega+{\mathcal{U}} \delta\Upsilon+\mathcal{D}_{v}\Omega\ \delta\Pi+\lambda^{-1}\delta{\hat{\mathcal{M}}}\\
         &=\frac{\partial_{r}V}{\eta}\delta {\Omega}+U\delta\Upsilon+\frac1\eta\left(r{\partial_{r}V}-{V}\right)\delta(\eta\lambda)+2{\mathcal{R}\theta_{l}}\frac{\delta\eta}{\eta}-\delta(2\mathcal{R}\theta_l)+\partial_{\phi}X\, ,
    \end{split}
\end{equation}
with
\begin{equation}
    X:=\lambda^{-1}\left[\Omega\mathtt{U}\delta\Pi-2\mathcal{R}\delta\mathtt{U}-2\delta\left(\frac{\partial_{\phi}\lambda}{\lambda}\right)\right], \qquad \mathtt{U}:=\lambda(U-{\cal U})=\frac{1}{{\cal R}}\left(\frac{\Upsilon}{2{\cal R}}+\frac{\partial_\phi\eta}{\eta}\right).
\end{equation}
The first line in \eqref{delta-slash-H-thermo} is written in terms of parameters more naturally defined at the asymptotic boundary $\cc_\infty$ or ${\cal I}$ whereas the second line is in terms of parameters more natural to the boundary at generic $r$, $\cc_r$. 

\subsection{Derivation of the first law,  a new viewpoint}\label{sec:Thermo-viewpoint}

The seminal Iyer-Wald paper \cite{Iyer:1994ys} presents a derivation of the first law of black hole thermodynamics within  covariant phase space formalism. This derivation is based on the expression for the symplectic form density $\boldsymbol{\omega} (\delta_1 g, \delta_2 g)$ computed  over the field variations generated by symmetry generators $\xi$, {$\boldsymbol{\omega} (\delta g, \delta_\xi g)$}. Since the symplectic density is closed, the charge variation $\slashed{\delta} Q_{\xi}:=\int_\Sigma  \boldsymbol{\omega} (\delta g, \delta_\xi g)$} where $\Sigma$ is a partial Cauchy surface, localizes over the two boundaries of $\Sigma$, the bifurcation surface of a black hole with a Killing horizon $\text{H}$ and asymptotic spatial infinity $i^0$. Since the symplectic form is closed, this yields equality,
\be\label{Wald-1st-law}
{\slashed{\delta} Q_\xi{|}_{_{{\text{H}}}}=\slashed{\delta} Q_\xi{|}_{_{{i^0}}}\, .}
\ee
The above is true for any $\xi$. If $\xi$ is taken to be the Killing vector generating the horizon $\zeta_{\tiny{\text{H}}}$, recalling that $\zeta_{\tiny{\text{H}}}=\sum_i \mu_i \zeta_i$ where $\zeta_i$ are asymptotic Killing vectors associated with ADM charges and $\mu_i$ are some constant (chemical potentials), the above yields the standard first law, once we use Wald's entropy formula \cite{Wald:1993nt, Iyer:1994ys}. The above derivation crucially depends on the information on the two boundaries of the Cauchy surface, the $T\delta S$ term comes from LHS of \eqref{Wald-1st-law} which is computed over $\text{H}$ and the RHS of \eqref{Wald-1st-law} yield the usual $\delta M-\Omega\delta J$ which is computed at $i^0$. {Recall that in general charge variation $\slashed{\delta} Q_\xi$ is ambiguous to the choice of $Y$-freedom. However, since in \eqref{Wald-1st-law} we equate $\slashed{\delta} Q_\xi$ at two different boundaries $\text{H}, i^0$, if the $Y$-term has the same value at both boundaries, it cancels off in the expression for the first law. Explicitly, Iyer-Wald derivation is invariant under this certain class of $Y$-terms.}

In \cite{Hajian:2015xlp} an alternative derivation of the first law was presented which uses the information over a single boundary, building on the fact that for ``exact symmetries'' (Killing vectors of the black hole background geometry) the charges are symplectic symmetries and they may be computed over any codimension 2 compact spacelike surface, see Chapter 5 of \cite{Grumiller:2022qhx} for a more detailed discussion. This derivation has the advantage  over the Iyer-Wald derivation in that it relies only on one boundary, however, it has the disadvantage that in general it depends on the $Y$-term. 

Here we present a derivation of the first law which combines all positive features and advantages of the three derivations:
\begin{enumerate}
    \item Null surface thermodynamics \cite{Adami:2021kvx}, in that it applies to geometries without exact Killing symmetries and importantly, it is a local first law, a first law which is true for any $v,\phi$ and that it applies to any null surface.
    \item Iyer-Wald formulation \cite{Iyer:1994ys}, in that it is independent of certain class of $Y$-terms.
    \item The derivation in \cite{Hajian:2015xlp}, in that  the first law may be computed for any arbitrary boundary $\cc_r$. 
\end{enumerate}  
In our new viewpoint, we use the same steps of null surface thermodynamics \cite{Adami:2021kvx}, but to remove the $Y$-term dependence we equate two equivalent expressions for $\slashed{\delta} {\cal H}$ {at the same boundary $\cc_r$. This extends this advantage of Iyer-Wald derivation to any choice of $Y$-freedom.} In the next subsection, we show how this derivation works for the {asymptotically} AdS$_3$ case.

\subsection{Local first law for causal surface thermodynamics}

We first note that the $r$-superscaling aspect $\Omega$, superrotation aspect $\Upsilon$ and the $r$-supertranslation aspect $2{\cal P}$ are integrable while the charge associated with $v$-supertranslations $\slashed{\delta} {\cal H}$ is not integrable. The first three integrable charges may hence  be respectively viewed as charge densities associated with $W=1, Y=1$ and $Z=1$. Explicitly, if we denote 
\be
\slashed{\delta} Q_\xi :=\int_{\cc_{r,v}} \ \slashed{\delta} {\cal Q} [T,Z,W,Y]\, ,
\ee
then
\be
{\cal Q}[0,1,0,0]=2{\cal P}=2\Omega e^{\Pi/2}\, , \qquad {\cal Q}[0,0,1,0]=\Omega\, , \qquad {\cal Q}[0,,0,1]=\Upsilon\, ,\qquad \slashed{\delta} {\cal Q}[1,0,0,0]=\slashed{\delta} {\cal H}\, .
\ee
The crucial observation made in \cite{Adami:2021kvx, Sheikh-Jabbari:2022xix} which constitutes the basis of our local first law derivation here is that $\slashed{\delta} {\cal H}$, as \eqref{delta-slash-H-thermo} explicitly shows, is written in terms of the other three charges and the associated chemical potentials (canonical conjugates to these charges). This observation replaces the relation between the (Killing) symmetry  vectors like $\zeta_{\tiny{\text{H}}}=\sum_i \mu_i \zeta_i$ which is the basis of first law derivations in \cite{Iyer:1994ys} or \cite{Hajian:2015xlp}.

To obtain the local first law, we start with the identity, 
\begin{equation}\label{first-law-starting-point}
    \begin{split}
         {\mathcal{U}} \delta\Upsilon+\lambda^{-1}\delta{\hat{\mathcal{M}}}-\mathcal{D}_{v}\Pi \, \delta\Omega+\mathcal{D}_{v}\Omega\ \delta\Pi=
         &\frac{\partial_{r}V}{\eta}\delta {\Omega}+U\delta\Upsilon+\frac{\left(r{\partial_{r}V}-{V}\right)}{\eta}\delta{\cal P} 
         +2{\mathcal{R}\theta_{l}}\frac{\delta\eta}{\eta}-\delta(2\mathcal{R}\theta_l)+\partial_{\phi}X\, .
    \end{split}
\end{equation}
The above lays the basis of our local first law. Compared to the analysis in \cite{Adami:2021kvx} it has the advantage and distinctive feature that the contribution of the $Y$-term to the charge variations drops out from the two sides of the equation and our final first law becomes $Y$-freedom invariant. This extends  the advantage of the Iyer-Wald derivation discussed above to a generic choice of $Y$-term. After rearrangement of some terms we obtain, 
\begin{equation}\label{firstlaw-midway}
\begin{split}
    \lambda^{-1}\delta{\hat{\mathcal{M}}}&=
    \left(\frac{\partial_{r}V}{\eta}+2\frac{\mathcal{D}_{v}{\cal P}}{{\cal P}}\right)\delta {\Omega}+(U-\mathcal{U})\delta\Upsilon+\left[\frac{1}{\eta}\left(r{\partial_{r}V}-{V}\right)-{2}\frac{\mathcal{D}_{v}\Omega}{{\cal P}}\right]\delta\mathcal{P}
        +2{\mathcal{R}\theta_{l}}\frac{\delta\eta}{\eta}-\delta(2\mathcal{R}\theta_l)+\partial_{\phi}X\\
        &=\left(\frac{\partial_{r}V}{\eta}+2\frac{\text{D}_{v}{\cal P}}{{\cal P}}\right)\delta {\Omega}+(U-\mathcal{U})\delta\hat\Upsilon+\left[\frac{1}{\eta}\left(r{\partial_{r}V}-{V}\right)-{2}\frac{\text{D}_{v}\Omega}{{\cal P}}\right]\delta\mathcal{P}
        +2{\mathcal{R}\theta_{l}}\frac{\delta\eta}{\eta}-\delta(2\mathcal{R}\theta_l)+\partial_{\phi}\hat{X}\\
\end{split}
\end{equation}
where $\hat{{\cal M}}, \hat{\Upsilon}$ are defined in \eqref{M-Upsilon-EoM''} and 
$$\hat{X}:=-2\lambda^{-1}\left[\mathcal{R}\delta\mathtt{U}+\delta\left(\frac{\partial_{\phi}\lambda}{\lambda}\right)\right].
$$ 
The difference between the first and second equalities in \eqref{firstlaw-midway} is that we have replaced the covariant derivative appropriate for the boundary at infinity ${\cal D}_v$ with $\text{D}_v$ appropriate for the boundary $\cc_r$, cf. \eqref{Dv-s}. 

{Next, we note that \eqref{firstlaw-midway} suggests to define the following thermodynamic variables}
\begin{equation}\label{Thermo-stuff}
\begin{split}
    &\mathtt{U} :=\omega_l=\lambda(U-{\cal U})=\frac{1}{{\cal R}}\left(\frac{\Upsilon}{2{\cal R}}+\frac{\partial_\phi\eta}{\eta}\right)\, , \quad  \quad \text{S}:=\frac{{\cal R}}{4G}\, , \\
    &2\pi \lambda^{-1} \text{T} := \kappa+\frac{\mathcal{D}_v\lambda}{\lambda}= \lambda^{-1}(-\mathtt{U}^2+\frac{1}{\ell^2}) {\cal R}\, ,\\
& 16\pi G\ \text{E}:= 
{\hat{\cal M}}+ 2\lambda{{\cal D}_v {\cal R}}+\lambda^2{V}+2{\cal S}[\int \lambda, \phi]
=\mathcal{M}+2\lambda \mathcal{R}\frac{\mathcal{D}_{v}{\cal P}}{{\cal P}}+\lambda^2 V=(\mathtt{U}^2+\frac{1}{\ell^2}) {\cal R}^2\, , \\  
& 16\pi G\ \text{J} :=\hat{\Upsilon}+  2\lambda \partial_\phi(\frac{{\cal R}}{\lambda})
= \Upsilon+2{\cal R}\frac{\partial_\phi\eta}{\eta}=2\mathcal{R}^2\omega_{l}=2{\cal R}^2\mathtt{U}\, .
\end{split}
\end{equation}
Note that $\kappa, \mathtt{U}$ have  geometric meanings \eqref{kappa-thetal-thetan}. The factors of $\lambda$ in $\text{T}, \mathtt{U}$,  the extra $\frac{\mathcal{D}_v\lambda}{\lambda}$ term in $\text{T}$ and the contribution from the Schwarzian derivative may be understood as $\lambda$ is the asymptotic local time scale.\footnote{It is crucial to show that physical observables like entropy and laws of black hole thermodynamics, like the first law, are observer independent. The forms of the chemical potential $\mathtt{U}, \text{T}$ given in \eqref{Thermo-stuff} and that they are of the form of difference of a quantity at the boundary and at infinity  can be used to argue for this observer independence. Had we added an electric charge to the system, {the gauge invariance} of the corresponding chemical potential could have been argued for in the same way. See \cite{Hajian:2022lgy} for more discussions.} We also comment that our viewpoint in deriving the first law through the identity \eqref{first-law-starting-point} removes two other disadvantages of the usual Iyer-Wald derivation \cite{Hajian:2015xlp, Grumiller:2022qhx}: The normalization of the Killing vectors $\zeta_{\tiny{\text{H}}}, \zeta_i$ and hence the definitions of the chemical potentials depend on the choice of coordinates and observers. In particular, the horizon angular velocity should be measured by the asymptotic static (non-rotating) observer. While this may be a natural choice for asymptotic flat spacetimes, it is not for asymptotic AdS cases. As we see, in our derivation the difference $\mathtt{U}:=\lambda(U-{\cal U})$ appears which is independent of the observer, recall \eqref{Asymptotic-AdS3-boundary-metric} and that $\lambda^{-1}$ is the asymptotic local time scale.

{Finally,  using the thermodynamic quantities \eqref{Thermo-stuff}, the first law \eqref{firstlaw-midway} takes the standard form:}
\be\label{first-law}
\boxed{{{\delta \text{E}=\text{T}\delta{\text{S}}+\mathtt{U}\delta\text{J}}}\, .}
\ee
{With thermodynamic quantities \eqref{Thermo-stuff} in hand, the Gibbs-Duhem equation is also readily read as,
\begin{equation}\label{GD}
    \boxed{\text{E}=\mathtt{U}\text{J}+\frac12\text{T} \text{S}\, .}
\end{equation}}
{We note that the first law may also be written as}
\be\label{first-law-newform}
\lambda^{-1} \delta \text{E} +{\cal U} \delta \text{J}={U}\delta \text{J}+ (\lambda^{-1}\text{T}) \delta {\text{S}}\, ,
\ee 
and its integrability condition over the solution space is,
\be\label{integrability-thermo}
\delta(\lambda^{-1})\wedge \delta \text{E} +\delta{\cal U}\wedge \delta \text{J}=\delta{U}\wedge \delta \text{J}+ \delta(\lambda^{-1}\text{T}) \wedge \delta {\text{S}}\, .
\ee 

\paragraph{BTZ example.}
{Now that we have obtained our local first law and the Gibbs-Duhem equation, it is illuminating to examine the seminal example of a BTZ black hole. In this regard, we note that a BTZ black hole solution with inner and outer radii $r_-, r_+$, corresponds to a solution with the following parameters in our conventions}
\begin{equation}
    \Lambda=-\frac{1}{\ell^2}\, ,\hspace{.5 cm} \eta=1\, , \hspace{.5 cm} \lambda=1\, , \hspace{.5 cm} \Omega=r_+\, , \hspace{.5 cm} \mathcal{U}=\frac{r_-}{\ell r_+}\, , \hspace{.5 cm} \Upsilon=2\frac{r_+ r_-}{\ell}\, , \hspace{.5 cm} \mathcal{M}=\frac{r_+^2+r_-^2}{\ell^2}\, .
\end{equation}
Our local thermodynamical quantities are hence, 
\be
\mathtt{U}=\frac{2r_+ r_-}{\ell{\cal R}^2}\, ,\qquad 16\pi G \text{E}=\frac{r_+^2 r_-^2}{\ell^2{\cal R}^2}+\frac{{\cal R}^2}{\ell^2}\, ,\qquad 8\pi G \text{J}=\frac{r_+ r_-}{\ell{\cal R}^2}\, ,\qquad 2\pi \text{T}=\frac{{\cal R}}{\ell^2}-\frac{r_+^2 r_-^2}{\ell^2{\cal R}^3}\, ,\qquad \text{S}=\frac{{\cal R}}{4G}\, ,
\ee
where ${\cal R}=r+r_+$. Note that our thermodynamic quantities are $r$ dependent and one can readily check that the first law 
\eqref{first-law} {and Gibbs-Duhem equation \eqref{GD}} hold at any $r$.

\paragraph{First law at the asymptotic boundary.} 
The first law \eqref{first-law} is true for any $r,v,\phi$, and in particular for large $r$, 
\be
\text{E}\simeq \frac{{\cal R}^2}{16\pi G\ell^2}\, ,\qquad  \text{J}\simeq \frac{{\cal R}}{8\pi G}\frac{\partial_\phi\eta}{\eta}\, ,\qquad  \mathtt{U}\simeq \frac{\partial_\phi\eta}{\eta{\cal R}}\, , \qquad 2\pi \text{T} \simeq \frac{{\cal R}}{\ell^2}\, , \qquad \text{S}=\frac{{\cal R}}{4G}\, .
\ee
The leading contribution to the first law takes the form $\delta \text{E}=\text{T}\delta {\text{S}}$ and the contribution from the angular momentum becomes subleading. Recalling the asymptotic metric \eqref{Asymptotic-AdS3-boundary-metric}, the leading contribution to the energy is proportional to the volume of the boundary metric (it is the Casimir energy) and the entropy is proportional to the area of the circle at constant $v$.

\section{Discussion and outlook}\label{sec:discussion}

We have studied $3d$ gravity in {the} presence of a generic causal (timelike or null) boundary ($\cc_r$ or ${\cal N}$) and constructed the solution phase space. For the generic case, this solution space has 4 boundary degrees of freedom (4 functions on $\cc_r$) labeled by 4 surface charges associated with symmetry generators. Two of these symmetry generators are diffeomorphisms at the boundary and two others are associated with supertranslation and super-scaling in the $r$ direction transverse to $\cc_r$. In the case with a generic timelike boundary and regardless of  spacetime being asymptotically AdS$_3$ or flat space,  we have shown by an explicit construction that the boundary degrees of freedom admit a relativistic, but non-conformal, $1+1$ dimensional fluid description. The key object in this hydrodynamic description is the causal boundary Brown-York energy-momentum tensor at $\cc_r$. 

In our understanding, this generic hydrodynamic description arises as a result of excising the part of spacetime ``behind the boundary'' which in our conventions corresponds to $r<0$ region. When we restrict the spacetime to $r\geq 0$ regions and to compensate for the $r<0$ part of spacetime, one should associate appropriate degrees of freedom to the boundary. These boundary degrees of freedom in general interact with the ``bulk modes'' and these interactions give rise to the effective hydrodynamic description. Of course, in the $3d$ case, where there are no bulk graviton modes, the latter is only a result of Einstein equations of the bulk projected at the boundary, see section \ref{sec2:soln-space-review} or \cite{Adami:2022ktn} for more discussions. More explicitly, the boundary modes and associated surface charges are giving an effective, coarse-grained description of the  would-be microscopic degrees of freedom  residing at the boundary. The hydrodynamic description is then nothing but the requirement that this microscopic boundary system is  coupled to the bulk modes in a way consistent with general covariance in the bulk. This hydrodynamic description does not specify what the boundary theory is. 

This  hydrodynamic picture for generic timelike boundary should be revisited for three special cases, where the boundary is approaching the asymptotic boundary of spacetime in AdS or flat spacetimes, as well as the cases with null boundary. We hence studied these three special cases in section \ref{sec:Three-Special-cases}. For the asymptotic AdS case, we first argued that in the symmetry algebra Diff$\oplus$Heisenberg, the Heisenberg part involves Weyl scaling plus its Heisenberg conjugate. In this case, the hydrodynamic description becomes that of a conformal $1+1$-dimensional fluid. The Heisenberg part of the algebra {does} not enter the hydrodynamic description, as the symplectic form \eqref{symplectic-form-hydrodynamic} manifests. For a generic null boundary case, as is manifested in \eqref{SF-null},  we have a hydrodynamic description as well as a complementary thermodynamic description discussed in earlier papers \cite{Adami:2021kvx, Sheikh-Jabbari:2022xix}. The hydrodynamic description of the last special case, the asymptotic boundary of flat space ${\cal I}$, was constructed as an infinite AdS radius limit \cite{Barnich:2012aw, Barnich:2013yka} of the relativistic conformal hydrodynamic appearing in the asymptotic boundary of AdS$_3$. This limit is essentially the Carrollian limit of the conformal hydrodynamics and leads to a conformal Carrollian fluid discussed in \cite{Ciambelli:2018xat, Ciambelli:2018wre, Ruzziconi:2019pzd, Freidel:2022vjq}. 	 Our analysis here also sheds light on the notion of {the} equilibrium limit of the theory at the boundary. While it has been established that Carroll geometries are natural intrinsic geometries of null surfaces, both at finite and infinite regions \cite{Chandrasekaran:2018aop, Chandrasekaran:2021hxc, Ashtekar:2021wld}, there {have} been some open questions regarding entropy and {the} nature of hydrodynamic quantities and our analyses here clarify them.

Our analysis has been performed on an extended solution space compared to earlier analyses. This extension made it possible to uncover a non-conformal fluid picture at a generic boundary. In comparison, the conformal fluid description one finds at the asymptotic AdS boundary involves only 2 out of 4 boundary fields {that} take part in the hydrodynamic description. We discussed that the 2 hydrodynamic degrees of freedom are ``codimension 1 modes'' whereas the other 2 modes which do not appear in the hydrodynamic description are ``codimension 2 modes'', as has been manifested in \eqref{r-independent-symplectic-form} and \eqref{symplectic-form-hydrodynamic}.  The codimension 2 modes are parameterizing the boundary mode and its canonical conjugate and do not enter into how bulk and boundary modes balance each other (what the hydrodynamic picture is about). This is compatible with the general hydrodynamic picture we discussed above. 

Among other things, in appendix \ref{Appendix} we  showed how the hydrodynamic descriptions at generic $\cc_r$ and the conformal fluid at the boundary of AdS are related in an explicit large $r$ limit. In this sense, our construction extends those in \cite{Chandrasekaran:2021hxc, Freidel:2022vjq} and shows there is a non-singular limit on the finite $r$ null boundary. 

For a generic null surface inside AdS$_3$ which may be the horizon of a BTZ black hole, our construction here establishes {the}  existence of a ``hydrodynamical holographic RG flow'' from a conformal fluid to a conformal Carroll fluid: As one moves from the AdS$_3$ causal boundary inward to a null surface in the bulk, the conformal hydrodynamic description at the AdS$_3$ boundary is transformed to the thermo/hydro dynamic description at the null surface. The hydrodynamic holographic RG flow  can connect the two places where hydrodynamic description of gravity arise: \textit{membrane paradigm} for {the}  (stretched) horizon of black holes \cite{Damour:1978cg, Thorne:1986iy, Price:1986yy} (see also Chapter 6 of \cite{Grumiller:2022qhx}) and the \textit{fluid/gravity correspondence} in the context of AdS/CFT duality \cite{Bhattacharyya:2007vjd, Hubeny:2011hd} and its generalizations \cite{Iqbal:2008by, Eling:2009pb, Nickel:2010pr}, which provide a  description of the hydrodynamic limit of a field theory residing at the AdS boundary through its gravitational dual. Our analyses here connect the two through an RG flow on the CFT side. It is desirable to explore this
hydrodynamic RG flow in a more general (e.g. higher dimensional) settings.

We discussed {at} the end of section \ref{sec:HCS--finite-r-AdS} that the non-conformal boundary fluid description for a generic boundary $\cc_r$ is not unique. We constructed an infinite family of such energy-momentum tensors related by the subclass of Weyl transformations which are functions of {the} trace of the causal boundary Brown-York energy-momentum tensor ${\cal T}$. We showed that this subclass {does} not change the hydrodynamic symplectic form and hence represents canonical transformations over the hydrodynamic phase space; i.e. these different hydrodynamic descriptions  are physically equivalent. 
The same non-uniqueness is expected to be there when we consider asymptotic AdS (or flat) boundary cases in the divergence-free frame. That is, there should be infinitely many such divergence-free frames in section \ref{sec:HCS--finite-r-AdS}. It is interesting to explore the family of divergence-free descriptions and the implications for the boundary theory. 

We also argued that the symmetry algebra generated by $\xi(T,Z,W, Y)$ \eqref{3d-NBS-KV-algebra''} at large $r$ for AdS case rearranges such that we get an algebra which is an extension of $2d$ Diff$\oplus$Weyl symmetry at the asymptotic boundary. On the other hand, consistency of the description requires that the boundary theory (which admits a relativistic conformal fluid description in the continuum limit) should have this symmetry as a gauge (local) symmetry. On the other hand, we know that string worldsheet theory is the most general $2d$ theory with Diff$\oplus$Weyl symmetry. Therefore, the boundary theory should be a string theory. It is also known that we do not have consistent string theory with only AdS$_3$ as its target space.  So, an indirect consequence of our analyses and discussions here is that to {upgrade} the effective, coarse-grained description we uncovered here to a fully-fledged microscopic (quantum) level, one should supplement the AdS$_3$ part with internal spaces (like $AdS_3\times S^3\times X_4$ spaces). In other words, pure AdS$_3$ quantum gravity is expected not to be a consistent theory. These results resonate with the conclusions in \cite{Maloney-Witten} and the follow-up works.

Besides the hydrodynamic description at a generic timelike boundary $\cc_r$, we showed that the system also admits a thermodynamic description. We argued for the latter by explicitly constructing a local first law of thermodynamics and Gibbs-Duhem relations  at $\cc_r$. This is an extension of our earlier results for generic null boundaries \cite{Adami:2021kvx}. This result may seem intriguing as the lore is that such a thermodynamic description is a feature of null surfaces, in particular (Killing) horizons of black holes. It is desirable to explore if the thermodynamics at {the} causal boundary discussed here can be extended to more interesting higher dimensional cases and if yes, how does it extend the picture discussed in Ted Jacobson's seminal paper \cite{Jacobson:1995ab}. Moreover, to derive this thermodynamic description we discussed a new viewpoint (cf. section \ref{sec:Thermo-viewpoint}). It is interesting to explore  implications of this new viewpoint for the seminal Iyer-Wald derivation \cite{Iyer:1994ys}.

In the $3d$ Einstein gravity setup we studied here there are no propagating bulk modes. It is interesting to reconsider our analysis for cases with bulk modes. There are two such examples: (1) How does our analysis here extend to higher dimensions; do we have a hydrodynamic description for a generic timelike boundary in $d>3$? (2)
How is the hydrodynamic in  Topologically Massive Gravity (TMG) \cite{Deser:1981wh} which admit bulk (chiral) gravitons? Analysis of the solution phase space for the TMG in {the} presence of a generic null boundary and its thermodynamical description has been studied \cite{Adami:2021sko, Taghiloo:2022hxc}. One may extend these analyses for a generic causal boundary and explore the hydro/thermo dynamical descriptions.

\section*{Acknowledgement}
We would like to thank Lars Andersson, Shing-Tung Yau and Mohammad Hassan Vahidinia  for  discussions and Arjun Baghci, Luca Ciambelli, Romain Ruzziconi and Celine Zwikel for comments on the draft. MMShJ would like to acknowledge SarAmadan grant No. ISEF/M/401332. The work of HA is supported by the National Natural Science Foundation of China under Grant No. 12150410311.

\appendix
\section{Hydrodynamics at infinity from hydrodynamic in the bulk}\label{Appendix}

As previously discussed, the hydrodynamic description involves only two charges associated with diffeomorphisms on the boundary, while the other two charges associated with $Z, W$ transformations do not enter this description. Therefore, to have a cleaner and simpler hydrodynamic description one may use $W$ transformation to set ${\cal P}=\eta\lambda$ equal to one. This fixing of $W$ symmetry is physically relevant specially when we discuss the hydrodynamics in the asymptotic regions either for AdS (cf. section \ref{Hydrodynamics at infinity:AdS-case}) or flat space (cf. section \ref{Hydrodynamics at infinity:flat-case}).  

\paragraph{Metric at ${\cal P}=1$.} Let us denote the value of quantity $X$ at ${\cal P}=1$ by $\tilde{X}$. In this case,  $\eta=\lambda^{-1}$ and the metric takes the form
\begin{equation}\label{metric-P=1}
\d s^2= -\tilde{V} \d v^2 + \frac{2}{\lambda}  \d v \d r + \tilde{{\cal R}}^2 \left( \d \phi + \tilde{U} \d v \right)^2\, ,
\end{equation}
where 
\be
\begin{split}
\tilde{{\cal R}}&=r+\Omega\, ,\\ 
    \tilde{V}&=V(\eta=\lambda^{-1})= \frac{1}{\lambda^2}\left(-\Lambda\tilde{{\cal R}}^2-{\cal M}+2\lambda\tilde{{\cal R}} \partial_\phi{\cal U}+\frac{\Upsilon^2}{4{\tilde{\cal R}}^2}-\frac{\Upsilon}{\tilde{{\cal R}}}\frac{\partial_\phi\lambda}{\lambda}\right)\, ,\\
\tilde{U}&=U(\eta=\lambda^{-1})={\cal U}-\frac{1}{\tilde{{\cal R}}}\frac{\partial_\phi\lambda}{\lambda^2}+\frac{\Upsilon}{2\lambda\tilde{{\cal R}}^2}\, ,
\end{split}
\ee
and 
\be
\begin{split}
    \tilde{\mathcal{M}}={\cal M}-2\lambda\Omega \partial_\phi{\cal U}-2\lambda {\cal D}_v \Omega + 4\left(\frac{\partial_\phi\lambda}{\lambda}\right)^2-2\frac{\partial^2_\phi\lambda}{\lambda}\, ,\qquad 
\tilde{\Upsilon}=\Upsilon-2\partial_\phi\Omega\, .
\end{split}
\ee
here we used  ${\cal D}_v{\cal P}|_{\mathcal{P}=1}=-\partial_\phi{\cal U}$.
\paragraph{Asymptotic expansion, AdS case.} Given the above one may compute the asymptotic values of $\theta_s,\omega_s, \kappa_t$ for ${\cal P}=1$ case. For the AdS$_3$ case, we may do so by simply setting ${\cal P}=1$ in \eqref{thetas-omegas-kappat}, to obtain
\begin{equation}
    \begin{split}
       \tilde{\kappa}_{t}+ \tilde{\theta}_{s}&=-\frac{\ell}{r^2}\left[\tilde{\mathcal{M}}-\left(\ell \lambda \partial_{\phi}{\mathcal{U}}\right)^2-\ell^2 \lambda\mathcal{D}_{v}\left(\lambda\partial_{\phi}\mathcal{U}\right)+\lambda\partial_{\phi}\left(\frac{\partial_{\phi}\lambda}{\lambda^2}\right)\right] +\mathcal{O}(r^{-3}) \\ &=-\frac{\ell}{r^2}\left[\tilde{\mathcal{M}}+\hat{t}\cdot \hat\nabla \hat\theta_t- \hat{k}\cdot\hat\nabla\hat\theta_k-\hat\theta_k^2\right] +\mathcal{O}(r^{-3})\, , \\
        \tilde{\omega}_{s}=& \frac{1}{2r^2 }\left[\tilde{\Upsilon}-2\ell^2\partial_{\phi}\left(\lambda\partial_{\phi}\mathcal{U}\right)\right]+\mathcal{O}(r^{-3})=\frac{1}{2r^2 }\left[\tilde{\Upsilon}+2\ell\hat{k} \cdot \hat{\nabla} \hat{\theta}_{t}\right]+\mathcal{O}(r^{-3})\, , \\
        {\tilde{\kappa}_{t}-\tilde\theta_{s}+\frac{2}{\ell}=}&{\frac{\ell\, \lambda }{r^2 }\left[\ell^2 \mathcal{D}_{v}\left(\lambda\partial_{\phi}\mathcal{U}\right)-\partial_{\phi}\left(\frac{\partial_{\phi}\lambda}{\lambda^2}\right)\right]+\mathcal{O}(r^{-3})}\\
        =&{\frac{\ell}{r^2 }\left[-\hat{t}\cdot \hat\nabla \hat\theta_t-\hat{\theta}_{t}^2+\hat{k}\cdot \hat\nabla\hat\theta_k+\hat\theta_k^2\right]+\mathcal{O}(r^{-3})}
        =-\frac{\ell}{2r^2}\hat{R}+\mathcal{O}(r^{-3})\, .
    \end{split}
\end{equation}
where $\hat{R}$ is the Ricci scalar of boundary metric \eqref{rel:metric-asy}, given in \eqref{R-hat}, and \eqref{hat-theta}  which yields as ${\hat{\theta}}_{t}=-\ell\lambda\partial_\phi{\cal U} ,\ {\hat{\theta}}_{k}=-\lambda^{-1}{\partial_\phi\lambda}$  
and 
$$\ell\lambda\mathcal{D}_{v}\hat\theta_t=\hat{t}\cdot \hat\nabla \hat\theta_t+\hat{\theta}_{t}^2\, , \hspace{1 cm} \partial_\phi{\hat{\theta}}_{t} =\hat{k}\cdot \hat\nabla{\hat{\theta}}_{t}\, , \hspace{1 cm}\partial_\phi{\hat{\theta}}_{k} =\hat{k}\cdot \hat\nabla{\hat{\theta}}_{k}\, .$$
Recalling \eqref{Brown-York-mu-nu} and \eqref{Tmunu-BY-compt}, and also \eqref{EM-tensor-hattext}
and \eqref{D-F-E-J}, we learn that
\be
r^2{\tilde{\mathcal E}}=\hat{\mathcal E}+\frac{c}{24\pi}(\hat{t}\cdot \hat\nabla \hat\theta_t- \hat{k}\cdot\hat\nabla\hat\theta_k-\hat\theta_k^2)\, ,\qquad r^2\tilde{\mathcal{J}}=\hat{\mathcal{J}}+\frac{c}{12\pi}\hat{k} \cdot \hat{\nabla} \hat{\theta}_{t}\, ,\qquad r^2\tilde{\mathcal{T}}=-\frac{c}{24\pi}\hat{R}\, ,\qquad c:=\frac{3\ell}{2 G}\, .
\ee
We therefore, obtain the divergence-free asymptotic energy-momentum tensor $\hat{\text{T}}^{ab}$ \eqref{EM-tensor-hattext}
and \eqref{D-F-E-J} in terms of the energy-momentum tensor ${\mathcal T}^{ab}$ \eqref{Brown-York-mu-nu} in the large $r$ limit, 
\begin{equation}
   \boxed{ \hat{\text{T}}^{ab}=\lim_{r\to \infty}(r^2\tilde{\mathcal{T}}^{ab})\, . }
\end{equation}
A similar analysis may be carried out for the null surface and the asymptotic boundary in flat space $\mathcal{I}$. This will relate to the energy-momentum tensors discussed in section \ref{sec:NBH-finite-r} and \ref{Hydrodynamics at infinity:flat-case}.

\addcontentsline{toc}{section}{References}
\bibliographystyle{fullsort.bst}
\bibliography{reference}

\begin{thebibliography}{10}

\bibitem{Adami:2021kvx}
H.~Adami, M.~M. Sheikh-Jabbari, V.~Taghiloo, and H.~Yavartanoo, ``{Null surface
  thermodynamics},'' {\em Phys. Rev. D} {\bf 105} (2022), no.~6, 066004,
  \href{http://www.arXiv.org/abs/2110.04224}{{\tt 2110.04224}}.

\bibitem{Landau:1953gs}
L.~D. Landau, ``{On the multiparticle production in high-energy collisions},''
  {\em Izv. Akad. Nauk Ser. Fiz.} {\bf 17} (1953) 51--64.

\bibitem{Maldacena:1997re}
J.~M. Maldacena, ``{The large $N$ limit of superconformal field theories and
  supergravity},'' {\em Adv. Theor. Math. Phys.} {\bf 2} (1998) 231--252,
\href{http://www.arXiv.org/abs/hep-th/9711200}{{\tt hep-th/9711200}}.

\bibitem{Aharony:1999ti}
O.~Aharony, S.~S. Gubser, J.~M. Maldacena, H.~Ooguri, and Y.~Oz, ``{Large N
  field theories, string theory and gravity},'' {\em Phys. Rept.} {\bf 323}
  (2000) 183--386,
\href{http://www.arXiv.org/abs/hep-th/9905111}{{\tt hep-th/9905111}}.

\bibitem{Bhattacharyya:2007vjd}
S.~Bhattacharyya, V.~E. Hubeny, S.~Minwalla, and M.~Rangamani, ``{Nonlinear
  Fluid Dynamics from Gravity},'' {\em JHEP} {\bf 02} (2008) 045,
  \href{http://www.arXiv.org/abs/0712.2456}{{\tt 0712.2456}}.

\bibitem{Hubeny:2011hd}
V.~E. Hubeny, S.~Minwalla, and M.~Rangamani, ``{The fluid/gravity
  correspondence},'' in {\em {Theoretical Advanced Study Institute in
  Elementary Particle Physics}: {String theory and its Applications: From meV
  to the Planck Scale}}, pp.~348--383.
\newblock 2012.
\newblock \href{http://www.arXiv.org/abs/1107.5780}{{\tt 1107.5780}}.

\bibitem{Bhattacharyya:2008mz}
S.~Bhattacharyya, R.~Loganayagam, I.~Mandal, S.~Minwalla, and A.~Sharma,
  ``{Conformal Nonlinear Fluid Dynamics from Gravity in Arbitrary
  Dimensions},'' {\em JHEP} {\bf 12} (2008) 116,
  \href{http://www.arXiv.org/abs/0809.4272}{{\tt 0809.4272}}.

\bibitem{Banerjee:2013qha}
R.~Banerjee, ``{Exact results in two dimensional chiral hydrodynamics with
  gravitational anomalies},'' {\em Eur. Phys. J. C} {\bf 74} (2014), no.~4,
  2824, \href{http://www.arXiv.org/abs/1303.5593}{{\tt 1303.5593}}.

\bibitem{Haack:2008cp}
M.~Haack and A.~Yarom, ``{Nonlinear viscous hydrodynamics in various dimensions
  using AdS/CFT},'' {\em JHEP} {\bf 10} (2008) 063,
  \href{http://www.arXiv.org/abs/0806.4602}{{\tt 0806.4602}}.

\bibitem{Campoleoni:2018ltl}
A.~Campoleoni, L.~Ciambelli, C.~Marteau, P.~M. Petropoulos, and K.~Siampos,
  ``{Two-dimensional fluids and their holographic duals},'' {\em Nucl. Phys. B}
  {\bf 946} (2019) 114692, \href{http://www.arXiv.org/abs/1812.04019}{{\tt
  1812.04019}}.

\bibitem{Ciambelli:2020ftk}
L.~Ciambelli, C.~Marteau, P.~M. Petropoulos, and R.~Ruzziconi,
  ``{Fefferman-Graham and Bondi Gauges in the Fluid/Gravity Correspondence},''
  {\em PoS} {\bf CORFU2019} (2020) 154,
  \href{http://www.arXiv.org/abs/2006.10083}{{\tt 2006.10083}}.

\bibitem{Ciambelli:2020eba}
L.~Ciambelli, C.~Marteau, P.~M. Petropoulos, and R.~Ruzziconi, ``{Gauges in
  Three-Dimensional Gravity and Holographic Fluids},'' {\em JHEP} {\bf 11}
  (2020) 092, \href{http://www.arXiv.org/abs/2006.10082}{{\tt 2006.10082}}.

\bibitem{Penna:2017vms}
R.~F. Penna, ``{BMS$_3$ invariant fluid dynamics at null infinity},'' {\em
  Class. Quant. Grav.} {\bf 35} (2018), no.~4, 044002,
  \href{http://www.arXiv.org/abs/1708.08470}{{\tt 1708.08470}}.

\bibitem{Penna:2017bdn}
R.~F. Penna, ``{Near-horizon BMS symmetries as fluid symmetries},'' {\em JHEP}
  {\bf 10} (2017) 049,
\href{http://www.arXiv.org/abs/1703.07382}{{\tt 1703.07382}}.

\bibitem{Ciambelli:2018xat}
L.~Ciambelli, C.~Marteau, A.~C. Petkou, P.~M. Petropoulos, and K.~Siampos,
  ``{Covariant Galilean versus Carrollian hydrodynamics from relativistic
  fluids},'' {\em Class. Quant. Grav.} {\bf 35} (2018), no.~16, 165001,
  \href{http://www.arXiv.org/abs/1802.05286}{{\tt 1802.05286}}.

\bibitem{Ciambelli:2018wre}
L.~Ciambelli, C.~Marteau, A.~C. Petkou, P.~M. Petropoulos, and K.~Siampos,
  ``{Flat holography and Carrollian fluids},'' {\em JHEP} {\bf 07} (2018) 165,
  \href{http://www.arXiv.org/abs/1802.06809}{{\tt 1802.06809}}.

\bibitem{Ruzziconi:2019pzd}
R.~Ruzziconi, ``{Asymptotic Symmetries in the Gauge Fixing Approach and the BMS
  Group},'' {\em PoS} {\bf Modave2019} (2020) 003,
  \href{http://www.arXiv.org/abs/1910.08367}{{\tt 1910.08367}}.

\bibitem{Freidel:2022vjq}
L.~Freidel and P.~Jai-akson, ``{Carrollian hydrodynamics and symplectic
  structure on stretched horizons},''
  \href{http://www.arXiv.org/abs/2211.06415}{{\tt 2211.06415}}.

\bibitem{Brown:1986nw}
J.~D. Brown and M.~Henneaux, ``{Central Charges in the Canonical Realization of
  Asymptotic Symmetries: An Example from Three-Dimensional Gravity},'' {\em
  Commun. Math. Phys.} {\bf 104} (1986)
207--226.

\bibitem{Adami:2020ugu}
H.~Adami, M.~M. Sheikh-Jabbari, V.~Taghiloo, H.~Yavartanoo, and C.~Zwikel,
  ``{Symmetries at null boundaries: two and three dimensional gravity cases},''
  {\em JHEP} {\bf 10} (2020) 107,
  \href{http://www.arXiv.org/abs/2007.12759}{{\tt 2007.12759}}.

\bibitem{Adami:2022ktn}
H.~Adami, P.~Mao, M.~M. Sheikh-Jabbari, V.~Taghiloo, and H.~Yavartanoo,
  ``{Symmetries at causal boundaries in 2D and 3D gravity},'' {\em JHEP} {\bf
  05} (2022) 189, \href{http://www.arXiv.org/abs/2202.12129}{{\tt 2202.12129}}.

\bibitem{Geiller:2021vpg}
M.~Geiller, C.~Goeller, and C.~Zwikel, ``{3d gravity in Bondi-Weyl gauge:
  charges, corners, and integrability},'' {\em JHEP} {\bf 09} (2021) 029,
  \href{http://www.arXiv.org/abs/2107.01073}{{\tt 2107.01073}}.

\bibitem{Sheikh-Jabbari:2022mqi}
M.~M. Sheikh-Jabbari, ``{On symplectic form for null boundary phase space},''
  {\em Gen. Rel. Grav.} {\bf 54} (2022), no.~11, 140,
  \href{http://www.arXiv.org/abs/2209.05043}{{\tt 2209.05043}}.

\bibitem{Grumiller:2022qhx}
D.~Grumiller and M.~M. Sheikh-Jabbari, {\em {Black Hole Physics: From Collapse
  to Evaporation}}.
\newblock Grad.Texts Math. Springer, 11, 2022.

\bibitem{Duval:2014uoa}
C.~Duval, G.~Gibbons, P.~Horvathy, and P.~Zhang, ``{Carroll versus Newton and
  Galilei: two dual non-Einsteinian concepts of time},''
\href{http://www.arXiv.org/abs/1402.0657}{{\tt 1402.0657}}.

\bibitem{Duval:2014uva}
C.~Duval, G.~Gibbons, and P.~Horvathy, ``{Conformal Carroll groups and BMS
  symmetry},''
\href{http://www.arXiv.org/abs/1402.5894}{{\tt 1402.5894}}.

\bibitem{Duval:2014lpa}
C.~Duval, G.~Gibbons, and P.~Horvathy, ``{Conformal Carroll groups},''
\href{http://www.arXiv.org/abs/1403.4213}{{\tt 1403.4213}}.

\bibitem{Henneaux:1979vn}
M.~Henneaux, ``{Geometry of Zero Signature Space-times},'' {\em Bull. Soc.
  Math. Belg.} {\bf 31} (1979) 47--63.

\bibitem{Ciambelli:2019lap}
L.~Ciambelli, R.~G. Leigh, C.~Marteau, and P.~M. Petropoulos, ``{Carroll
  Structures, Null Geometry and Conformal Isometries},'' {\em Phys. Rev. D}
  {\bf 100} (2019), no.~4, 046010,
  \href{http://www.arXiv.org/abs/1905.02221}{{\tt 1905.02221}}.

\bibitem{Bagchi:2022iqb}
A.~Bagchi, D.~Grumiller, and M.~M. Sheikh-Jabbari, ``{Horizon Strings as 3d
  Black Hole Microstates},'' \href{http://www.arXiv.org/abs/2210.10794}{{\tt
  2210.10794}}.

\bibitem{Adami:2021sko}
H.~Adami, M.~M. Sheikh-Jabbari, V.~Taghiloo, H.~Yavartanoo, and C.~Zwikel,
  ``{Chiral Massive News: Null Boundary Symmetries in Topologically Massive
  Gravity},'' {\em JHEP} {\bf 05} (2021) 261,
  \href{http://www.arXiv.org/abs/2104.03992}{{\tt 2104.03992}}.

\bibitem{Taghiloo:2022kmh}
V.~Taghiloo, H.~Adami, M.~M. Sheikh-Jabbari, H.~Yavartanoo, and C.~Zwikel,
  ``{Symmetries at Null Boundaries: 3-dimensional Einstein gravity},'' {\em
  PoS} {\bf Regio2021} (2022) 008.

\bibitem{Ciambelli:2022vot}
L.~Ciambelli, ``{From Asymptotic Symmetries to the Corner Proposal},''
  \href{http://www.arXiv.org/abs/2212.13644}{{\tt 2212.13644}}.

\bibitem{Balasubramanian:1999re}
V.~Balasubramanian and P.~Kraus, ``A stress tensor for anti-de {S}itter
  gravity,'' {\em Commun. Math. Phys.} {\bf 208} (1999) 413--428,
\href{http://www.arXiv.org/abs/hep-th/9902121}{{\tt hep-th/9902121}}.

\bibitem{York:1972sj}
J.~W. York, Jr., ``Role of conformal three geometry in the dynamics of
  gravitation,'' {\em Phys. Rev. Lett.} {\bf 28} (1972)
1082--1085.

\bibitem{Gibbons:1976ue}
G.~W. Gibbons and S.~W. Hawking, ``Action integrals and partition functions in
  quantum gravity,'' {\em Phys. Rev.} {\bf D15} (1977)
2752--2756.

\bibitem{Wald:1984rg}
R.~M. Wald, {\em {General Relativity}}.
\newblock Chicago Univ. Pr., Chicago, USA, 1984.

\bibitem{Chandrasekaran:2021hxc}
V.~Chandrasekaran, E.~E. Flanagan, I.~Shehzad, and A.~J. Speranza,
  ``{Brown-York charges at null boundaries},''
  \href{http://www.arXiv.org/abs/2109.11567}{{\tt 2109.11567}}.

\bibitem{Alessio:2020ioh}
F.~Alessio, G.~Barnich, L.~Ciambelli, P.~Mao, and R.~Ruzziconi, ``{Weyl charges
  in asymptotically locally AdS$_3$ spacetimes},'' {\em Phys. Rev. D} {\bf 103}
  (2021), no.~4, 046003, \href{http://www.arXiv.org/abs/2010.15452}{{\tt
  2010.15452}}.

\bibitem{Henneaux:2021yzg}
M.~Henneaux and P.~Salgado-Rebolledo, ``{Carroll contractions of
  Lorentz-invariant theories},'' {\em JHEP} {\bf 11} (2021) 180,
  \href{http://www.arXiv.org/abs/2109.06708}{{\tt 2109.06708}}.

\bibitem{Bagchi:2022eav}
A.~Bagchi, A.~Banerjee, S.~Dutta, K.~S. Kolekar, and P.~Sharma, ``{Carroll
  covariant scalar fields in two dimensions},''
  \href{http://www.arXiv.org/abs/2203.13197}{{\tt 2203.13197}}.

\bibitem{Campoleoni:2022ebj}
A.~Campoleoni, M.~Henneaux, S.~Pekar, A.~P\'erez, and P.~Salgado-Rebolledo,
  ``{Magnetic Carrollian gravity from the Carroll algebra},'' {\em JHEP} {\bf
  09} (2022) 127, \href{http://www.arXiv.org/abs/2207.14167}{{\tt 2207.14167}}.

\bibitem{Bekaert:2015xua}
X.~Bekaert and K.~Morand, ``{Connections and dynamical trajectories in
  generalised Newton-Cartan gravity II. An ambient perspective},'' {\em J.
  Math. Phys.} {\bf 59} (2018), no.~7, 072503,
  \href{http://www.arXiv.org/abs/1505.03739}{{\tt 1505.03739}}.

\bibitem{Mars:1993mj}
M.~Mars and J.~M.~M. Senovilla, ``{Geometry of general hypersurfaces in
  space-time: Junction conditions},'' {\em Class. Quant. Grav.} {\bf 10} (1993)
  1865--1897, \href{http://www.arXiv.org/abs/gr-qc/0201054}{{\tt
  gr-qc/0201054}}.

\bibitem{Donnay:2019jiz}
L.~Donnay and C.~Marteau, ``{Carrollian Physics at the Black Hole Horizon},''
\href{http://www.arXiv.org/abs/1903.09654}{{\tt 1903.09654}}.

\bibitem{Bagchi:2023ysc}
A.~Bagchi, K.~S. Kolekar, and A.~Shukla, ``{Carrollian Origins of Bjorken
  Flow},'' \href{http://www.arXiv.org/abs/2302.03053}{{\tt 2302.03053}}.

\bibitem{Adami:2021nnf}
H.~Adami, D.~Grumiller, M.~M. Sheikh-Jabbari, V.~Taghiloo, H.~Yavartanoo, and
  C.~Zwikel, ``{Null boundary phase space: slicings, news \& memory},'' {\em
  JHEP} {\bf 11} (2021) 155, \href{http://www.arXiv.org/abs/2110.04218}{{\tt
  2110.04218}}.

\bibitem{Sheikh-Jabbari:2022xix}
S.~Sheikh-Jabbari, H.~Adami, V.~Taghiloo, and H.~Yavartanoo, ``{Null Surface
  Thermodynamics},'' {\em PoS} {\bf Regio2021} (2022) 034.

\bibitem{Fiorucci:2020xto}
A.~Fiorucci and R.~Ruzziconi, ``{Charge Algebra in Al(A)dS$_n$ Spacetimes},''
  \href{http://www.arXiv.org/abs/2011.02002}{{\tt 2011.02002}}.

\bibitem{Ruzziconi:2020wrb}
R.~Ruzziconi and C.~Zwikel, ``{Conservation and Integrability in
  Lower-Dimensional Gravity},'' {\em JHEP} {\bf 04} (2021) 034,
  \href{http://www.arXiv.org/abs/2012.03961}{{\tt 2012.03961}}.

\bibitem{Troessaert:2013fma}
C.~Troessaert, ``{Enhanced asymptotic symmetry algebra of $AdS$$_{3}$},'' {\em
  JHEP} {\bf 08} (2013) 044,
\href{http://www.arXiv.org/abs/1303.3296}{{\tt 1303.3296}}.

\bibitem{Campoleoni:2022wmf}
A.~Campoleoni, L.~Ciambelli, A.~Delfante, C.~Marteau, P.~M. Petropoulos, and
  R.~Ruzziconi, ``{Holographic Lorentz and Carroll frames},'' {\em JHEP} {\bf
  12} (2022) 007, \href{http://www.arXiv.org/abs/2208.07575}{{\tt 2208.07575}}.

\bibitem{Barnich:2012aw}
G.~Barnich, A.~Gomberoff, and H.~A. Gonzalez, ``{The Flat limit of three
  dimensional asymptotically anti-de Sitter spacetimes},'' {\em Phys.Rev.} {\bf
  D86} (2012) 024020,
\href{http://www.arXiv.org/abs/1204.3288}{{\tt 1204.3288}}.

\bibitem{Barnich:2013yka}
G.~Barnich and H.~A. Gonzalez, ``{Dual dynamics of three dimensional
  asymptotically flat Einstein gravity at null infinity},'' {\em JHEP} {\bf
  1305} (2013) 016,
\href{http://www.arXiv.org/abs/1303.1075}{{\tt 1303.1075}}.

\bibitem{Barnich:2006av}
G.~Barnich and G.~Compere, ``{Classical central extension for asymptotic
  symmetries at null infinity in three spacetime dimensions},'' {\em
  Class.Quant.Grav.} {\bf 24} (2007) F15--F23,
\href{http://www.arXiv.org/abs/gr-qc/0610130}{{\tt gr-qc/0610130}}.

\bibitem{Bagchi:2012cy}
A.~Bagchi and R.~Fareghbal, ``{BMS/GCA Redux: Towards Flatspace Holography from
  Non-Relativistic Symmetries},'' {\em JHEP} {\bf 1210} (2012) 092,
\href{http://www.arXiv.org/abs/1203.5795}{{\tt 1203.5795}}.

\bibitem{Taghiloo:2022hxc}
V.~Taghiloo, ``{Null surface thermodynamics in topologically massive
  gravity},'' {\em Eur. Phys. J. C} {\bf 83} (2023), no.~2, 182,
  \href{http://www.arXiv.org/abs/2205.10909}{{\tt 2205.10909}}.

\bibitem{Iyer:1994ys}
V.~Iyer and R.~M. Wald, ``Some properties of {N}{\"o}ther charge and a proposal
  for dynamical black hole entropy,'' {\em Phys. Rev.} {\bf D50} (1994)
  846--864,
\href{http://arXiv.org/abs/gr-qc/9403028}{{\tt gr-qc/9403028}}.

\bibitem{Wald:1993nt}
R.~M. Wald, ``Black hole entropy is the {N}{\"o}ther charge,'' {\em Phys. Rev.}
  {\bf D48} (1993) 3427--3431,
\href{http://arXiv.org/abs/gr-qc/9307038}{{\tt gr-qc/9307038}}.

\bibitem{Hajian:2015xlp}
K.~Hajian and M.~M. Sheikh-Jabbari, ``{Solution Phase Space and Conserved
  Charges: A General Formulation for Charges Associated with Exact
  Symmetries},'' {\em Phys. Rev. D} {\bf 93} (2016), no.~4, 044074,
  \href{http://www.arXiv.org/abs/1512.05584}{{\tt 1512.05584}}.

\bibitem{Hajian:2022lgy}
K.~Hajian, M.~M. Sheikh-Jabbari, and B.~Tekin, ``{Gauge invariant derivation of
  zeroth and first laws of black hole thermodynamics},'' {\em Phys. Rev. D}
  {\bf 106} (2022), no.~10, 104030,
  \href{http://www.arXiv.org/abs/2209.00563}{{\tt 2209.00563}}.

\bibitem{Chandrasekaran:2018aop}
V.~Chandrasekaran, {\'E}.~{\'E}. Flanagan, and K.~Prabhu, ``{Symmetries and
  charges of general relativity at null boundaries},'' {\em JHEP} {\bf 11}
  (2018) 125,
\href{http://www.arXiv.org/abs/1807.11499}{{\tt 1807.11499}}.

\bibitem{Ashtekar:2021wld}
A.~Ashtekar, N.~Khera, M.~Kolanowski, and J.~Lewandowski, ``{Non-Expanding
  Horizons: Multipoles and the Symmetry Group},'' {\em JHEP} {\bf 01} (2022)
  028, \href{http://www.arXiv.org/abs/2111.07873}{{\tt 2111.07873}}.

\bibitem{Damour:1978cg}
T.~Damour, ``{Black Hole Eddy Currents},'' {\em Phys. Rev. D} {\bf 18} (1978)
  3598--3604.

\bibitem{Thorne:1986iy}
K.~S. Thorne, R.~Price, and D.~Macdonald, {\em Black Holes: The Membrane
  Paradigm}.
\newblock Yale University Press,
1986.
\newblock

\bibitem{Price:1986yy}
R.~H. Price and K.~S. Thorne, ``{Membrane Viewpoint on Black Holes: Properties
  and Evolution of the Stretched Horizon},'' {\em Phys. Rev.} {\bf D33} (1986)
915--941.

\bibitem{Iqbal:2008by}
N.~Iqbal and H.~Liu, ``{Universality of the hydrodynamic limit in AdS/CFT and
  the membrane paradigm},'' {\em Phys.Rev.} {\bf D79} (2009) 025023,
\href{http://www.arXiv.org/abs/0809.3808}{{\tt 0809.3808}}.

\bibitem{Eling:2009pb}
C.~Eling, I.~Fouxon, and Y.~Oz, ``{The Incompressible Navier-Stokes Equations
  from Membrane Dynamics},'' {\em Phys. Lett. B} {\bf 680} (2009) 496--499,
  \href{http://www.arXiv.org/abs/0905.3638}{{\tt 0905.3638}}.

\bibitem{Nickel:2010pr}
D.~Nickel and D.~T. Son, ``{Deconstructing Holographic Liquids},'' {\em New J.
  Phys.} {\bf 13} (2011) 075010, \href{http://www.arXiv.org/abs/1009.3094}{{\tt
  1009.3094}}.

\bibitem{Maloney-Witten}
A.~Maloney and E.~Witten, ``{Quantum Gravity Partition Functions in Three
  Dimensions},'' {\em JHEP} {\bf 1002} (2010) 029,
\href{http://www.arXiv.org/abs/0712.0155}{{\tt 0712.0155}}.

\bibitem{Jacobson:1995ab}
T.~Jacobson, ``{Thermodynamics of space-time: The Einstein equation of
  state},'' {\em Phys. Rev. Lett.} {\bf 75} (1995) 1260--1263,
\href{http://www.arXiv.org/abs/gr-qc/9504004}{{\tt gr-qc/9504004}}.

\bibitem{Deser:1981wh}
S.~Deser, R.~Jackiw, and S.~Templeton, ``{Topologically Massive Gauge
  Theories},'' {\em Annals Phys.} {\bf 140} (1982)
372--411.

\end{thebibliography}

\providecommand{\href}[2]{#2}\begingroup\raggedright\endgroup

\end{document}